%% file: sample631.tex
\documentclass[twocolumn]{aastex631}

\usepackage{CJK}
\usepackage{bm} 
\usepackage{amsmath}
\usepackage{multirow}
\newcommand*\diff{\mathop{}\!\mathrm{d}}
\newcommand{\Secref}[1]{\hyperref[#1]{Section~\ref*{#1}}}
\newcommand{\Appref}[1]{\hyperref[#1]{Appendix~\ref*{#1}}}

\shorttitle{Direct spectroscopy with JWST NIRSpec IFU}
\shortauthors{Ruffio et al.}

\graphicspath{{./}{figures/}}
\submitjournal{AJ}
\accepted{May 22, 2024}

\begin{document}
\begin{CJK*}{UTF8}{gbsn}

% \title{The transformative capability of JWST NIRSpec for direct spectroscopy of high contrast exoplanets: application to the brown dwarf HD 19467 B}
% \title{JWST-TST High Contrast: Direct spectroscopy of high contrast companions with the JWST NIRSpec IFU: application to the brown dwarf HD 19467 B}
\title{JWST-TST High Contrast: Achieving direct spectroscopy of faint substellar companions next to bright stars with the NIRSpec IFU}

\newcommand{\petit}{\texttt{petitRADTRANS}}
\newcommand{\emcee}{\texttt{emcee}}
\newcommand{\pmn}{\texttt{PyMultiNest}}
\newcommand{\moog}{\texttt{MOOG}}
\newcommand{\teff}{T$_{\rm eff}$}
\newcommand{\feh}{\rm{[Fe/H]}}

\newcommand{\caltech}{Department of Astronomy, California Institute of Technology, Pasadena, CA 91125, USA}
\newcommand{\gps}{Division of Geological \& Planetary Sciences, California Institute of Technology, Pasadena, CA 91125, USA}
\newcommand{\ucsc}{Department of Astronomy \& Astrophysics, University of California, Santa Cruz, CA95064, USA}
\newcommand{\keck}{W. M. Keck Observatory, 65-1120 Mamalahoa Hwy, Kamuela, HI, USA}
\newcommand{\ucla}{Department of Physics \& Astronomy, 430 Portola Plaza, University of California, Los Angeles, CA 90095, USA}
\newcommand{\jpl}{Jet Propulsion Laboratory, California Institute of Technology, 4800 Oak Grove Dr.,Pasadena, CA 91109, USA}
\newcommand{\ucsd}{Department of Astronomy \& Astrophysics,  University of California, San Diego, La Jolla, CA 92093, USA}
\newcommand{\ifahonolulu}{Institute for Astronomy, University of Hawai`i, 2680 Woodlawn Drive, Honolulu, HI 96822, USA}
\newcommand{\berkeley}{Department of Astronomy, University of California at Berkeley, CA 94720, USA}
\newcommand{\ames}{NASA Ames Research Centre, MS 245-3, Moffett Field, CA 94035, USA}
\newcommand{\stsci}{Space Telescope Science Institute, 3700 San Martin Drive, Baltimore, MD 21218, USA}

%%%%%%%%%%%%%%%%%
%Main project co-authors and general TST high contrast collaborators

\correspondingauthor{Jean-Baptiste Ruffio}
\email{jruffio@ucsd.edu}

\author[0000-0003-2233-4821]{Jean-Baptiste Ruffio}
\affiliation{\ucsd}

\author[0000-0002-3191-8151]{Marshall D. Perrin}
\affiliation{\stsci}

\author[0000-0002-9803-8255]{Kielan K. W. Hoch}
\affiliation{\stsci}

\author[0000-0003-2769-0438]{Jens Kammerer}
\affiliation{European Southern Observatory, Karl-Schwarzschild-Str. 2, 85748 Garching, Germany}
\affiliation{\stsci}

\author[0000-0002-9936-6285]{Quinn M. Konopacky}
\affiliation{\ucsd}

\author{Laurent Pueyo}
\affiliation{\stsci}

\author[0000-0003-2429-5811]{Alex Madurowicz}
\affiliation{\ucsc}

\author[0000-0003-4203-9715]{Emily Rickman}
\affiliation{European Space Agency (ESA), ESA Office, Space Telescope Science Institute, 3700 San Martin Dr, Baltimore, MD 21218, USA}

\author[0000-0002-9807-5435]{Christopher A. Theissen}
\affiliation{Department of Astronomy \& Astrophysics, University of California, San Diego, La Jolla, CA 92093, USA}

\author[0000-0003-2429-5811]{Shubh Agrawal}
\affiliation{Department of Physics and Astronomy, University of Pennsylvania, Philadelphia, PA 19104, USA}

\author[0000-0002-7162-8036]{Alexandra Z. Greenbaum}
\affiliation{IPAC, Mail Code 100-22, California Institute of Technology, 1200 E. California Blvd., Pasadena, CA 91125, USA}

\author[0000-0002-5500-4602]{Brittany E. Miles}
\affiliation{Steward Observatory, University of Arizona, 933 N. Cherry Ave., Tucson, AZ 85721, USA}
\altaffiliation{51 Pegasi b Fellow}

\author[0000-0002-7129-3002]{Travis S. Barman}
\affiliation{Lunar and Planetary Laboratory, University of Arizona, Tucson, AZ 85721, USA}

\author[0000-0001-6396-8439]{William O. Balmer}
\affiliation{William H. Miller III Department of Physics and Astronomy, Johns Hopkins University, Baltimore, MD 21218, USA}
\affiliation{\stsci}

\author[0000-0002-3414-784X]{Jorge~Llop-Sayson}
\affiliation{California Institute of Technology, 1200 E California Blvd., Pasadena, CA 91125}

\author[0000-0001-8627-0404]{Julien H. Girard} %Coronagraphy	Subject
\affiliation{\stsci}

\author[0000-0002-4388-6417]{Isabel Rebollido} %Coronagraphy	Project	
\affiliation{Centro de Astrobiolog\'{i}a (CAB, CSIC-INTA), ESAC Campus Camino Bajo del Castillo, s/n, Villanueva de la Ca\~{n}ada, E-28692 Madrid, Spain} %

\author[0000-0003-2753-2819]{R\'emi Soummer} %Coronagraphy	Subject
\affiliation{\stsci}

%%%%%%%%%%%%%%%%%
%Alphabetical: Other GTO (TST and NIRCam) collaborators

\author[0000-0002-0832-710X]{Natalie H. Allen} %Transits	Project
\affiliation{William H. Miller III Department of Physics and Astronomy, Johns Hopkins University, Baltimore, MD 21218, USA}
\altaffiliation{NSF Graduate Research Fellow}

\author[0000-0003-2861-3995]{Jay Anderson} %Proper Motions	Program
\affiliation{\stsci}

\author[0000-0002-5627-5471]{Charles A. Beichman}
\affiliation{NASA Exoplanet Science Institute/IPAC, Jet Propulsion Laboratory, California Institute of Technology, 1200 E California Boulevard, Pasadena, CA 91125, USA}

\author[0000-0003-3858-637X]{Andrea Bellini} %Proper Motions	Subject
\affiliation{\stsci}

\author{Geoffrey Bryden}
\affiliation{\jpl}

\author[0000-0001-9513-1449]{N\'estor Espinoza} %Transits	Subject	
\affiliation{\stsci}
\affiliation{William H. Miller III Department of Physics and Astronomy, Johns Hopkins University, Baltimore, MD 21218, USA}

\author[0000-0002-5322-2315]{Ana Glidden} %Transits	Project
\affiliation{Department of Earth, Atmospheric, and Planetary Sciences, Massachusetts Institute of Technology, Cambridge, MA 02139, USA}
\affiliation{Kavli Institute for Astrophysics and Space Research, Massachusetts Institute of Technology, Cambridge, MA 02139, USA}

\author[0000-0001-5732-8531]{Jingcheng Huang} % Transits	Project
\affiliation{Department of Earth, Atmospheric, and Planetary Sciences, Massachusetts Institute of Technology, Cambridge, MA 02139, USA}

\author[0000-0002-8507-1304]{Nikole K. Lewis} %Transits	Program
\affiliation{Department of Astronomy and Carl Sagan Institute, Cornell University, 122 Sciences Drive, Ithaca, NY 14853, USA}

\author[0000-0001-9673-7397]{Mattia Libralato} % Proper Motions	Subject
\affiliation{AURA for the European Space Agency (ESA), Space Telescope Science Institute, 3700 San Martin Drive, Baltimore, MD 21218, USA}

\author[0000-0002-2457-272X]{Dana R. Louie} % Transits	Project
\altaffiliation{NASA Postdoctoral Program Fellow}
\affiliation{NASA Goddard Space Flight Center, Greenbelt, MD 20771, USA}

\author[0000-0001-8368-0221]{Sangmo Tony Sohn} % Proper Motions	Subject
\affiliation{\stsci}

\author[0000-0002-6892-6948]{Sara Seager} % Transits	Program
\affiliation{Department of Physics and Kavli Institute for Astrophysics and Space Research, Massachusetts Institute of Technology, Cambridge, MA 02139, USA}
\affiliation{Department of Earth, Atmospheric, and Planetary Sciences, Massachusetts Institute of Technology, Cambridge, MA 02139, US}
\affiliation{Department of Aeronautics and Astronautics, MIT, 77 Massachusetts Avenue, Cambridge, MA 02139, USA}

\author[0000-0001-7827-7825]{Roeland P. van der Marel} %Proper Motions	Program
\affiliation{\stsci}
\affiliation{William H. Miller III Department of Physics and Astronomy, Johns Hopkins University, Baltimore, MD 21218, USA}

\author[0000-0003-4328-3867]{Hannah R. Wakeford} %Transits	Subject
\affiliation{University of Bristol, HH Wills Physics Laboratory, Tyndall Avenue, Bristol, UK}

\author[0000-0002-1343-134X]{Laura L. Watkins} % Proper Motions	Subject
\affiliation{AURA for the European Space Agency (ESA), Space Telescope Science Institute, 3700 San Martin Drive, Baltimore, MD 21218, USA}

\author[0000-0001-7591-2731]{Marie~Ygouf}
\affiliation{\jpl}

\author{C. Matt Mountain}
\affiliation{Association of Universities for Research in Astronomy, 1331 Pennsylvania Avenue NW Suite 1475, Washington, DC 20004, USA}

%% Note that the \and command from previous versions of AASTeX is now
%% depreciated in this version as it is no longer necessary. AASTeX 
%% automatically takes care of all commas and "and"s between authors names.

%% AASTeX 6.31 has the new \collaboration and \nocollaboration commands to
%% provide the collaboration status of a group of authors. These commands 
%% can be used either before or after the list of corresponding authors. The
%% argument for \collaboration is the collaboration identifier. Authors are
%% encouraged to surround collaboration identifiers with ()s. The 
%% \nocollaboration command takes no argument and exists to indicate that
%% the nearby authors are not part of surrounding collaborations.

%% Mark off the abstract in the ``abstract'' environment. 
\begin{abstract}
The JWST NIRSpec integral field unit (IFU) presents a unique opportunity to observe directly imaged exoplanets from $3-5\,\mu\mathrm{m}$ at moderate spectral resolution ($R\sim2,700$) and thereby better constrain the composition, disequilibrium chemistry, and cloud properties of their atmospheres. In this work, we present the first NIRSpec IFU high-contrast observations of a substellar companion that requires starlight suppression techniques.
We develop specific data reduction strategies to study faint companions around bright stars, and assess the performance of NIRSpec at high contrast. 
First, we demonstrate an approach to forward model the companion signal and the starlight directly in the detector images, which mitigates the effects of NIRSpec's spatial undersampling. We demonstrate a sensitivity to planets that are $3\times10^{-6}$ fainter than their stars at $1^{\prime\prime}$, or  $3\times10^{-5}$ at $0.3^{\prime\prime}$.
Then, we implement a reference star point spread function (PSF) subtraction and a spectral extraction that does not require spatially and spectrally regularly sampled spectral cubes. This allows us to extract a moderate resolution ($R\sim2,700$) spectrum of the faint T-dwarf companion HD~19467~B from $2.9-5.2\,\mu\mathrm{m}$ with signal-to-noise ratio (S/N) $\sim10$ per resolution element.
Across this wavelength range, HD~19467~B has a flux ratio varying between $10^{-5}-10^{-4}$ and a separation relative to its star of $1.6^{\prime\prime}$. 
A companion paper by Hoch et al. more deeply analyzes the atmospheric properties of this companion based on the extracted spectrum.
Using the methods developed here, NIRSpec's sensitivity may enable direct detection and spectral characterization of relatively old ($\sim1\,\mathrm{Gyr}$), cool ($\sim250\,\mathrm{K}$), and closely separated ($\sim3-5\,\mathrm{au}$) exoplanets that are less massive than Jupiter.
\end{abstract}

%% Keywords should appear after the \end{abstract} command. 
%% The AAS Journals now uses Unified Astronomy Thesaurus concepts:
%% https://astrothesaurus.org
%% You will be asked to selected these concepts during the submission process
%% but this old "keyword" functionality is maintained in case authors want
%% to include these concepts in their preprints.
% --- Exoplanet detection methods (489)
% 

%% From the front matter, we move on to the body of the paper.
%% Sections are demarcated by \section and \subsection, respectively.
%% Observe the use of the LaTeX \label
%% command after the \subsection to give a symbolic KEY to the
%% subsection for cross-referencing in a \ref command.
%% You can use LaTeX's \ref and \label commands to keep track of
%% cross-references to sections, equations, tables, and figures.
%% That way, if you change the order of any elements, LaTeX will
%% automatically renumber them.
%%
%% We recommend that authors also use the natbib \citep
%% and \citet commands to identify citations.  The citations are
%% tied to the reference list via symbolic KEYs. The KEY corresponds
%% to the KEY in the \bibitem in the reference list below. 

\input{content}

%% IMPORTANT! The old "\acknowledgment" command has be depreciated. It was
%% not robust enough to handle our new dual anonymous review requirements and
%% thus been replaced with the acknowledgment environment. If you try to 
%% compile with \acknowledgment you will get an error print to the screen
%% and in the compiled pdf.

\begin{acknowledgments}
\textbf{Acknowledgments:} This paper reports work carried out in the context of the JWST Telescope Scientist Team (\url{https://www.stsci.edu/~marel/jwsttelsciteam.html}; PI: M. Mountain). Funding is provided to the team by NASA through grant 80NSSC20K0586. Based on observations with the NASA/ESA/CSA JWST, associated with program GTO-1414 (PI: Marshall Perrin), obtained at the Space Telescope Science Institute, which is operated by AURA, Inc., under NASA contract NAS 5-03127.

Material presented in this work is supported by the National Aeronautics and Space Administration under Grants/Contracts/Agreements No. 80NSSC21K0573 (J.-B.R., Q.M.K., T.S.B., and K.K.W.) issued through the Astrophysics Division of the Science Mission Directorate. Any opinions, findings, and conclusions or recommendations expressed in this work are those of the author(s) and do not necessarily reflect the views of the National Aeronautics and Space Administration.
This work benefited from NASA's Nexus for Exoplanet System Science (NExSS) research coordination network sponsored by NASA's Science Mission Directorate.
A.G. acknowledges support from the Robert R Shrock Graduate Fellowship.
D.R.L.'s research activities were supported by an appointment to the NASA Postdoctoral Program at the NASA Goddard Space Flight Center, administered by Oak Ridge Associated Universities under contract with NASA.
N.H.A. acknowledges support by the National Science Foundation Graduate Research Fellowship under Grant No. DGE1746891.
This work was authored by employees of Caltech/IPAC under Contract No. 80GSFC21R0032 with the National Aeronautics and Space Administration.

This research has made use of the SVO Filter Profile Service (\url{http://svo2.cab.inta-csic.es/theory/fps/}) supported from the Spanish MINECO through grant AYA2017-84089.
\end{acknowledgments}

%% To help institutions obtain information on the effectiveness of their 
%% telescopes the AAS Journals has created a group of keywords for telescope 
%% facilities.
%
%% Following the acknowledgments section, use the following syntax and the
%% \facility{} or \facilities{} macros to list the keywords of facilities used 
%% in the research for the paper.  Each keyword is check against the master 
%% list during copy editing.  Individual instruments can be provided in 
%% parentheses, after the keyword, but they are not verified.

\vspace{5mm}
\facilities{JWST (NIRSpec); JWST (NIRCam)}

%% Similar to \facility{}, there is the optional \software command to allow 
%% authors a place to specify which programs were used during the creation of 
%% the manuscript. Authors should list each code and include either a
%% citation or url to the code inside ()s when available.

\software{ \texttt{Python}, \texttt{numpy} \citep{harris2020array}, \texttt{scipy} \citep{2020SciPy-NMeth}, \texttt{Matplotlib}  \citep{Hunter2007}, astropy \citep{Astropy2013}, , \texttt{pandas} \citep{pandas}, \texttt{h5py}\footnote{\url{http://www.h5py.org/}}, \texttt{mkl}, \texttt{PyAstronomy} \citep{pya}, \texttt{astroquery} \citep{Ginsburg2019AJ....157...98G}, \texttt{multiprocessing}, \texttt{whereistheplanet}\footnote{\url{https://www.whereistheplanet.com/}} \citep{whereistheplanet}, \texttt{jwst} \citep{Bushouse2023},  \texttt{webbpsf} \citep{Perrin2012SPIE.8442E..3DP,Perrin2014SPIE.9143E..3XP}, \texttt{stdatamodels}, \texttt{SpaceKLIP} \citep{Kammerer2022SPIE12180E..3NK}, BREADS\footnote{\url{https://github.com/jruffio/breads}} \citep{Ruffio2021AJ....162..290R,Agrawal2023AJ....166...15A}, ChatGPT 3.5\footnote{\url{https://chat.openai.com}} (for minor coding suggestions)}

%% Appendix material should be preceded with a single \appendix command.
%% There should be a \section command for each appendix. Mark appendix
%% subsections with the same markup you use in the main body of the paper.

%% Each Appendix (indicated with \section) will be lettered A, B, C, etc.
%% The equation counter will reset when it encounters the \appendix
%% command and will number appendix equations (A1), (A2), etc. The
%% Figure and Table counter will not reset.

%\appendix
\input{appendix}

%% For this sample we use BibTeX plus aasjournals.bst to generate the
%% the bibliography. The sample631.bib file was populated from ADS. To
%% get the citations to show in the compiled file do the following:
%%
%% pdflatex sample631.tex
%% bibtext sample631
%% pdflatex sample631.tex
%% pdflatex sample631.tex

\bibliography{bibliography}{}
\bibliographystyle{aasjournal}

%% This command is needed to show the entire author+affiliation list when
%% the collaboration and author truncation commands are used.  It has to
%% go at the end of the manuscript.
%\allauthors

%% Include this line if you are using the \added, \replaced, \deleted
%% commands to see a summary list of all changes at the end of the article.
%\listofchanges

\end{CJK*}
\end{document}

%% file: content.tex
\section{Introduction}
\label{sec:intro}
\subsection{Moderate-resolution spectroscopy for high-contrast imaging}

Moderate-resolution integral field spectroscopy ($1,000<R<10,000$) has proven to be a powerful technique for studying the atmospheres of directly imaged exoplanets. 
Moderate spectral resolution makes it possible to resolve the distinct molecular features of a companion that has a cool atmosphere compared to its star.
Even when classical subtraction of the stellar point spread function (PSF) is challenging, the signal of faint sub-stellar companions can be disentangled from the bright host stars using cross-correlation-inspired techniques. 
\citet{Konopacky2013} first demonstrated the technique with the moderate-resolution detection of carbon monoxide (CO) and water (H$_{2}$O) absorption lines in the atmosphere of HR 8799 c. 
Moderate resolution spectroscopy has been used on multiple high-contrast sub-stellar companions with ground-based telescopes in the near infrared ($<2.4\,\mu\mathrm{m}$): HR 8799 bcd \citep{Konopacky2013,Barman2015ApJ...804...61B,PetitditdelaRoche2018A&A...616A.146P,Ruffio2019AJ....158..200R, Ruffio2021AJ....162..290R}, $\beta$ Pictoris b \citep{Hoeijmakers2018A&A...617A.144H}, $\kappa$ Andromedae B \citep{Wilcomb2020AJ....160..207W}, HIP 65426 b \citep{Petrus2021A&A...648A..59P}, TYC 8998-760-1 b \citep{Zhang2021Natur.595..370Z}, PDS 70b \citep{Cugno2021A&A...653A..12C}, VHS 1256 b \citep{Hoch2022AJ....164..155H}, HD 284149 AB b \citep{Hoch2023AJ....166...85H}. 
Thanks to the prominent features of CO and H$_{2}$O at $K$ band, these observations have provided a way to measure the atmospheric carbon-to-oxygen (C/O) ratio, potentially shedding light on the formation pathways of exoplanets \citep[e.g.][]{Oberg2011ApJ...743L..16O,Molliere2022ApJ...934...74M,Hoch2023AJ....166...85H}. In the case of TYC 8998-760-1 b, moderate-resolution spectroscopy enabled the detection of the isotopologue $^{13}$CO \citep{Zhang2021Natur.595..370Z}. Radial velocity measurements from moderate-resolution spectra of the orbiting companion can also provide a better handle on their orbital parameters such as eccentricity \citep{DoO2023AJ....166...48D}. Conversely, the lack of detectable molecular features can also inform about the presence of dust in the surroundings of a planet, as was the case for the young and still forming PDS 70 b \citep{Cugno2021A&A...653A..12C}. However, molecular features are not the only leverage for detecting planets as moderate resolution integral field spectroscopy led to the detection of the second planet in this system PDS~70~c from its H$\alpha$ emission line \citep{Haffert2019NatAs...3..749H}.

These observations were made possible by moderate-resolution ($R\sim4,000$) integral field spectrographs (IFS), namely Keck/OSIRIS \citep{Larkin2006NewAR..50..362L} and VLT/SINFONI \citep{Eisenhauer2003SPIE.4841.1548E,Bonnet2004Msngr.117...17B}, which were not originally designed for direct imaging of exoplanets.  
These observations have shown how moderate spectral resolution can be used for high-contrast science even without coronagraphs.
Furthermore, \citet{Agrawal2023AJ....166...15A} demonstrated an improved sensitivity with Keck/OSIRIS at the smallest projected separations ($<300\,\mathrm{mas}$) compared to classical high-contrast imaging techniques.
Using similar data reduction techniques, higher spectral resolution ($R>10,000$) is also possible and enables measurements otherwise not possible such as more precise radial velocity and spin measurements \citep{Snellen2014Natur.509...63S,Wang2021AJ....162..148W}. Even without integral field spectroscopy, high resolution spectroscopy can be used for detection with new concepts such as vortex fiber nulling \citep{Echeverri2023JATIS...9c5002E}.
High-contrast high-resolution spectroscopy has also been proposed as a way to directly detect Earth-sized exoplanets with the future extremely large class telescopes \citep{Hawker2019MNRAS.484.4855H,Kasper2021Msngr.182...38K, Houlle2021A&A...652A..67H}.
\citet{Landman2023A&A...675A.157L} demonstrated that the optimal resolution for exoplanet detection could be around $R\sim2,000$.

JWST provides a transformative capability for studying exoplanet atmospheres by enabling moderate-resolution spectroscopy ($R\sim2,700$) beyond $3\,\mu\mathrm{m}$. By providing spectral coverage over most of the emitted light of directly imaged exoplanets, JWST will improve our ability to accurately measure disequilibrium chemistry, composition, and cloud properties of directly imaged exoplanets. 
For example, MIRI and NIRSpec will allow the detection of molecules including H$_{2}$O, CO, CH$_{4}$, CO$_{2}$, NH$_{3}$, or H$_{2}$S, many of which are inaccessible from the ground \citep{Patapis2022A&A...658A..72P, Malin2023A&A...671A.109M,Miles2023ApJ...946L...6M}. This capability was inaugurated as part of the Early Release Science (ERS) program that targeted the widely separated planetary mass companion VHS 1256 b from $1-20\,\mu\mathrm{m}$ \citep{Miles2023ApJ...946L...6M}. 
VHS 1256 b was prioritized as the ERS target in part because of its wide separation and faint primary makes it a comparatively easy target that does not require starlight suppression techniques. In JWST cycle 1, several GO and GTO programs have begun applying the NIRSpec IFU towards spectroscopy of slightly more challenging targets at contrasts of $10^{-2}$--$10^{-3}$ (e.g. TWA 27 B in program 1270, \cite{Luhman2023ApJ...949L..36L}; TYC 8998-760-1 b and c in program 2044, Hoch et al. in prep).  The work we present here aims for the first time at the more substantial challenge of companions at $10^{-4}$ contrast and even higher contrast.

\subsection{The context and goals of this program}
\label{sec:context}

In this work, we present the first high-contrast spectroscopic observations with the NIRSpec integral field unit (IFU) in which the signal of a companion needs to be disentangled from its host star.
The cycle 1 GTO program 1414 (PI: Perrin) was designed with complementary scientific and technical goals to achieve: i) atmospheric characterization of a benchmark brown dwarf companion with measured dynamical mass, at dramatically greater precision and sensitivity than possible from the ground, and ii) an assessment of the performance and optimal observing strategy for high-contrast science with the NIRSpec IFU. 
This work focuses on the latter goal, but also includes the spectral extraction for the brown dwarf companion. The atmospheric characterization is described in a separate companion paper (Hoch et al. in prep.).

The target selected, the old and cold T-dwarf HD~19467~B \citep{Crepp2014ApJ...781...29C}, is a roughly 70 Jupiter mass object orbiting a solar-mass star. 
Its apparent separation is $1.6^{\prime\prime}$ from the host star, with a flux ratio varying between $10^{-5}-10^{-4}$ from $3-5\,\mu\mathrm{m}$. The speckles at this separation are similar in intensity to the companion. It makes HD~19467~B well suited for testing high-contrast techniques with the NIRSpec IFU.
This same brown dwarf companion was also the target of NIRCam coronagraphy very early in cycle 1 \citep[GTO 1189; PI: Roellig;][]{Greenbaum2023}, which now enables cross-validation of measurements between the two instruments.

Program 1414 was designed to test and compare multiple strategies for achieving high contrast with the NIRSpec IFU. Commonly used observing strategies for high-contrast imaging and spectroscopy include: Reference Differential Imaging \citep[RDI;][]{Lafreniere2009}, Angular Differential Imaging \citep[ADI;][]{Schneider1998SPIE.3356..222S,Liu2004, marois_angular_2006}, Spectral Differential Imaging \citet[SDI;][]{Marois2000}, and methods that leverage the moderate to high spectral resolution of the data such as cross-correlation techniques \citep[e.g.][]{Konopacky2013}. In order to test these strategies and the impact of saturation on the final sensitivity, the observing plan included pairs of observations to allow tests of ADI as well as RDI both with and without the bright saturating star within the IFU field of view. 
Unfortunately, the observations partially failed due to a guide star acquisition error. A repeat of the full sequence including the failed ones was approved, and is expected in January 2024.
However, the partial dataset is already highly informative for determining the high-contrast sensitivity of NIRSpec and identifying the current limitations of the instrument. 
In this initial work, we explore two complementary techniques: the first leverages the distinct spectral signature of the planet compared to its host star, and the second uses RDI point spread function (PSF) subtraction adapted to work on the IFU spectral ``point cloud'' without interpolation into datacubes. The point cloud refers to the native detector pixel sampling of the observation, as described below.

This paper is part of a series to be presented by the JWST Telescope Scientist Team (JWST-TST)\footnote{\url{https://www.stsci.edu/~marel/jwsttelsciteam.html}}, led by M. Mountain and convened in 2002 following a competitive NASA selection process. In addition to providing scientific support for observatory development through launch and commissioning, the team was awarded 210 hours of Guaranteed Time Observer (GTO) time. This time is being used for studies in three different subject areas: (a) Transiting Exoplanet Spectroscopy (lead: N. Lewis; e.g., \citealt{Grant2023ApJ...956L..29G}); (b) Exoplanet and Debris Disk High-Contrast Imaging (lead: M. Perrin; e.g., \citealt{Rebollido2024AJ....167...69R}); and (c) Local Group Proper Motion Science (lead: R. van der Marel; e.g., \citealt{Libralato2023ApJ...950..101L}). A common theme of these investigations is the desire to pursue and demonstrate science for the astronomical community at the limits of what is made possible by the exquisite optics and stability of JWST. 
The high-contrast portion of the TST investigation includes efforts studying exoplanetary systems and circumstellar disks using the full range of high contrast modes with JWST NIRCam, MIRI, and NIRSpec. The programs within this area were crafted to rapidly advance knowledge of high-contrast strategies and best practices with JWST early in the mission, while yielding significant astrophysical insights into a wide range of circumstellar systems.

\subsection{The need for a high-contrast-optimized data reduction strategy for JWST NIRSpec}
\label{sec:limitations}

High-contrast science is particularly challenging because it requires subtracting, or modelling, the stellar host point spread function very accurately to uncover the unbiased companion spectrum. Limitations of the standard NIRSpec IFU data reduction strategy warranted the development of a different approach to analyzing JWST IFU data.

A particular challenge with the NIRSpec IFU is that it is spatially undersampled. In fact, it is one of the most spatially undersampled modes of JWST. For example, the full-width-at-half-maximum (FWHM) of the NIRSpec PSF at $3\,\mu\mathrm{m}$ equals its spatial pixel (spaxel) size of $0.1^{\prime\prime}$; it is not Nyquist-sampled at any wavelength. As a result, a prominent artifact in NIRSpec IFU spectra of the current pipeline outputs is a low-frequency quasi-periodic flux fluctuation. This artifact can have a peak-to-valley amplitude up to 50\% of the continuum in a single spaxel of a single exposure (See Figure 9 in \citealt{Law2023}, which depicts this for MIRI MRS; the issue equally affects NIRSpec IFU data). 
This artifact is due to the interplay between the Drizzle-based cube extraction \citep{Fruchter2002,Law2023}, the spatial undersampling, and the curvature of the spectral traces on the detector. Specifically, the trace is not perfectly horizontal on the detector so the flux of a point source is periodically either concentrated on a pixel row or split between neighboring rows as the light slowly moves vertically as a function of wavelength. 
The issue of spectral oscillations due to curved traces on the detector is not unique to NIRSpec and was for example also seen with Spitzer \citep{Smith2007PASP..119.1133S,Lebouteiller2010PASP..122..231L}.

Existing mitigation strategies include combining several dither positions and integrating the flux over a wide aperture centered on the point source in the final spectral cube to average this effect. Using MIRI Medium Resolution Spectroscopy (MRS), \citet{Law2023} shows that the oscillations can be reduced to $<5\%$ of the continuum with an aperture extraction radius of at least 0.5 times the PSF FWHM, and reduced to $<1\%$ with an extraction radius at least 1.5 times the PSF FWHM.
However, as discussed in more detail in Section \ref{sec:sampling}, the effectiveness of the current dithering is field- and wavelength-dependent as the spatial dimensions are not uniformly sampled by the various dithering patterns. Using aperture photometry to reduce these systematics is also undesirable in some cases. In a high-contrast speckle dominated regime, aperture photometry leads to a higher level of starlight contamination in the companion spectrum compared to PSF fitting, and effectively reducing the spatial resolution of the IFU. Furthermore, for cases in which we want to perform PSF subtractions, that subtraction ought to be performed for each spaxel, so averaging over apertures is inherently not a desirable approach. 
A second systematic noise floor is set by uncorrected bad pixels in NIRSpec extractions, which is currently at a similar amplitude as the undersampling oscillations.

These systematics prevent a sufficiently accurate subtraction of the stellar PSF for high-contrast imaging in current standard pipeline output products. With a systematic floor at 1\% of the continuum, a typical directly imaged planet that might be 100 times fainter than the speckle field (a.k.a., diffracted starlight) at its separation would have a S/N per spectral bin limited to unity.

We therefore have developed an alternative approach to NIRSpec IFU data reduction that enables the detection and characterization of high-contrast companions down to the stellar photon noise regime. This is made possible by forward modeling the astrophysical scene directly in the flux-calibrated NIRSpec detector images, without interpolation into datacubes. The algorithms and software leverage the framework developed for Keck/OSIRIS in \citet{Ruffio2021AJ....162..290R} and \cite{Agrawal2023AJ....166...15A}.
A similar approach was used in \citet{DeGraaf2023arXiv230809742D} to forward model the detector images with the NIRSpec multi-object spectroscopy (MOS) mode.

We note in passing that, though the mathematical approaches differ significantly, conceptually the core idea of forward modeling the undersampled data directly in the detector frame is adjacent to well-established techniques for achieving highly precise astrometry and photometry in undersampled Hubble, or Spitzer IRAC images, by fitting models to data in the detector frame \citep{Anderson2000PASP..112.1360A,Anderson2016wfc..rept...12A,Esplin2016AJ....151....9E}.
In both cases, highly precise measurements with minimal systematics are enabled by fitting models to the information content present across many undersampled pixels. 
A specific example with Hubble of managing undersampled imaging data by combining multiple dither positions for high contrast is \citet{Rajan2015ApJ...809L..33R}.

\subsection{Outline}

The NIRSpec IFU observations of HD~19467~B are presented in Section \ref{sec:obs} and the initial data reduction steps are described in Section \ref{sec:datareduc}. 
In Section \ref{sec:FM}, we propose a strategy to forward model the companion signal and the starlight directly in the detector images. This method addresses the spatial undersampling and bad pixel issues in NIRSpec.  
We also demonstrate the sensitivity of the NIRSpec IFU for the direct detection of exoplanets as a function of their effective temperature.
The cost of this proposed forward model is the effective loss of the spectral continuum of the companion. 
In order to recover the full spectrum of HD~19467~B including its continuum, we implement a reference star PSF subtraction routine described in Section \ref{sec:RDI}. This RDI implementation does not require regularly sampled spectral cubes, but it remains limited by the large interpolation errors of spatially resampling the reference star observations onto the science dataset.
In Section \ref{sec:FMRDI}, we propose a regularization framework that combines the flexibility of the forward model with prior knowledge of the PSF profile to combine the best of both worlds. Improving RDI for the NIRSpec IFU will be needed to go beyond this proof of concept.
In Section \ref{sec:discussion}, we discuss our results and make recommendations for future high-contrast observations with the NIRSpec IFU. 
We conclude in Section \ref{sec:conclusion}.

\begin{figure*}
  \centering
  \includegraphics[trim={0cm 0cm 0cm 0cm},clip,width=1\linewidth]{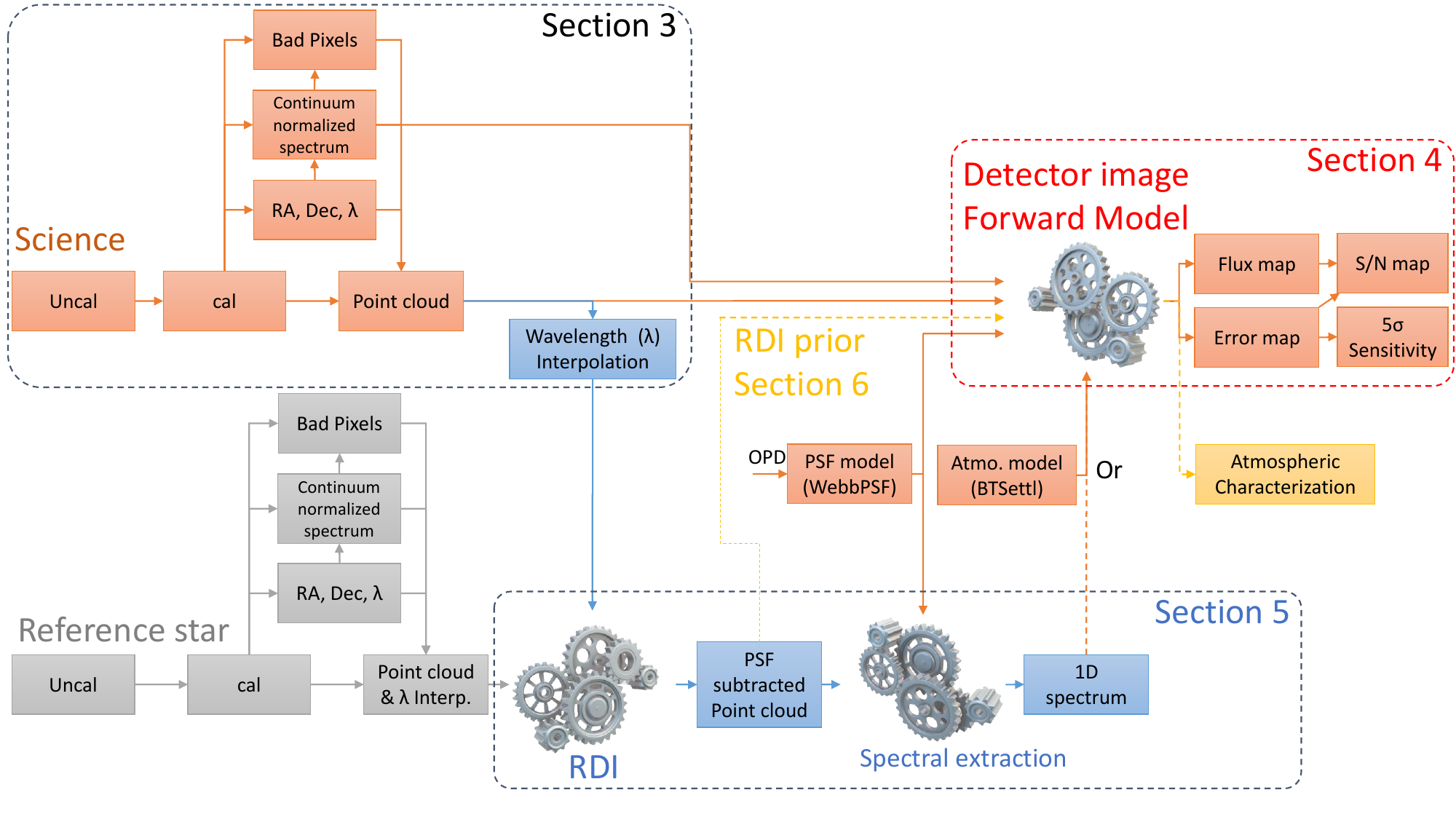}
  \caption{Illustration of the data processing steps showing conceptually the information flow between the complementary techniques used. Section \ref{sec:datareduc} describes the processing and reduction of uncalibrated (uncal) data, for both science and reference targets. The point cloud results from combining flux-calibrated detector images (cal) with the corresponding spatial (RA, Dec) and spectral ($\lambda$) coordinates of each pixel.
  The output spectral information from Section \ref{sec:datareduc} flows into both the Forward Model, described in section \ref{sec:FM}, and the RDI PSF subtraction, described in section  \ref{sec:RDI}. The outputs of the forward model are the flux, S/N, and detection sensitivity maps.
  The output of the RDI PSF subtraction is the companion spectrum.
  Section \ref{sec:FMRDI} presents steps toward more tightly integrating these methods through the use of RDI outputs to set priors on the forward model.}
  \label{fig:flowchart}
\end{figure*}

This outline is illustrated in Figure \ref{fig:flowchart} including the data flow and main data processing steps.

\section{NIRSpec Observations}
\label{sec:obs}

\begin{figure*}
  \centering
  \includegraphics[trim={1cm 1.0cm 1cm 0.5cm},clip,width=1\linewidth]{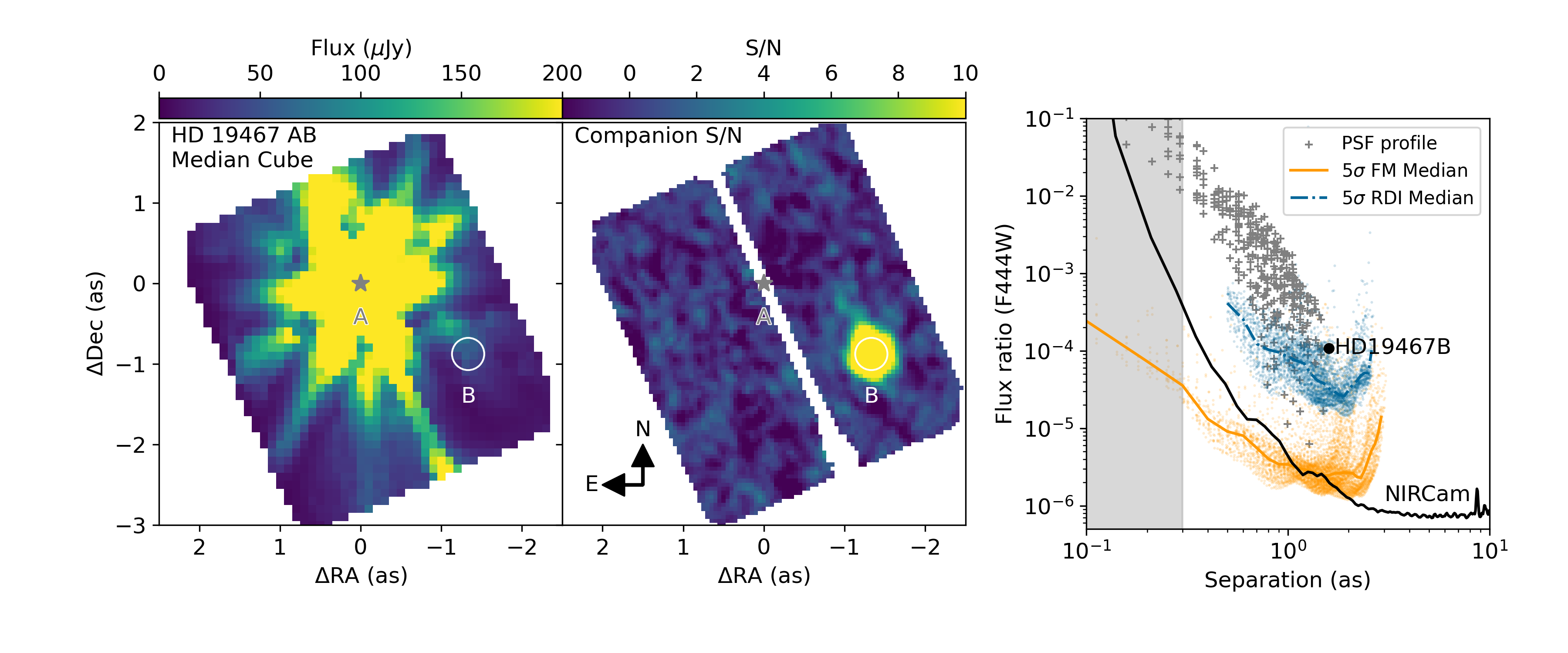}
  \caption{Detection of the high-contrast companion HD~19467~B with the JWST NIRSpec IFU. (Left) Median spectral cube image from the JWST science calibration pipeline. (Middle) Detection map representing the companion signal-to-noise ratio (S/N) at each sky coordinate. The S/N is calculated using a joint forward model of the starlight and the companion signal in the detector images, which can be compared to a cross-correlation S/N. HD~19467~B is detected with $S/N=232$ after combining all exposures. (Right) Detection sensitivity ($5\sigma$) expressed as a companion-to-star flux ratio of various detection methods: forward model (FM), reference differential imaging (RDI), and the published ADI+RDI NIRCam F444W sensitivity from \citet{Carter2023} as the black line. We scaled the NIRCam contrast curve with the square root of the exposure time and stellar brightness to match the NIRSpec dataset, although we note that the NIRCam detection limits have not yet been shown to be photon noise limited at small projected separations. The scatter in the NIRSpec detection limits do not represent its uncertainty, but they are the results of the 2D spatial variations of the NIRSpec PSF. The region within $0.3^{\prime\prime}$ is greyed out because the current implementation of the method is not valid there, but it may theoretically be achieved in the future. The scatter plot of the PSF profile (grey) represents the raw PSF intensity values in each spaxel relative to the peak. In this plot, we demonstrate that the NIRSpec IFU reaches deep high-contrast sensitivity ($10^{-6}-10^{-5}$) at small projected separation ($0.3-2^{\prime\prime}$) with moderate-resolution spectroscopy. 
  }
  \label{fig:summarydetec}
\end{figure*}

The observations used in this work are from cycle 1 GTO program 1414 (PI: Marshall Perrin) obtained on January 25 (UT) 2023 and detailed in Table \ref{tab:observations}. 
The first two science observations (Obs. 2 \& 3), intended for the ADI test and RDI with star offset, failed due to guide star acquisition error. The currently available data only includes the second roll angle with the saturated star within the field of view (Obs. 4), as well as both reference observations (Obs. 1 \& 5). In this work, we only analyze obs. 4 \& 5.
It consists of one 35 min JWST NIRSpec IFU \citep{Jakobsen2022,Boker2022} observation of the late T-type brown-dwarf companion HD~19467~B using NIRSpec's G395H grating and F290LP filter. The astrophysical scene is illustrated in Figure \ref{fig:summarydetec}.
The reference star, taken immediately after, is used to subtract the stellar point spread function in the science observation before the spectral extraction of the companion spectrum (see Section \ref{sec:RDI}). The reference star is however not used to detect the companion with the forward model in Section \ref{sec:FM}. 

Both science and reference stars are too bright for target acquisition, so the absolute pointing accuracy of JWST after guide star acquisition of $0.1''$ was used.  Relatedly, choosing to omit target acquisition also allows the NIRSpec grating wheel mechanism to remain unmoved at the G395H position throughout the entire sequence of observations. This eliminates any potential effects from the small non-repeatability in position of that mechanism, which could potentially have a minor effect on PSF subtractions but can be entirely avoided in this way.  The small (0.25'') cycling dither pattern with nine positions was chosen to improve the spatial sampling, while ensuring the science target remained in the IFU FOV within the pointing accuracy. 
The reference star HD~18511 is brighter ($K=4.3$, \citet{Cutri2003}; A2, \citep{Houk1988}) than the science host HD~19467 ($K=5.4$, \citet{Cutri2003}; G3V, \citet{Houk1988}), so the exposure time was set shorter to reach a similar flux level. 
The core of the stellar PSF fully saturates up to $0.2''$ at $3.0\,\mu\mathrm{m}$ ($<0.1''$ at $5.0\,\mu\mathrm{m}$), with partial saturation of the ramps up to $0.5''$. 
The saturation also causes charges to transfer across most of the detector chip, but primarily affecting pixels in the same slices as the saturated PSF core \citep{Boker2022}.

\begin{figure*}
  \centering
  \includegraphics[trim={0cm 0cm 0cm 0cm},clip,width=1\linewidth]{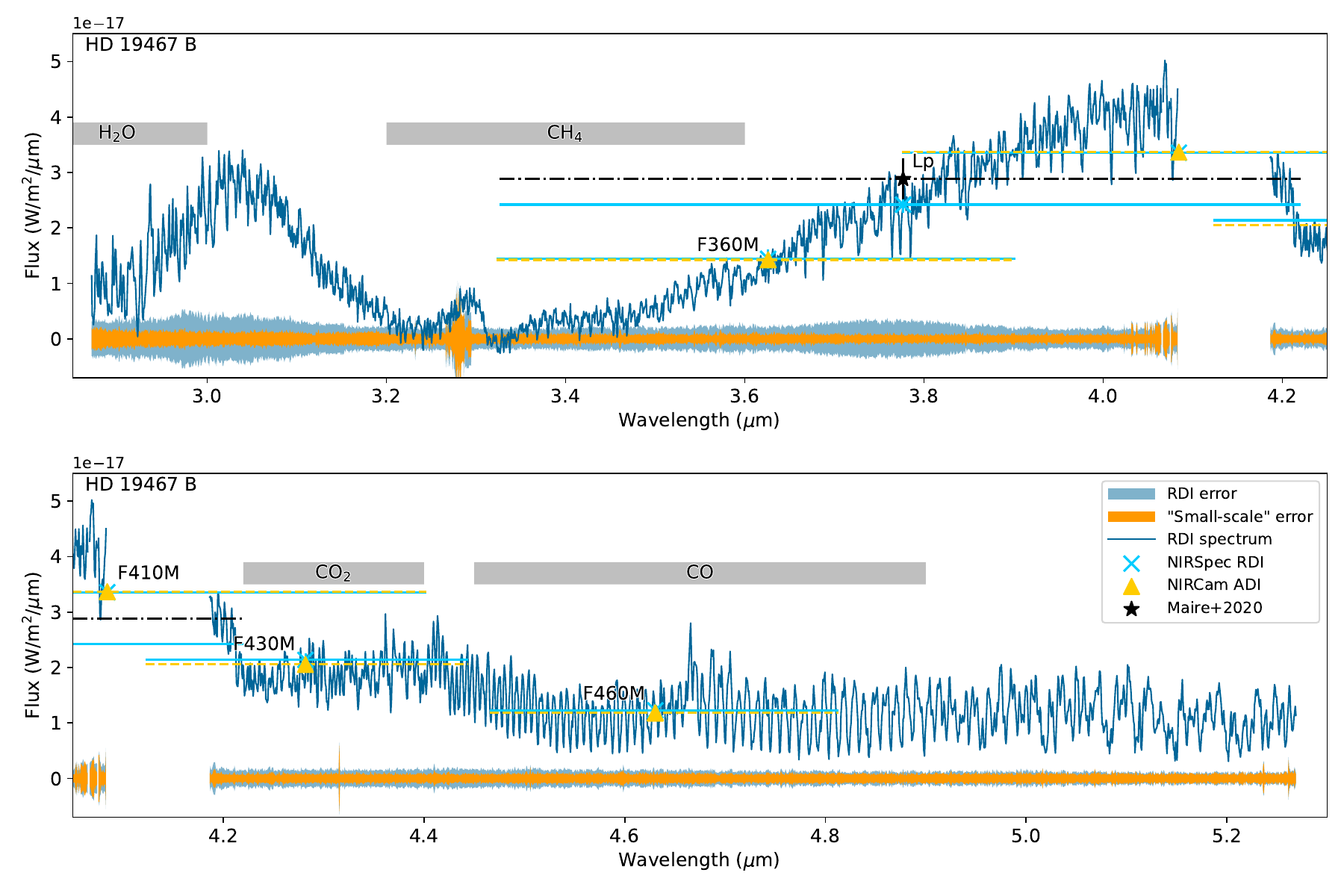}
  \caption{The spectrum of HD~19467~B after reference-star differential imaging (RDI) and \texttt{WebbPSF} spectral extraction (Section \ref{sec:RDI}). The RDI flux error is shown as the blue shaded region, which is estimated from the standard deviation of residual speckles spectra in an annulus around the companion. These errors are in practice correlated, so we conceptually separate the noise into two terms: the noise correlated on small spectral scales from interpolations ($<2$ pixels) and the noise correlated on larger scales from residuals speckles. The standard deviation of the small-scale fraction of the total variance is labeled as the small-scale error. The photometry derived from the NIRSpec RDI spectrum is consistent with our reanalysis of the NIRCam photometry originally presented in \citet{Greenbaum2023} as well as the Lp data point from \citet{Maire2020} within calibration systematic uncertainties.}
  \label{fig:summaryRDIspec}
\end{figure*}

The wavelength range of the spectral gap between the two NIRSpec detectors around $4\,\mu\mathrm{m}$ varies depending on position in the field of view: the layout of the IFU spectra projected onto the detectors results in each slice from the IFU slicer having a distinct wavelength gap. Therefore, in principle, the gap can be reduced by taking multiple offset exposures such that the target is observed across widely separated slices.  The only successful science observation (obs. 4) was aimed at keeping both the host and the companion in the field of view, using the small (0.25'' extent) dither pattern only, and therefore the gap in the spectrum remained large (Figure \ref{fig:summaryRDIspec}). However, the first failed science observation (obs. 2) would have placed the companion at the center of the IFU field, in a set of slices with the gap shifted by about half its width. This means that the gap across all datasets combined will be reduced when the program is fully completed. Science programs with particular needs for wavelength coverage around the gap region can use target placement within the IFU field to optimize wavelength coverage.

\begin{table*}[!ht]
    \centering
\begin{tabular}{lllllllll } 
 \hline
Obs. & Status &  Target & Grating Filter & Groups & $T_{\mathrm{int}}$ per exp. & Dithers & Total $T_{\mathrm{int}}$ & Description \\
 \hline
1&OK - not used & HD 18511& G395H/F290LP & 5 & 87.5 s & 9 & 13 min & PSF star for obs. 2 \\
2&GSA FAILED & HD 19467& G395H/F290LP & 15 & 233.4 s & 9 & 35 min & offset out of FOV \\
3&GSA FAILED & HD 19467& G395H/F290LP & 15 & 233.4 s & 9 & 35 min & in field, roll angle 1\\
4&OK - this work& HD 19467& G395H/F290LP & 15 & 233.4 s & 9 & 35 min & in field, roll angle 2\\
5&OK - this work& HD 18511& G395H/F290LP & 5 & 87.5 s & 9 & 13 min &  PSF star for obs. 4\\
\hline
\end{tabular}
\caption{Observations of the cycle 1 GTO program 1414 (PI: Perrin) that occurred on January 25 (UT) 2023. The readout pattern is \texttt{NRSIRS2RAPID}. The nine dithers followed the small cycling pattern starting from the first position. Observation 2 and 3 failed due to a problem with the guide star acquisition (GSA).  This data can be found in MAST: \dataset[https://doi.org/10.17909/q524-zn59]{https://doi.org/10.17909/q524-zn59}.}
\label{tab:observations}
\end{table*}

\section{Data reduction}
\label{sec:datareduc}

\subsection{Detector image flux calibration}

NIRSpec IFU data is typically reduced as follows using the JWST data pipeline  \citep{Bushouse2023}: 1) the stage 1 pipeline is used to process uncalibrated up-the-ramp data to generate rate maps (\textit{*\_rate.fits}) in DN/s, 2) the stage 2 pipeline is used to flux calibrate the 2D images (\textit{*\_cal.fits}), 3) spectral cubes (\textit{*\_s3d.fits}) are produced by the stage 2, or stage 3, pipeline to obtain spatially and spectrally regularly-sampled data using a 3D implementation of the Drizzle algorithm \citep{Fruchter2002,Law2023}, 4) spectra of point sources can be obtained using aperture photometry of the final cubes. 
As explained in Section \ref{sec:limitations}, we do not use the spectral cubes in this work. Instead, we will use the flux-calibrated detector images as the starting point of further analyses.

The uncalibrated NIRSpec detector images were generated using the version 2022{\_}4a (SDP{\_}VER) of the JWST Science Data Processing (SDP) subsystem. The science calibration pipeline version ``1.12.5'' (CAL{\_}VER) stages 1 and 2 were used to produce the flux-calibrated detector images.
The version of the Calibration Reference Data System (CRDS) selection software was 11.17.14 (CRDS{\_}VER) and 
the CRDS context version is jwst{\_}1185.pmap (CRDS{\_}CTX) \citep{Greenfield2016}.

NIRSpec detector images feature systematic vertical strips due to the 1/f noise. The charge transfer from the saturated central star also creates a broad background feature covering most of the detector. We address both issues by implementing an intermediate step between the stage 1 and stage 2 pipeline to remove these background features in the rate maps described in Appendix \ref{app:cleanrate}. The method is analogous to the NSClean algorithm \citep{Rauscher2024PASP..136a5001R}, but uses a column-wise spline to fit the background column-wise instead of a Fourier decomposition.

Note that the pipeline's ``cube build'' step works in part by extracting from the ``cal'' files the fluxes at each detector pixel's unique ${x, y, \lambda}$ and then interpolating (drizzling) those into a regular cube sampling. The set of fluxes on the irregular ${x, y, \lambda}$ sampling defined by the IFU projection onto the detector is called the ``point cloud''\footnote{see \url{https://jwst-pipeline.readthedocs.io/en/latest/jwst/cube_build/main.html}}. Though we do not use pipeline-produced cubes, we make use of the point cloud concept and terminology. 
The point cloud is conceptually isomorphic to the set of all valid detector pixels. In other words, the detector images and the point cloud are exactly equivalent, and switching from one to the other does not involve any interpolation.

\subsection{Additional preprocessing steps}

We further process each individual flux-calibrated image (``cal'') as described below:
\begin{description}
    \item[Bad pixel identification and masking] Any pixel marked as ``do not use'' in the data quality (``DQ'') extension of the ``cal'' file are masked in any subsequent steps. Pixels with comparatively very large estimated flux errors in the ``ERR'' extension of ``cal'' files proved to be outliers despite their indicated larger error. We therefore identify these as bad pixels using a row-by-row sigma clipping of the error map. Each row of the flux error map is high-pass filtered using a median filter and a 50-pixel sliding window. Any pixels deviating by more than 50 times the median absolute deviation of the residuals of each row is marked as bad. Additional bad pixels are identified and masked after the continuum normalized spectrum is derived as explained in Section \ref{sec:contnormspec}.
    \item[Wavelength grid definition] The wavelength of each pixel is extracted from the ``WAVELENGTH'' extension of the ``cal'' file. We use this original wavelength information when forward modeling the detector images. Not interpolating the pixels on a regular wavelength grid is indeed always preferable to avoid introducing new interpolation systematics in the analysis. 
    However, fitting a PSF is more tractable on a regularly sampled wavelength grid, which is what we choose to do when implementing the PSF subtraction and the spectral extraction.  
    For these cases, we define a fixed wavelength sampling for each of the two NIRSpec detectors that ranges from the minimum to the maximum value in the wavelength map with a bin size equal to the median wavelength difference between two horizontally neighboring pixels on the detector. The sampling is kept the same throughout this work for all the datasets and it will be referred to as $\{ \lambda_i \}$. The minimum, maximum, and bin size for the shorter wavelength detector (NRS1) is $2.8595\,\mu\mathrm{m}$, $4.1013\,\mu\mathrm{m}$, and $6.8\times10^{-4}\,\mu\mathrm{m}$, while it is $4.0813\,\mu\mathrm{m}$, $5.2787\,\mu\mathrm{m}$, and $6.7\times10^{-4}\,\mu\mathrm{m}$ for the longer wavelength detector (NRS2). 
    \item[Pixel sky coordinates] The ($\Delta \mathrm{RA}_\mathrm{pix}$,$\Delta \mathrm{Dec}_{\mathrm{pix}}$) relative positions of each pixel in the detector images are calculated relative to the host star ($\mathrm{RA}_\mathrm{star}$,$\mathrm{Dec}_{\mathrm{star}}$) in sky coordinates using the WCS headers in the FITS file and the related tools in the JWST science calibration pipeline. The pixel relative coordinates are converted to offsets in arcseconds accounting for the scaling by the cosine of the declination for the right ascension direction: \\
    $\Delta \mathrm{RA}_\mathrm{pix} = (\mathrm{RA}_\mathrm{pix}- \mathrm{RA}_\mathrm{star})*\cos(\mathrm{Dec}_{\mathrm{pix}})$, \\
    $\Delta \mathrm{Dec}_\mathrm{pix} = \mathrm{Dec}_{\mathrm{pix}}-\mathrm{Dec}_{\mathrm{star}}$.
    \item[Charge transfer masking]  The saturation of the host star leads to charge transfer to neighboring pixels on the detector, which creates a bright bar artifact in the other pixels in the same IFU slices as shown in Figure \ref{fig:summarydetec} or Figure 7 in \citet{Boker2022}. A more detailed explanation can be found in Appendix~\ref{app:cleanrate}. To avoid any biases from the charge transfer, we mask a $0.3''$ wide bar across the data (i.e. roughly discard the three slices centered on the star). We approximately align the bar mask with the star correcting for any offset between the predicted and measured position of the star. 
\end{description}

All these steps were implemented in the NIRSpec instrument class \texttt{jwstnirpsec\_cal} in the open-source package \texttt{breads}, the Broad Repository for Exoplanet Analysis, Discovery, and Spectroscopy\footnote{\url{https://github.com/jruffio/breads} commit hash \texttt{fb83691}} \citep{breads}. The Python scripts used throughout this work to process data and generate figures is publicly available in a Github repository\footnote{\url{https://github.com/jruffio/HD_19467_B} commit hash b853a66} \citep{HD19467Bscripts}.

\subsection{Continuum normalized spectrum of the host star}
\label{sec:contnormspec}

\begin{figure*}
  \centering
  \includegraphics[trim={0cm 0cm 0cm 0cm},clip,width=1\linewidth]{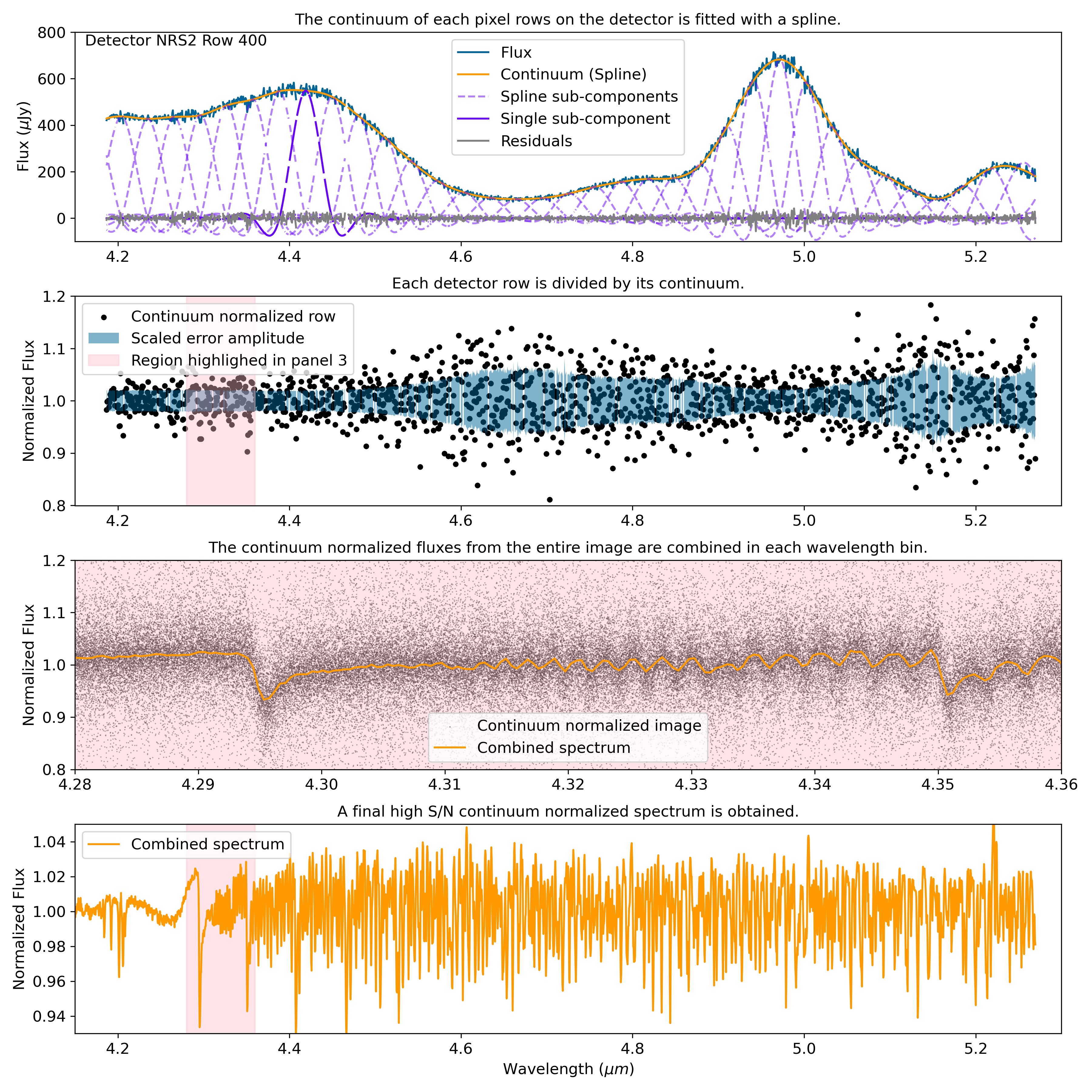}
  \caption{Empirical derivation of a continuum normalized spectrum of the host star HD~19467. This corresponds to a single exposure (\texttt{jw01414004001\_02101\_00001\_nrs2\_cal.fits}) with an integration time of $233.4\,\mathrm{s}$ of the NRS2 NIRSpec detector. For clarity, the third panel from the top zooms in a narrow interval of the wavelength highlighted in red. The wavelength bins $\delta \lambda$ are defined such that $\lambda / \delta \lambda \sim 10,000$.}
  \label{fig:contnormspec}
\end{figure*}

The primary purpose of this section is to derive the best possible empirical spectrum of the star that can be used to model the starlight (i.e., speckles) when fitting for the companion. It is a cornerstone of the forward model defined in Section \ref{sec:FM}.
This spectrum will also be used to identify remaining bad pixels before any subsequent analysis, which is why it is introduced first.

As noted above, extracted spectra of point sources from datacubes are currently limited to a S/N per resolution element well below the photon-noise limit in the bright star regime due to the spatial undersampling of the NIRSpec IFU and the current management of bad pixels in the JWST science calibration pipeline (Section \ref{sec:limitations}). 

We show that it is possible to obtain a photon-noise limited spectrum at the cost of normalizing its continuum, using the method outlined below and illustrated in Figure \ref{fig:contnormspec}. 
A continuum-normalized spectrum is interesting because it is sufficient in several applications where reaching the photon noise limit is most important. For example, a continuum-normalized spectrum can be sufficient for measuring radial velocities or studying the relative amplitude of spectral features.

We use the fact that it is not necessary to extract the 1D spectrum of a point source before normalizing its continuum. 
We can indeed normalize the detector images by dividing the continuum of the spectrum row by row on the detector. To understand why this works, one could imagine normalizing a spectral cube by estimating and dividing the continuum one spaxel at a time. Each spaxel would now include a continuum normalized spectrum of the star, albeit with different levels of noise depending on the original flux level. These normalized spaxels could then be combined with a weighted mean into a master normalized spectrum of the star. We are doing something similar, albeit in detector space. Spaxels are equivalent to traces on the detector, so we could equivalently divide the continuum of each trace. One issue is that traces are slightly curved on the detector, which means that they do not exactly fall on individual pixel rows. However, this curvature only translates into a slow modulation of the continuum when considering a single row. This does not affect the continuum normalization as long as the curvature is mild enough and the slow modulation can be followed by the low pass filter.
This row-by-row normalization has proven to be an effective way to address the spatial undersampling and the bad pixels of NIRSpec, and importantly does not require any spatial interpolations.
The scale of the continuum fluctuations within a row is defined by the typical spatial extent of a speckle (i.e. the diffraction limit of the telescope), the movement of the speckles due to the magnification of the PSF with wavelength, and the curvature of the dispersion axis on the NIRSpec detector.
We found the spline model introduced in \citet{Agrawal2023AJ....166...15A} and \citet{Ruffio2023AJ....165..113R} to be a good model to fit the continuum and normalize it. 
For a general purpose application, it would be possible to use a simpler continuum normalization. However, the spline model is the foundation of the forward model defined in Section \ref{sec:FM}, which is used to recover the companion signal, so it is necessary to use the same continuum normalization at this stage.
The continuum of each row is modeled as a linear combination of modes defined for a set of $K=40$ nodes regularly spaced in wavelength space corresponding to a spacing of $\sim0.03\,\mu\mathrm{m}$. 

The linear model for each row of data is defined as 
\begin{equation}
     {\bm d}_{\mathrm{row}} = {\bm M}_{\mathrm{row}}{\bm \phi}_{\mathrm{row}} + {\bm n}_{\mathrm{row}},
     \label{eq:rowFM}
\end{equation}
where ${\bm d}_{\mathrm{row}}$ is the data vector of size $N_{\mathrm{row}}$ made out of the valid pixels in a given row of the detector. (An example of ${\bm d}_{\mathrm{row}}$ is illustrated as the blue line in the top panel of Figure \ref{fig:contnormspec}). 
The linear parameter ${\bm \phi}_{\mathrm{row}}$ has $K$ parameters corresponding to the values of the continuum at the $K$ spline nodes. Following the method in \citet{Agrawal2023AJ....166...15A}, ${\bm M}_{\mathrm{row}}=[{\bm c}_1,\dots,{\bm c}_K]$ is a matrix of shape $(N_{\mathrm{row}},K)$ in which each column ${\bm c}_i$ corresponds to the spline component centered on each node as illustrated with dashed-purple lines in the top panel of Figure \ref{fig:contnormspec}.
The photon noise is represented by the random vector ${\bm n}_{\mathrm{row}}$, which is modeled by a multivariate Gaussian distribution with a diagonal covariance matrix ${\bm \Sigma}_{\mathrm{row}}$. The diagonal elements of ${\bm \Sigma}_{\mathrm{row}}$ ${\bm s}_{\mathrm{row}}$ are defined from the estimated flux errors for each pixel. The flux errors are calculated by the JWST science calibration pipeline and saved as a FITS file extension.

Improving upon the models used in \citet{Agrawal2023AJ....166...15A} and \citet{Ruffio2023AJ....165..113R}, we add a Gaussian regularization term on the parameters ${\bm \phi}_{\mathrm{row}}$ to improve the numerical stability of the inversion of the linear least square problem. We discuss in Section \ref{sec:FMRDI} how this prior can also be used to better constrain the spectral continuum of a companion. 
We explain how this regularization changes the derivation of the covariance and the marginalized likelihood in Appendix \ref{app:newmaths}, which were originally derived in Appendix D.3 and D.6 in \citet{Ruffio2019AJ....158..200R} without regularization.
The Gaussian prior on ${\bm \phi}_{\mathrm{row}}$ is defined by a mean vector ${\bm \mu}_{{\bm \phi}}$ and vector of standard deviations ${\bm s}_{{\bm \phi}}$.
We fit the data a first time using the median flux of the row as the prior mean and standard deviation. Subsequently, in a second and final fit, we set both the mean and standard deviation of the prior to the best-fit values ${\bm \phi}_{\mathrm{row}}$ of the first fit.
Using the properties of Gaussian conjugate priors and linear models (see Appendix \ref{app:newmaths}), the problem can be rewritten in an identical format:
\begin{equation}
     {\bm d}^{\prime}_{\mathrm{row}} = {\bm M}^{\prime}_{\mathrm{row}}{\bm \phi}_{\mathrm{row}} + {\bm n}^{\prime}_{\mathrm{row}},
     \label{eq:rowFMreg}
\end{equation}
by concatenating the vectors and matrices as follows
\begin{equation}
    {\bm d}^{\prime}_{\mathrm{row}} =
    \begin{bmatrix}
    {\bm d}_{\mathrm{row}} \\
    {\bm \mu}_{{\bm \phi}}
    \end{bmatrix} 
,
    {\bm M}^{\prime}_{\mathrm{row}} =
    \begin{bmatrix}
    {\bm M}_{\mathrm{row}} \\
    I_{K}
    \end{bmatrix}.
\end{equation}
$I_{K}$ is the identity matrix of size $K$.
The new diagonal covariance matrix is defined by the diagonal elements $[{\bm s}_{\mathrm{row}}, {\bm s}_{{\bm \phi}}]$.
The advantage of this approach is that the regularization can be implemented with minimal changes to the code.

We define the best-fit model continuum in a row as ${\bm m}_{\mathrm{row}}$.
After the continuum has been fitted, it is divided from the data as ${\bm d}_{\mathrm{row}}/{\bm m}_{\mathrm{row}}$ assuming element-wise division (panel 2 in Figure \ref{fig:contnormspec}). This process is done for each row of the detector, leading to a continuum normalized detector image.
We apply the following criteria to select the pixels that will be used to derive the final 1D continuum normalized spectrum.
Pixels with normalized residuals $({\bm d}_{\mathrm{row}}-{\bm m}_{\mathrm{row}})/{\bm s}_{\mathrm{row}}$ that are 10 times the median absolute deviation in their row are masked.
Then, we also remove pixels with ${\bm m}_{\mathrm{row}}/{\bm s}_{\mathrm{row}} < 5$ to avoid including pixels that are dominated by the background. Out of the remaining pixels, we only consider the $50\%$ brighter, which leads to panel 3 in Figure \ref{fig:contnormspec}. The pixels meeting that criterion from all rows across the detector are then combined: The final continuum-normalized 1D spectrum is derived using a weighted mean in small wavelength bins of width $\delta \lambda$ such that $\lambda / \delta \lambda \sim 10,000$. The higher resolution sampling of this empirical spectrum is chosen to reduce interpolation errors in later steps.
The quality of the resulting spectrum is insensitive to spatial undersampling and bad pixel interpolation issues allowing it to reach the photon noise limit even in the bright star regime. This remains true even when the core of the PSF is saturated, as is the case here. These spectra can however be sensitive to the curvature of the traces on the detector if the low pass filter is not flexible enough to follow steeper fluctuations of the continuum in a given row.

Finally, we use this continuum normalized spectrum to robustly and empirically identify bad pixels in NIRSpec IFU images. We do this by first modifying the forward model of the data to also include the stellar features. This is done by imprinting the stellar spectral features in the spline modes directly such that the model is no longer a smooth continuum, but a continuum-modulated version of the stellar spectrum.
Specifically, we multiply each column of the model matrix ${\bm M}_{\mathrm{row}}$ by $\mathcal{S}({\bm w}_{\mathrm{row}})$ element-wise: $c_i \mathcal{S}({\bm w}_{\mathrm{row}})$, with ${\bm w}_{\mathrm{row}}$ the wavelength of each pixel in this row and $\mathcal{S}$ a function that linearly interpolates the continuum normalized spectrum of the star. We fit the more accurate forward model in the same way to each row of the detector and identify bad pixels through sigma clipping. We use the same method as before by identifying pixels in the residuals of a row that deviate by more than 10 times the median absolute deviation of the normalized residuals. 

The derived continuum-normalized spectrum and improved map of bad pixels are saved and used as inputs into the forward model described below. 

\subsection{Interpolating the data on a regular wavelength grid}
\label{sec:wvinterp}

\begin{figure*}
  \centering
  \includegraphics[trim={0cm 0cm 0cm 0cm},clip,width=1\linewidth]{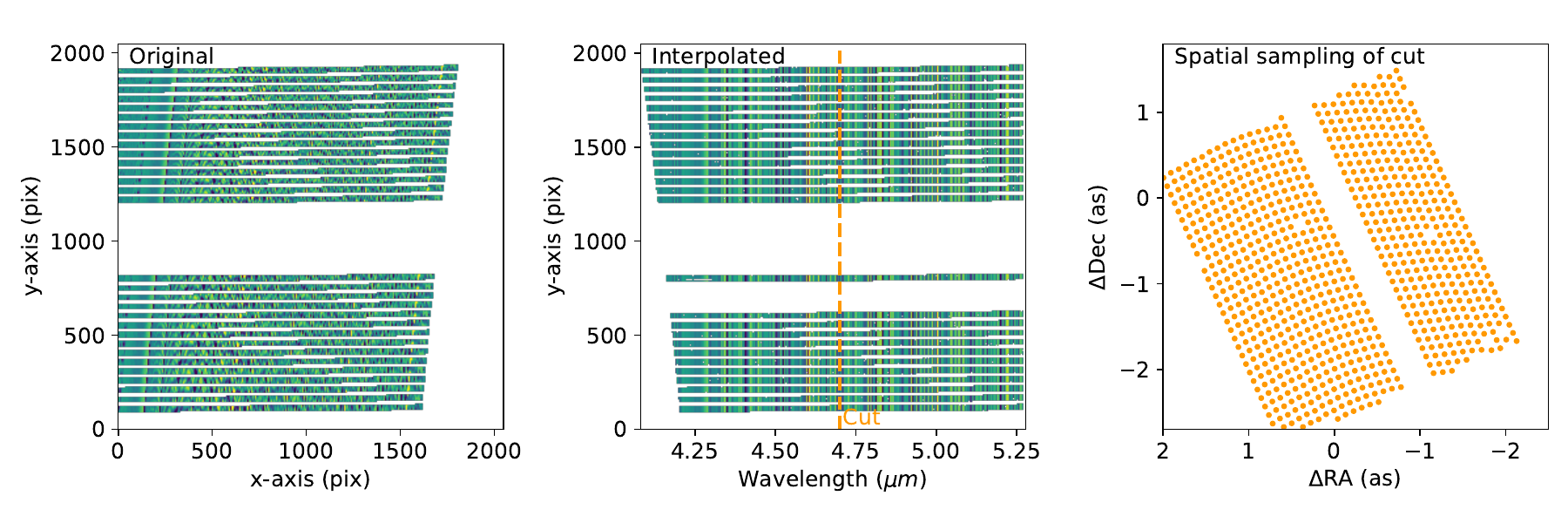}
  \caption{Interpolation of a NIRSpec detector image onto a regular wavelength grid (\texttt{jw01414004001\_02101\_00001\_nrs2\_cal.fits}). (Left) Original flux-calibrated detector image featuring the slices of the NIRSpec IFU. The slices around the saturated star are masked. (Middle) Interpolated image where each column is at a constant wavelength. (Right) Spatial sampling (2D point cloud) of a vertical cut of the interpolated image.}
  \label{fig:regwvsinterp}
\end{figure*}

The dispersion direction on the NIRSpec detector is slightly curved along the horizontal axis of the detector, so pixels in the vertical direction do not have a constant wavelength.
We therefore interpolate the detector images row by row onto the regular grid of wavelength $\{ \lambda_i \}$ to obtain monochromatic 2D slices of the point cloud and simplify the PSF fitting process. In the following, the fitted PSF will be either a PSF model from \texttt{WebbPSF} or a reference star with its own point cloud. We linearly interpolate not only the flux, but also the corresponding $\Delta \mathrm{RA}_\mathrm{pix}$, $\Delta \mathrm{Dec}_{\mathrm{pix}}$, and flux error maps for each exposure. 
The interpolation of the spatial coordinates is necessary because of the curvature of the spectral traces on the detector, which means that the spatial coordinates slowly vary along a pixel row. When we refer to the pixel sampling in the spatial and spectral direction, we actually mean the vertical and horizontal directions on the detector, which are not perfectly aligned with the true spatial and spectral direction, but very close.
We do not interpolate bad pixels so any rectified pixels neighboring a bad pixel are also marked as bad. In the rectified images, each column has a constant wavelength, but the spatial sampling remains irregular as shown in Figure \ref{fig:regwvsinterp}. 
The worst of the NIRSpec undersampling is in the vertical detector directions (i.e., spatial) so interpolating over the horizontal detector directions (i.e., spectral) comes with a limited penalty on the systematics compared to the spatial direction. Indeed, defining the sampling as the ratio between the FWHM and the pixel width, the sampling varies between $1$--$1.6$ across the $3$--$5\,\mu\mathrm{m}$ range in the spatial direction and $1.7$--$2.7$ in the spectral direction assuming $R\sim2,700$.

\subsection{Issue with dithering and spatial sub-pixel sampling}
\label{sec:sampling}

\begin{figure*}
  \centering
  \includegraphics[trim={0cm 0cm 0cm 0cm},clip,width=1\linewidth]{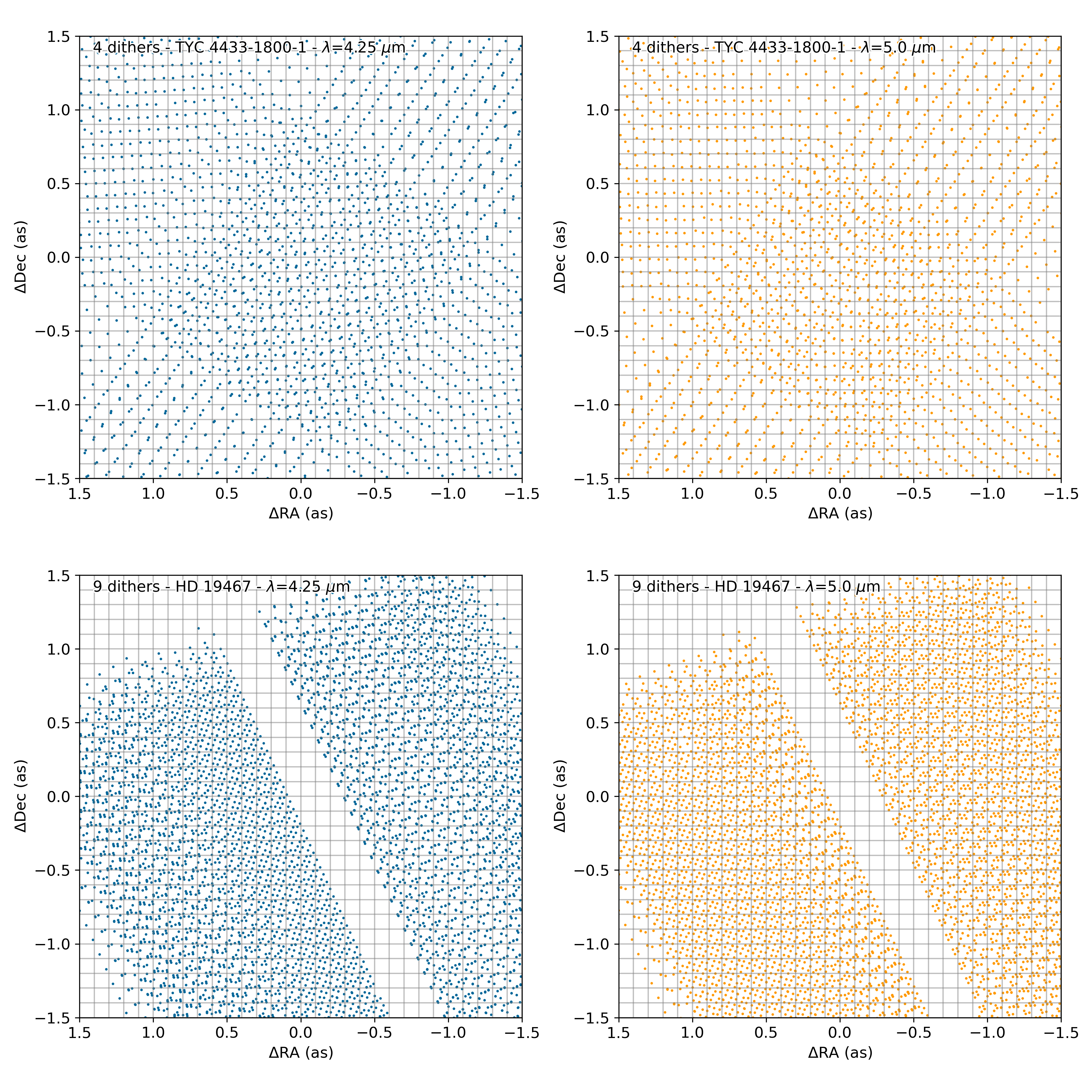}
  \caption{Non-uniform spatial sampling after combining multiple dithers, due to interactions between the IFU slice offsets and the dither pattern offsets. Compare with Figure \ref{fig:regwvsinterp} right panel which shows the spatial sampling of a single dither position. (Upper left) Example of spatial sampling after combining a 4-point nod at $4.25\,\mu\mathrm{m}$. This is the photometric standard dataset featuring TYC 4433-1800-1 (program 1128). (Upper Right) Illustration of the different sampling for the same dataset at $5\,\mu\mathrm{m}$. (Bottom panels) Similar illustrations for the HD~19467 science dataset presented in this work with 9 dithers from the small cycling pattern.}
  \label{fig:sampling}
\end{figure*}

During analyses of these data, we found that using the 9-point cycling dither pattern did not result in as uniform an improvement in the spatial sampling as had been desired. 
Although the 2D sampling of the point cloud is not regular, the density of points in the spatial dimensions of a single exposure is mostly uniform in sky coordinates with a typical separation of $\sim0.1''$ between any neighboring data points.
Ideally, a four-point dither strategy should improve the sampling by a factor two down to $\sim0.05''$ between data points, and the nine-point dither should in principle improve by a factor of three. However, we show in Figure \ref{fig:sampling} that the spatial sampling density resulting from the current dithering patterns is not uniform in the field of view due to sub-pixel offsets between slices. 
For parts of the field of view, some pixel phases never get sampled despite the various dither positions. Consequently, the spatial sampling of a science target might therefore not be significantly improved by the dithering strategy depending on its position in the field of view. Additionally, the sampling density is shown to vary with wavelength. A larger number of dithers is necessary compared to the recommended value of 4 to better address the spatial undersampling of the NIRSpec IFU.

\subsection{Spectral Extraction by Fitting a simulated PSF from \texttt{WebbPSF}}
\label{sec:webbpsffit}

There exist two general strategies to extract the flux from detector images: aperture photometry and PSF fitting, which might be referred to as box and optimal extraction in the context of spectral extraction. Hybrid strategies can also exist in practice.
Aperture photometry has the advantage of a simpler implementation and does not require an accurate model PSF. It is however particularly sensitive to masked bad pixels which need to be interpolated over to account for their unknown flux contribution. The finite size of the aperture also needs to be calibrated to account for the missing flux in the tail of the PSF. PSF fitting provides the most precise (\textit{i.e.}, small error bars) method to extract the flux because it optimally weighs down pixels with lower S/N. It can however be less accurate (\textit{i.e.}, biased). It is also typically more robust to masked bad pixels because they are simply removed from the $\chi^2$ calculation. However, PSF fitting requires an accurate model to avoid systematic biases, which can be difficult to obtain. 

The spatial undersampling of the NIRSpec IFU adds to the challenge as small-aperture extractions lead to spurious oscillations in the spectra \citep{Law2023}. The need for large-aperture extractions to address this issue limits the ability to extract spectra of closely separated or high-contrast targets. PSF fitting could however be used to retrieve the individual spectra of close binaries and blended stars by jointly fitting multiple PSFs.
In the context of high-contrast science, aperture photometry tends to underperform in the speckle-dominated regime. Indeed, fitting a PSF ensures a minimal contamination of the underlying speckles in the final companion spectrum.

\begin{figure*}
  \includegraphics[trim={0.75cm 0.75cm 1cm 0cm},clip,width=1\linewidth]{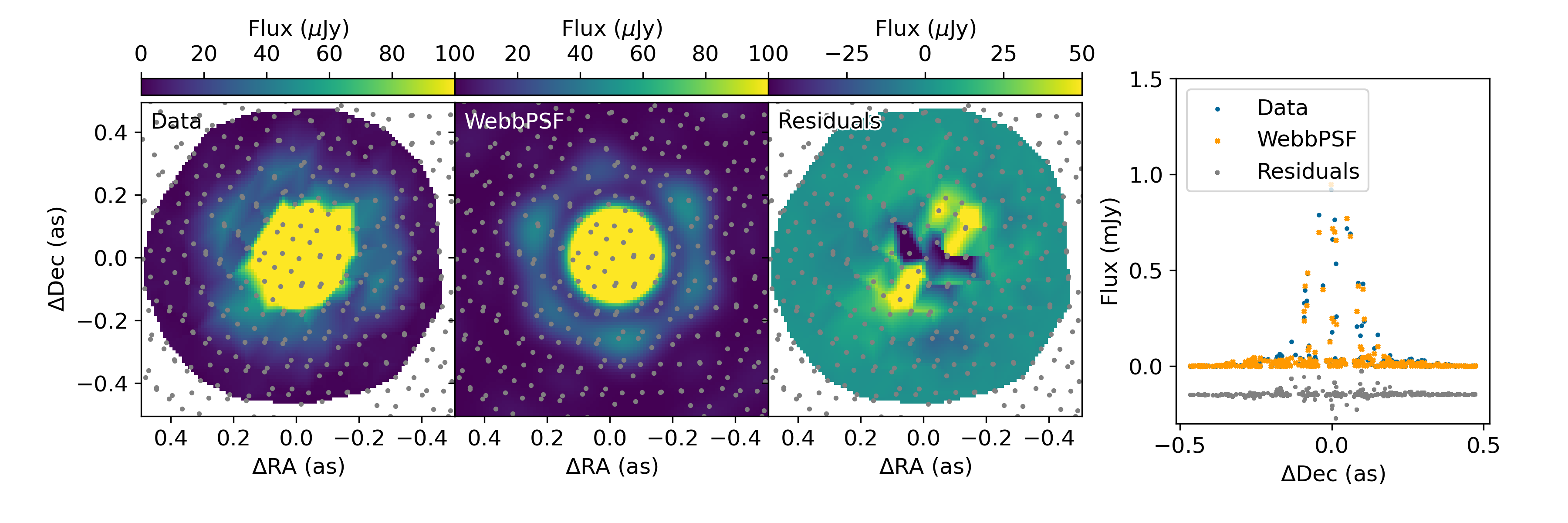}
  \caption{Fitting a simulated PSF from \texttt{WebbPSF} to the 2D point cloud. (Left) Data at a fixed wavelength: $4.7\,\mu\mathrm{m}$. The scatter plot indicates the spatial sampling of the point cloud data being fitted. The image is a linear interpolation of the 2D point cloud only used for visualization purposes. The data itself is not interpolated in the fitting routine. (Middle Left) Image of the super-sampled model PSF that is sampled onto the 2D point cloud of the science data. Unlike the data, the PSF model is interpolated in the fitting routine, which is fine because the model is numerically well sampled. (Middle right) Linear interpolation of the 2D point-cloud residuals: data minus best-fit model. (Right) Illustration of the 2D point cloud as a function of the declination showing the data, model, and residuals. The residuals were offset for clarity. This is data of the A0 star TYC 4433-1800-1 (program 1128) with a 4-point nod.}
  \label{fig:fitwebbpsf}
\end{figure*}

We argue that PSF fitting in the point cloud is the most promising alternative to leverage the full potential of the NIRSpec IFU by addressing the bad pixel and spatial undersampling issues.
We propose an implementation of this technique that uses a simulated PSF from \texttt{WebbPSF} \citep{Perrin2012SPIE.8442E..3DP,Perrin2014SPIE.9143E..3XP} to model data from the point cloud after it has been interpolated on a regular wavelength grid (see Section \ref{sec:wvinterp}) \textit{but without any spatial interpolation}. We combine all the dither positions to obtain a better sampled 2D point cloud before fitting the PSF.
Figure \ref{fig:fitwebbpsf} illustrates the data, model and residuals after fitting for the flux and centroid using a weighted $\chi^2$. 
The model PSF is generated at each wavelength of $\{\lambda_i\}$. 
We make the simplifying assumption that the PSF is spatially invariant across the IFU FOV. 
We choose a pixel scale of $0.1''$, a field of view of the simulated PSF to $6''$ to accomodate shifting the PSF model relative to the $3''$ IFU, and an oversampling of $10$. 
A super-sampled effective PSF \citep[ePSF][]{Anderson2000PASP..112.1360A} that includes the pixel area broadening is computed by convolving the model PSF from \texttt{WebbPSF} with a $0.1''$ square top hat. However other detector systematics such as the charge diffusion are not included. The model PSF is calculated in the instrument reference frame, but the data is fitted in sky coordinates. We therefore rotate the model PSF coordinates accordingly. The angle from North to the V3 axis (positive toward East) can be found in the \texttt{ROLL{\_}REF} fits keyword ($65.01^{o}$ for obs. 4). The angle from V3 to the vertical y-axis of NIRSpec is found in the \texttt{V3I{\_}YANG} fits keyword and is equal to $138.97^{o}$.

To extract a spectrum, we iterate over wavelengths, fitting a normalized PSF model to the point cloud subset at each wavelength.  The resulting flux values at each wavelength then directly yield the spectrum. 

We remind readers that the point cloud is isomorphic to the set of valid detector pixels (i.e. pixels within the IFU FOV spectral traces and not masked out as bad) via only a switch in the coordinate system for labeling those pixels. Thus, fitting a PSF in point cloud space to a particular wavelength is equivalent to fitting the PSF directly to the detector pixels illuminated by that wavelength. Doing so with a point cloud that combines values from the several dithers, as we do here, is equivalent to fitting the PSF to the relevant detector pixels for all the dithered images simultaneously.

\subsection{Validation of the spectral extraction}
\label{sec:specextract}

\begin{figure*}
  \centering
  \includegraphics[trim={0cm 0cm 0cm 0cm},clip,width=1\linewidth]{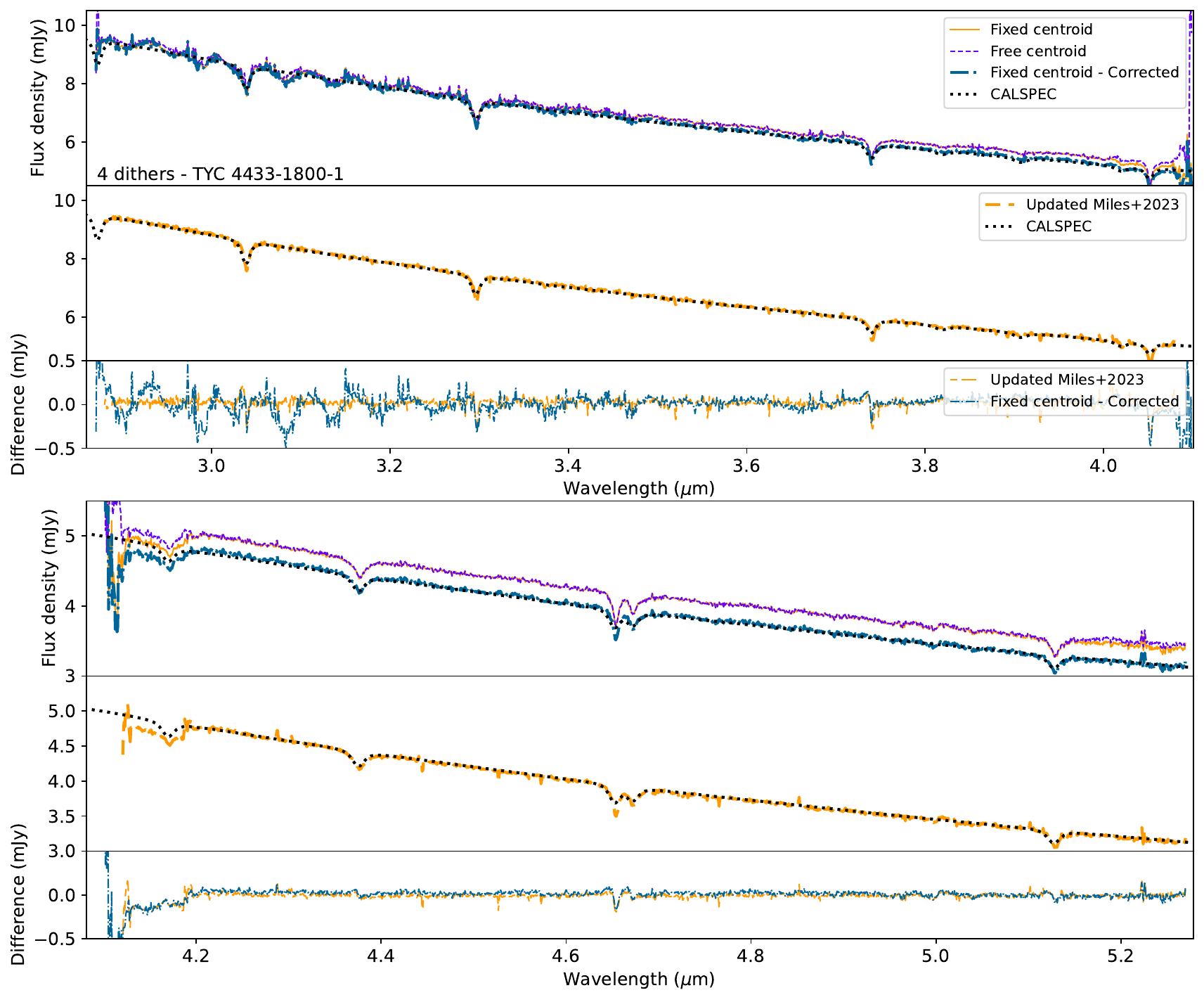}
  \caption{The spectrum of the A0 standard star TYC 4433-1800-1. The data comes from the photometric calibration program 1128 with a 4-point nod. 
  For each wavelength range, the three panels respectively include from top to bottom: the spectral extraction featured in this work, a spectral cube-based extraction, the difference between the two methods and a reference photometric calibration.
  (Top) The spectrum labeled ``free centroid'' is extracted by fitting the centroid and amplitude of a model PSF from \texttt{WebbPSF} to a 2D point cloud at each wavelength. ``fixed centroid'' means that only the flux was fitted for and the centroid was fixed to its median value in each detector. The ``Fixed centroid - Corrected'' spectrum was scaled by a linear trend to match the reference CALSPEC spectrum of TYC 4433-1800-1 \citep{Bohlin2014PASP..126..711B, Bohlin2022stis.rept....7B}. 
  (Middle) We show an updated extraction of the spectrum of TYC 4433-1800-1 originally published in \citet{Miles2023ApJ...946L...6M} obtained in private communication from the high-contrast ERS collaboration. This spectrum was extracted using aperture photometry of the spectral cubes built by the JWST science calibration pipeline. The \texttt{WebbPSF}-extracted spectrum remains impacted by the spatial undersampling at the shortest wavelengths. Future improvements in the calculation of the PSF model should improve these systematic errors.
  (Bottom) Difference of each spectrum with the CALSPEC spectrum. 
  }
  \label{fig:calspec}
\end{figure*}

\begin{figure*}
  \centering
  \includegraphics[trim={0cm 0cm 0cm 0cm},clip,width=0.5\linewidth]{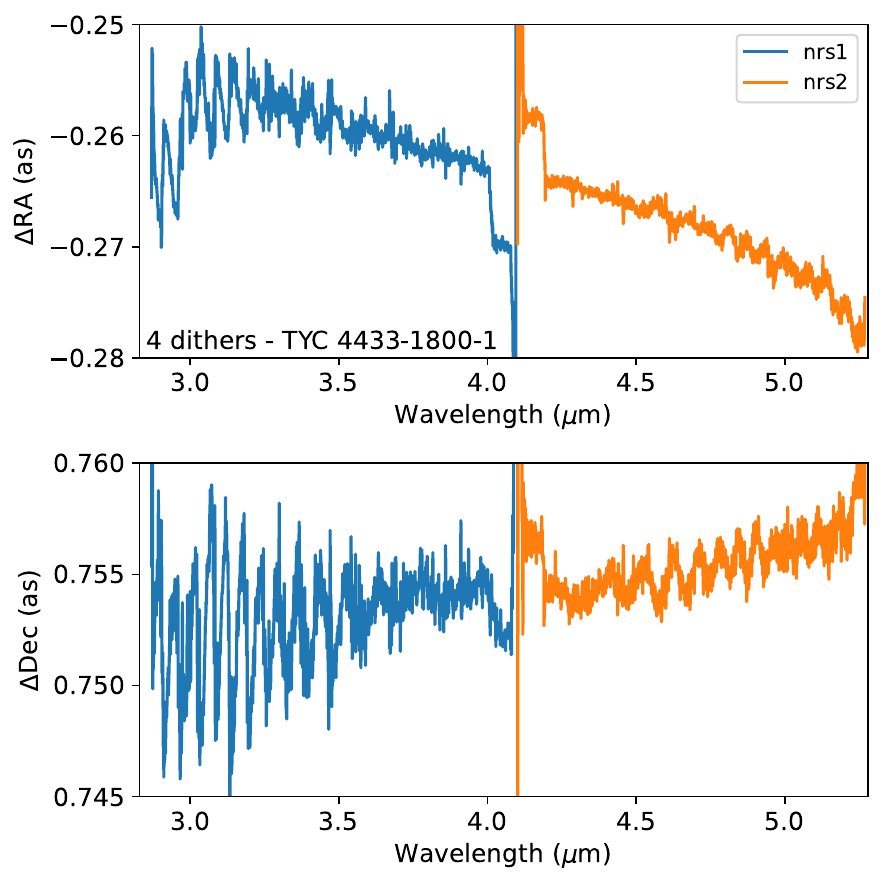}
  \caption{Best-fit centroid of TYC 4433-1800-1 extracted by fitting a model PSF from \texttt{WebbPSF} to the 2D point cloud at each wavelength.  The centroid features a $0.02^{\prime\prime}$ trend across the $3-5\,\mu\mathrm{m}$ wavelength range. The high-frequency quasi periodic fluctuations are likely caused by systematics coming from the spatial undersampling of the data combined with the imperfect PSF model. Note that the average offsets in RA and Dec not being zero are not a concern for this early commissioning dataset; it is the systematic trend versus wavelength that we draw attention to. The corresponding spectrum of the star is shown in Figure \ref{fig:calspec}.}
  \label{fig:centroid}
\end{figure*}

We validate the PSF-fitting spectral extraction method on the A0 standard star TYC 4433-1800-1 by comparing the retrieved spectrum to its CALSPEC counterpart \citep{Bohlin2014PASP..126..711B, Bohlin2022stis.rept....7B}. We use the dataset from the JWST photometric calibration program 1128
in the same grating and filter configuration (G395H/F290LP) as the HD~19467~B data. The sequence has only 4 dither positions resulting in the spatial sampling illustrated in the top panels of Figure \ref{fig:sampling}.
We extract the spectrum twice: once fitting at each wavelength for both the flux and the PSF centroid location, and the second time only fitting for the flux while fixing the centroid to its median position in each detector. The spectra are shown in Figure \ref{fig:calspec} and the estimated centroid of the star as a function of wavelength is shown in Figure \ref{fig:centroid}. 
% The extracted spectrum features a much lower level of high-frequency noise compared to the aperture flux extraction used in \citet{Miles2023ApJ...946L...6M}.
There remain systematics impacting the spectrum extracted from the point cloud described below. 

First, the apparent position of the star moves by $\sim0.02^{\prime\prime}$  across the $3-5\,\mu\mathrm{m}$ wavelength range, which is $\sim1/5$ of an NIRSpec IFU spaxel. The cause of this distortion is unknown, but plausibly it would be caused by imperfect calibration of the shape of the curved spectral traces. By fixing the star position to its median in each detector, the difference in the estimated flux compared to the free centroid fit is $<3\%$. The worst of the difference is localized within $0.1\,\mu\mathrm{m}$ of the edge of the spectral range on each detector, so we consider this approximation to be acceptable for now. However, assuming the same centroid across both detectors ($3-5\,\mu\mathrm{m}$) is not good as it leads to a flux difference up to $10\%$.

Another prominent systematic in the extracted spectrum is a $\sim4\%$ peak-to-valley oscillation at the shortest wavelengths ($<3.5\,\mu\mathrm{m}$). This feature can be explained by the combination of the imperfect \texttt{WebbPSF} simulations and the worse spatial sampling of the PSF at shorter wavelengths, exacerbated by having only 4 dithers on this calibration dataset. Although these oscillations are similar to the ones appearing in reconstructed spectral cubes, they are neither a fundamental limitation of the data nor the methodology. These oscillations should be significantly reduced, or removed, as more effort is put in improving the accuracy of the PSF model. However, we strongly recommend using a larger number of dither positions for NIRSpec IFU observations to limit this type of systematic. 

There appears to be a flux calibration systematic in the retrieved spectrum when compared to the CALSPEC reference spectrum.
The measured spectrum deviates from CALSPEC by a chromatic factor that follows an approximately linear trend from $3-5\,\mu\mathrm{m}$ up to $10\%$ of the continuum. 
% This is likely due to imperfect PSF normalization or aperture corrections at each wavelength (see Section \ref{sec:webbpsffit}), which we expect should be improved with iterations of the PSF modeling. 
This is likely due to the fact that our flux extraction method differs from the one used during the NIRSpec flux calibration.
The systematic is calibrated out by multiplying extracted spectra by a best-fit linear trend: $-0.03128\times\lambda/(1\,\mu\mathrm{m})+1.09007$. 
The corrected spectrum for TYC 4433-1800-1 is shown in purple in Figure \ref{fig:calspec}. After compensation for that scaling, the extracted spectrum has excellent agreement with the CALSPEC reference spectrum for this source. 

At the time of this writing, the IFU flux calibration should be considered accurate to $\sim5\%$ for the NIRSpec \texttt{FFlat} calibration file \texttt{jwst\_nirspec\_fflat\_0102.fits} from the CRDS database.
% This current calibration uncertainty remains the dominant source of systematics in this work for absolute flux accuracy. 

We conclude that the performance of the point cloud spectral extraction is satisfactory because the extraction systematics (e.g., bad pixels, oscillations) are not the dominant source of noise at high contrast in the subsequent work. 
The WebbPSF models for the NIRSpec IFU PSF are expected to further improve from ongoing work, including a cycle 2 PSF calibration program.  Current PSF models are already sufficient to show that forward modeling the point cloud is the most promising path to address the spatial sampling limitations of the NIRSpec IFU.

\subsection{A cross-check with NIRCam coronagraphy}
\label{sec:nircam}

We perform an additional cross check relative to NIRCam coronagraphy of the same target \citep{Greenbaum2023}. HD~19467 and its brown dwarf companion were observed with JWST/NIRCam \citep{Rieke2003,Horner2004,Rieke2023} bar mask coronagraphy on 2022 August 12 (program 1189; PI Thomas Roellig). Details on the observational setup can be found in \citet{Greenbaum2023}. For this cross check, we performed a re-reduction of the NIRCam data using updated pipelines and calibration files, which are significantly improved over the earlier versions available at the time of publication of \citet{Greenbaum2023}. We note that the reference star observation for this program failed, so we limit our re-reduction to use angular differential imaging \citep[ADI,][]{marois_angular_2006} using the two available roll angles of the science target. We specifically do not attempt a reference star subtraction using synthetic reference PSF images (synRDI) as was done by \citet{Greenbaum2023}.

For re-reduction, we employ the \texttt{spaceKLIP}\footnote{\url{https://github.com/kammerje/spaceKLIP}} community pipeline \citep{Kammerer2022,Carter2023} for reducing and analyzing JWST coronagraphy data. In summary, \texttt{spaceKLIP} uses the JWST data reduction pipeline to perform ramp fitting and the flux calibration of the individual images, and then uses custom routines to clean bad pixels, recenter the images on the position of the host star (which is attenuated by the coronagraphic mask), align the images, and prepare them for PSF subtraction using Karhunen-Lo\`eve image processing \citep[KLIP,][]{Soummer2012}. The PSF subtraction (here KLIP ADI) is performed using  \texttt{pyKLIP}\footnote{\url{https://bitbucket.org/pyKLIP/pyklip/src/master/}} \citep{Wang2015} and finally the companion properties are extracted by forward-modeling the companion's PSF with the \texttt{WebbPSF\_ext}\footnote{\url{https://github.com/JarronL/webbpsf_ext}} tool \citep{Leisenring_WebbPSF_Extensions_2021,Girard2022} and the \texttt{pyKLIP} forward-modeling framework \citep{Pueyo2016}. The details of the different \texttt{spaceKLIP} processing steps are discussed in \citet{Kammerer2022} and \citet{Carter2023}. However, several improvements have been made since then to enhance the photometric precision of the extraction. Firstly, the wavelength-dependent transmission of the coronagraphic mask (COM) substrate \citep{Krist2010} is now included, yielding an increase in measured companion flux of $\sim5$--10\% in the wavelength regime considered here. Secondly, we correct for aperture losses due to the limited size of the \texttt{WebbPSF\_ext} PSF models, which increases the measured companion flux by another $\sim10$--15\% if compared to the \texttt{spaceKLIP} version used in \citet{Carter2023}.

We obtain revised values for the photometry of the brown dwarf in the six observed filters (See Table~\ref{tab:photometry}). 
The updated $5\sigma$ sensitivity curves for these observations are shown in Appendix \ref{app:NIRCamContrast}.
Our revised photometry is $\sim30$--100\% brighter than the values presented in \citet{Greenbaum2023}. The initial photometric calibration used in \citet{Greenbaum2023} based on ground test data was inaccurate, and this was improved by in-flight photometric calibrations which became available in fall 2023.
We confirm the assumed coronagraphic mask throughput of $0.92$ at the position of HD~19467~B from \citet{Greenbaum2023} as we find values of 0.914--0.918 depending on the filter.
Besides the inacurrate calibrations, part of the difference in the estimated photometry could also be caused by aperture losses due to the limited size of the PSF models from \texttt{WebbPSF}. 
The individual photometric systematic terms are detailed as follows: $\sim 1\%$ from the uncertainty of the flux calibration of the JWST science calibration pipeline, $\sim 2\%$ from numerical inaccuracies in the forward-modeled PSFs from \texttt{WebbPSF}, and $\sim2\%$ from the uncertainty on the coronagraphic mask throughput due to the uncertainty on the companion/mask position. We will therefore assume a total $5\%$ systematic uncertainty for the NIRCam photometry that needs to be added in quadrature to the statistical uncertainties presented in Table~\ref{tab:photometry}.

The revised NIRCam photometry agrees within statistical uncertainties with the extracted NIRSpec RDI spectrum from this work and the VLT/NACO measurements from \citet{Maire2020} (see Figure \ref{fig:summaryRDIspec}).

\begin{table*}[!ht]
    \begin{tabular}{l c c l c c c}
        Filter & Flux & Flux& Systematic error& $\Delta\rm{mag}$ & $\rm{TP}_{\rm{MSK}}$ & $\rm{TP}_{\rm{COM}}$ \\
         & ($\mu$Jy) & ($10^{-17}\,\mathrm{W}/\mathrm{m}^2/\mu\mathrm{m}$)& ($10^{-17}\,\mathrm{W}/\mathrm{m}^2/\mu\mathrm{m}$) & & & \\
        \hline
        \multicolumn{7}{c}{NIRCam coronagraphy}\\
        F250M & $15.6\pm1.1$ & $0.74\pm0.05$ &  0.04 (5\%) & $13.41\pm0.08$ & 1.000 & 0.963\\
        F300M & $31.5\pm0.3$ & $1.06\pm0.01$ &  0.05 (5\%) & $12.33\pm0.01$ & 1.000 & 0.898\\
        F360M & $62.4\pm0.5$ & $1.42\pm0.01$ &  0.07 (5\%) & $11.21\pm0.01$ & 0.999 & 0.951\\
        F410M & $187.1\pm1.3$ & $3.37\pm0.02$ &  0.17 (5\%) & $9.78\pm0.01$ & 0.966 & 0.958\\
        F430M & $125.7\pm1.1$ & $2.06\pm0.02$ &  0.10 (5\%) & $10.10\pm0.01$ & 0.943 & 0.947\\
        F460M & $84.4\pm1.0$ & $1.18\pm0.01$ &  0.06 (5\%) & $10.33\pm0.01$ & 0.918 & 0.901\\
        \hline
        \multicolumn{7}{c}{NIRSpec IFU (from RDI)}\\
        \hline
        F360M & $63.3\pm7.6$ & $1.44\pm0.17$ &  0.07 (5\%) & $11.15^{+0.14}_{-0.12}$ & - & -\\
        F410M\tablenotemark{*} & $186.8\pm7.4$ & $3.36\pm0.13$ &  0.17 (5\%) & $9.75^{+0.04}_{-0.04}$ & - & -\\
        F430M\tablenotemark{*} & $130.7\pm7.6$ & $2.14\pm0.12$ &  0.11 (5\%) & $10.04^{+0.06}_{-0.06}$ & - & -\\
        F460M & $87.8\pm7.9$ & $1.23\pm0.11$ &  0.06 (5\%) & $10.26^{+0.10}_{-0.09}$ & - & -\\
        F356W\tablenotemark{*} & $65.1\pm5.0$ & $1.52\pm0.12$ &  0.08 (5\%) & $11.18^{+0.09}_{-0.08}$ & - & -\\
        F444W\tablenotemark{*} & $142.8\pm6.8$ & $2.19\pm0.11$ &  0.11 (5\%) & $9.91^{+0.05}_{-0.05}$ & - & -\\
        Lp\tablenotemark{*} & $115.1\pm6.7$ & $2.42\pm0.14$ &  0.12 (5\%) & $10.43^{+0.07}_{-0.06}$ & - & -\\
        \hline
    \end{tabular}
    \caption{Estimated photometry of HD~19467~B. (Top) Revised \emph{JWST}/NIRCam photometry from the data originally presented in \citet{Greenbaum2023}. The quoted uncertainties are the statistical uncertainties ($1\sigma$) from the MCMC fit of the forward-modeled NIRCam PSF to the data. Information on systematic uncertainties can be found in Section~\ref{sec:nircam} for NIRCam and Section~\ref{sec:RDIextract} for NIRSpec. $\rm{TP}_{\rm{MSK}}$ and $\rm{TP}_{\rm{COM}}$ denote the throughput of the coronagraphic mask and the COM substrate, respectively. (Bottom) Photometry derived from the NIRSpec IFU RDI spectrum of HD~19467~B shown in Figure \ref{fig:summaryRDIspec}.}
    \label{tab:photometry}
\tablenotetext{*}{These filters partially overlap with the wavelength gap between the two detectors in NIRSpec. The photometry was therefore computed after interpolating the spectrum in the gap with a BT-Settl model.}
\end{table*}

\section{Forward model of the NIRSpec detectors}
\label{sec:FM}

\subsection{Context}

In Section \ref{sec:specextract}, we demonstrated the spectral extraction of a point source in the NIRSpec field of view. The goal of this section is to study a faint point source next to a bright one by reconstructing the astrophysical scene in the detector images. The main challenge is to define a flexible model of the starlight that can accurately reproduce the chromatic and spatial fluctuations of the speckle field, and use that model to quantify the contrast limits on detection of faint companions.
The idea behind a forward model is to characterize a companion without an intermediate spectral extraction step. The atmospheric inference is performed using a likelihood that is defined directly on the observed data instead of a 1D spectrum.

For the reasons discussed above, reconstructing a regularly-sampled spectral cube from the detector images is a challenging task even when combining multiple dithers. Interpolating the detector pixels to build a spectral cube comes with a heavy penalty in terms of the systematic noise floor. 
We therefore propose to fully model the astrophysical scene and fit the data directly in the NIRSpec detector images to leverage the most information out of the data. 
We fit the companion signal and the starlight for each detector and each dither position separately. The full dataset is only later combined. 

\begin{figure*}
  \centering
  \includegraphics[trim={0cm 0cm 0cm 0cm},clip,width=1\linewidth]{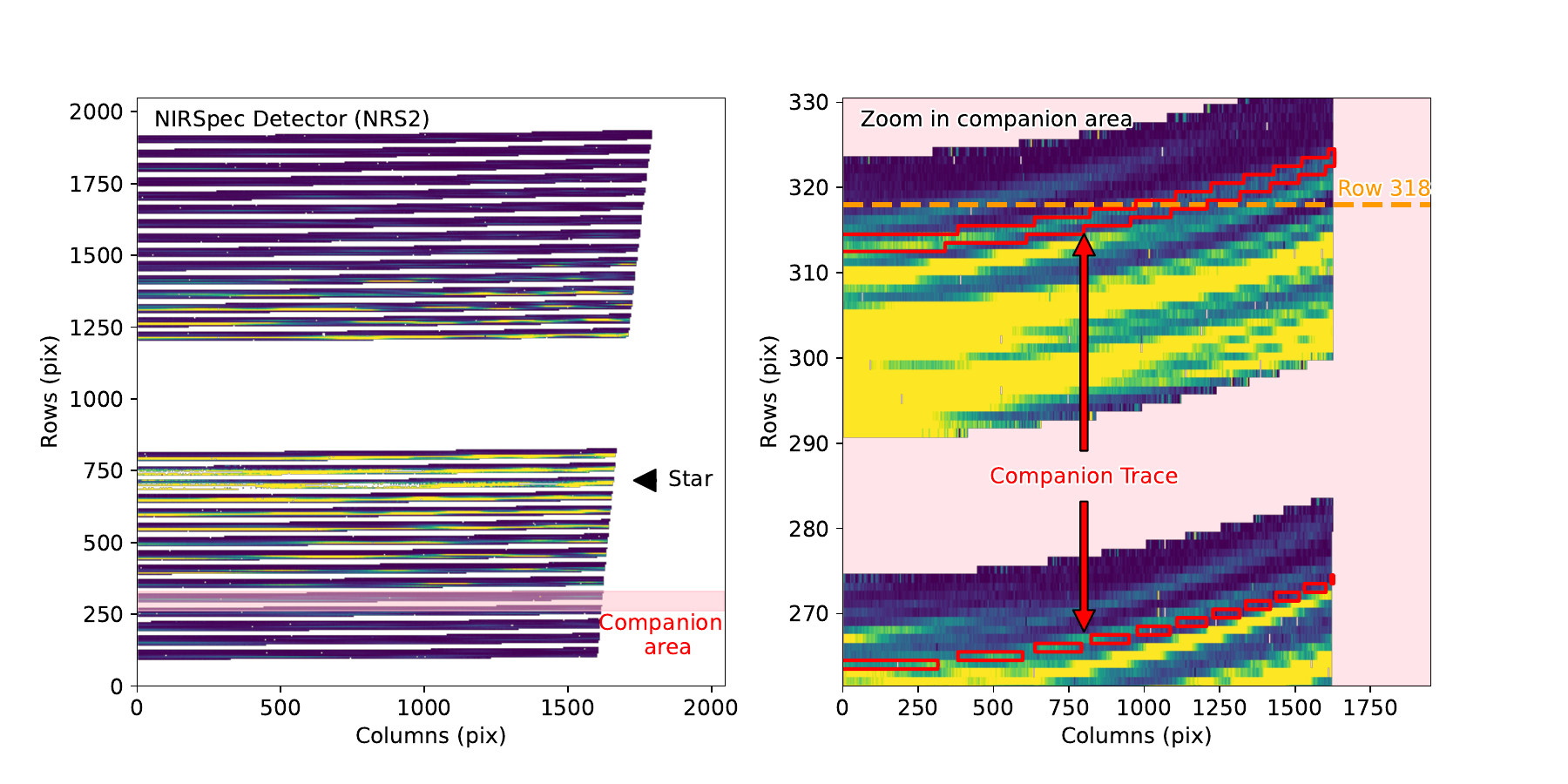}
  \caption{Topology of the companion signal in a NIRSpec detector image (file \texttt{jw01414004001\_02101\_00001\_nrs2\_cal.fits}). (Left) Full NRS2 detector image. The arrow points at the region of the star that is also partially saturated. The area of the IFU slices that includes the bulk of the companion signal is shaded in pink. (Right) Zoom in the highlighted companion area. The pixels that are within $0.1^{\prime\prime}$ of the companion position are painted in red. The dashed orange cut is the row data shown in Figure \ref{fig:fm}.}
  \label{fig:FM_im}
\end{figure*}

Figure \ref{fig:FM_im} illustrates the topology of the companion HD~19467~B signal on an example image of NIRSpec's longer-wavelength detector NRS2 (File \texttt{jw01414004001\_02101\_00001\_nrs2\_cal.fits}). A full description of the configuration of the IFU spectra on the two NIRSpec detectors can be found in Figure 6 of \citep{Boker2022}.
For a given companion position, we identify all the pixels that are within a radius of $0.1^{\prime\prime}$ of its coordinates. Choosing this radius is a trade off between computational tractability and the desire to account for as much of the companion flux as possible. The curvature of the spectral traces leads the companion signal to cross $R$ rows on the detector across two neighboring slices of the IFU, with $R=23$ in the example of Figure \ref{fig:FM_im}. The goal of the forward model is to jointly reproduce these 23 rows as accurately as possible with a minimal number of parameters.

\begin{figure*}
  \centering
  \includegraphics[trim={0cm 0cm 0cm 0cm},clip,width=1\linewidth]{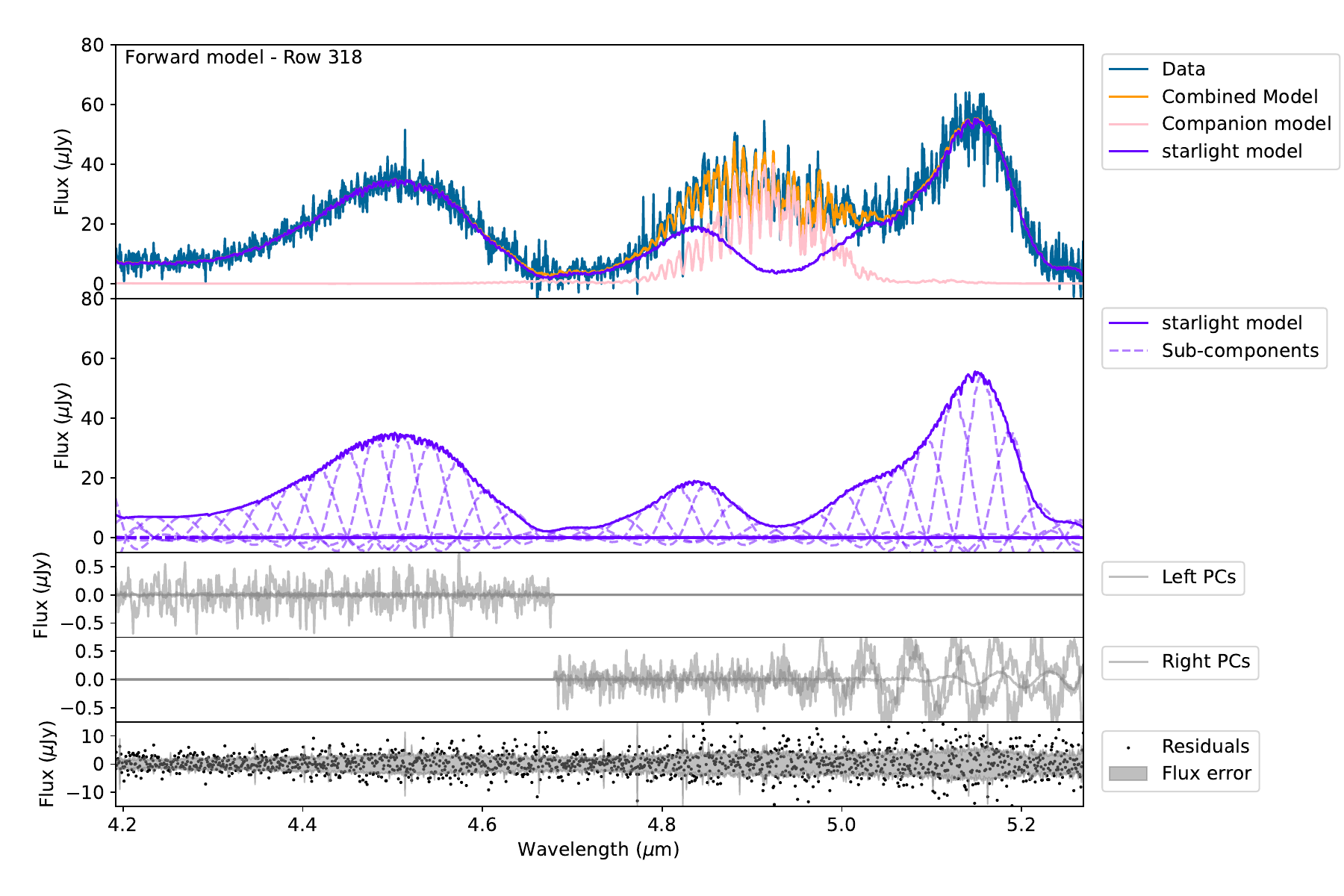}
  \caption{Joint forward model of the starlight and companion signal in NIRSpec detector images. (Top panel) Data and model for a single row of the detector image shown in Figure \ref{fig:FM_im}. ($2^{\mathrm{nd}}$ panel) Decomposition of the starlight model as a linear combination of 40 spline sub-components. ($3^{\mathrm{nd}}$ and $4^{\mathrm{th}}$ panels) Model contribution from the left and right principal components (PC). (Bottom panel) Residuals of the forward model fit compared to the flux error from the JWST science calibration pipeline.}
  \label{fig:fm}
\end{figure*}

We already described how the starlight can be fitted row by row on the detector, in Section \ref{sec:contnormspec}, where we derived an empirical continuum-normalized spectrum of the star, which was used to model the starlight in each row by modulating its continuum with the spline model. 
Here, we generalize this model to fit multiple rows of the detector at once jointly with a model of the companion. It uses the same framework developed in \citet{Ruffio2021AJ....162..290R} and \citet{Agrawal2023AJ....166...15A}, but adapting the forward model to work in detector images and adding a regularization on the starlight continuum.
Figure \ref{fig:fm} is a partial illustration of the data focusing on row number 318. The position of this row is marked in Figure \ref{fig:FM_im} for reference. The quality of the data is highlighted by the fact that the molecular features in the companion spectrum can be distinctly identified in a single exposure of a flux-calibrated detector image without starlight subtraction nor spectral extraction (see Figure~\ref{fig:fm}). We now show how this framework can be used in detection mode to look for faint objects, like a planet, around a bright star.

\subsection{Definition of the model}
\label{sec:FMdef}

For a fixed companion position in the FOV, we forward model all the $R$ rows intersecting the companion trace (Figure \ref{fig:FM_im}). We model the data as a linear combination of a companion model ${\bm p}$, a starlight contribution modulated by the spline ${\bm M}_{\mathrm{spl}}$, and additional terms ${\bm M}_{\mathrm{pc}}$ modeling residual features. The latter takes the form of principal components of the residuals following the method used in \citet{Ruffio2021AJ....162..290R}. 
$K$ is the number of spline nodes in each row as introduced in Section \ref{sec:contnormspec}.
The full model takes the form of the block matrix ${\bm M}$:
\begin{equation}
     {\bm d} = {\bm M}{\bm \phi} + {\bm n},\,\mathrm{with}\,
    {\bm M} =
    \begin{bmatrix}
    {\bm p} & {\bm M}_{\mathrm{spl}}  & {\bm M}_{\mathrm{pc}}   \\
     0 & I_{RK}     & 0 
    \end{bmatrix}
\label{eq:fm}
\end{equation}
The identity matrix $I_{RK}$ of size $\sim R\times K$ enables the regularization of the starlight continuum. The effect of the regularization on the linear model inversion compared to the mathematical framework presented in \citet{Ruffio2019AJ....158..200R} is discussed in Appendix~\ref{app:newmaths}.
The other blocks in ${\bm M}$ are described in more detail in the following sections.

The data vector ${\bm d}$ of pixel fluxes, the flux errors ${\bm s}$, and the linear parameters are defined as a concatenation of vectors for each row:
\begin{equation}
     {\bm d} = 
    \begin{bmatrix}
    {\bm d}_{\mathrm{row,1}} \\
    \vdots \\
    {\bm d}_{\mathrm{row,R}} \\
    {\bm \mu}_{{\bm \phi},1} \\
    \vdots \\
    {\bm \mu}_{{\bm \phi},R}
    \end{bmatrix},
     {\bm s} = 
    \begin{bmatrix}
    {\bm s}_{\mathrm{row,1}} \\
    \vdots \\
    {\bm s}_{\mathrm{row,R}} \\
    {\bm s}_{{\bm \phi},1} \\
    \vdots \\
    {\bm s}_{{\bm \phi},R}
    \end{bmatrix},
    {\bm \phi} = 
    \begin{bmatrix}
    \phi_{\mathrm{comp}} \\
    {\bm \phi}_{\mathrm{row,1}} \\
    \vdots \\
    {\bm \phi}_{\mathrm{row,R}} \\
    {\bm \phi}_{\mathrm{pc,1}} \\
    \vdots \\
    {\bm \phi}_{\mathrm{pc,R}} 
    \end{bmatrix}.
\end{equation}
The data vector ${\bm d}$ of pixel fluxes and errors ${\bm s}$ are approximately $R\times (N+K)$ long. The number of valid pixels $N<2048$ in each row ${\bm d}_{\mathrm{row,i}}$ varies depending on the number of bad pixels masked out in previous steps. The first $\sim R\times N$ elements of the data vector are the valid pixel fluxes from each row ${\bm d}_{\mathrm{row,i}}$, and associated errors ${\bm s}_{\mathrm{row,i}}$. 
The remaining $\sim R\times K$ elements are for regularizing the spline parameters modeling the starlight. The ${\bm \mu}_{{\bm \phi},i}$ are the target value of the spline parameters modeling the starlight in row $i$ as introduced in Section \ref{sec:contnormspec}. The ${\bm s}_{{\bm \phi},i}$ are the vectors of standard deviations of the Gaussian prior for the same parameters.
The vector ${\bm s}$ defines the square root of the diagonal elements of the covariance matrix ${\bm \Sigma}$ of the noise in the data. The noise is dominated by photon noise so a diagonal covariance matrix is justified.

There are approximately $1+R\times (K+Q)\sim1000$ free linear parameters in ${\bm \phi}$ that are fitted for. A subset of these parameters might become irrelevant if all the pixels that it would affect are already masked. In this case, the parameter is removed to avoid columns full of zeros in the matrix ${\bm M}$ that would prevent the inversion of the linear least square problem. 
The first linear parameter $\phi_{\mathrm{comp}}$ is the amplitude of the companion model ${\bm p}$ described in Section \ref{sec:compmodel}, which is the only linear parameter that is scientifically relevant. All other linear parameters can be considered nuisance parameters. The vectors ${\bm \phi}_{\mathrm{row,i}}$ are the amplitudes of the starlight continuum at each $K=40$ nodes in a row $i$, which is described in more detail in Section \ref{sec:starmodel}. The vectors ${\bm \phi}_{\mathrm{pc,1}}$ are amplitude of the $Q=6$ principal components terms for each row. The calculation of the principal component is explained in Section \ref{sec:pcmodel}.

\subsection{Companion model}
\label{sec:compmodel}

The companion signal is modeled from a simulated PSF from \texttt{WebbPSF}, $\mathcal{W}_{\mathrm{psf}}({\bm x}_{i},{\bm y}_{i},{\bm w}_{i})$, and a model spectrum of its atmosphere, $\mathcal{P}({\bm w}_{i})$, where ${\bm x}_{i}$, ${\bm y}_{i}$, and ${\bm w}_{i}$ are respectively the $\Delta \mathrm{RA}$, $\Delta \mathrm{Dec}$, and wavelength vectors for the row $i$ in the image. The choice of atmospheric models is discussed later in this section.
The contribution of the companion signal to a single row of the detector is illustrated with a pink line in Figure \ref{fig:fm}.
With the data being already flux calibrated, the absolute flux of the companion can be directly estimated. By normalizing the model spectrum $\mathcal{P}$ in a user-defined spectral filter, $\phi_{\mathrm{comp}}$ now directly represents the absolute flux in the chosen band. We choose to express the NIRSpec detection sensitivity using the NIRCam F444W filter to simplify the comparison with NIRSpec. However, in this context, F444W is only a reference filter for expressing the flux of the companion, but the entire wavelength range of F290LP is used in the fit.

The estimated uncertainty of the companion flux $\phi_{\mathrm{comp}}$ resulting from the fit defines the sensitivity of the observation at any position in the FOV. The companion flux divided by its uncertainty defines its detection S/N.
The full PSF model cube generated in Section \ref{sec:webbpsffit}, which is spatially super sampled and calculated at each wavelength, is $\sim10\,\mathrm{GB}$ for each detector. 
We therefore approximate the companion PSF to limit the computational resources needs of the code.
The simulated PSF is only calculated at the central wavelength of the detector $w_0$, but the input coordinates are scaled to model the magnification of the PSF at each wavelength.
As a result, the companion model vector is defined from the concatenation of the model in each row as follow:
\begin{equation}
     {\bm p} = 
    \begin{bmatrix}
    \mathcal{P}({\bm w}_{1})\mathcal{W}_{\mathrm{psf}}(w_0\,{\bm x}_{i}/{\bm w}_{i},\,w_0{\bm y}_{i}/{\bm w}_{i},w_0) \\
    \vdots \\
    \mathcal{P}({\bm w}_{R})\mathcal{W}_{\mathrm{psf}}(w_0\,{\bm x}_{i}/{\bm w}_{i},w_0\,{\bm y}_{i}/{\bm w}_{i},w_0)
    \end{bmatrix},
\end{equation}
assuming element-wise multiplication and division of vectors.

The most important caveat related to this specific implementation of the forward model is that the estimated flux of the companion is model dependent. 
One way to illustrate the problem is that the peak to valley of the moderate-resolution spectral features is first scaled to fit the data irrespective of its continuum. Then, the starlight model (Section \ref{sec:starmodel}) compensates for any differences in the remaining continuum level. The starlight model is typically flexible enough to accommodate any continuum level of the companion. 
Without a strong prior, the starlight model is even allowed to become negative if it is the preferred solution for the data. 
This is why the continuum information of the companion is effectively lost in this implementation of the forward model.
In theory, the amplitude of the spectral features of the star contains information about the speckle intensity, but the spectral features for HD~19467 are too shallow ($<5\%$ from Figure \ref{fig:contnormspec}) compared to the photon noise to carry any significant constraining power.
It remains possible that a bright and cooler star would contain enough signal in its own molecular lines to constrain the amplitude of the speckles independently of the companion model, but it is not the case for HD~19467. 
However, we propose a method to independently constrain the speckle intensity using the continuum prior and a reference PSF in Section \ref{sec:FMRDI}.

The accuracy of the estimated flux is therefore entirely dependent on the accuracy of the atmospheric model. More specifically, it relies on the ability of the model to relate the amplitude of the spectral features with the amplitude of the continuum. The possible inconsistency between the continuum levels of the model and the wavelength ranges of the two NIRSpec detectors also complicates our ability to accurately combine the reductions of the two detectors.
It is desirable to be able to compare the NIRSpec IFU sensitivity with the NIRCam coronagraphic mode. This is an issue because the NIRSpec forward model is sensitive to the moderate-resolution spectral features and insensitive to the continuum, while the NIRCam imager is sensitive to the continuum and insensitive to the spectral features.

Given the limited accuracy of the currently publicly available atmospheric models across such a broad wavelength range ($3-5\,\mu\mathrm{m}$; e.g. see \citealt{Petrus2023arXiv231203852P}), we prefer to 
use the empirical RDI spectrum extracted in Section \ref{sec:RDI} instead.
By choosing the spectrum of the companion itself to compute the sensitivity, we ensure the accuracy of the calculations.
The choice of the model does not significantly change the detection S/N, but it can affect our understanding of how bright a companion truly is. 

\subsection{Starlight model}
\label{sec:starmodel}

The starlight portion of the model ${\bm M}_{\mathrm{spl}}$ is a generalization of the single row model introduced in Section \ref{sec:contnormspec} (Equation \ref{eq:rowFM}) to the $R$ rows that are overlapping with the companion signal on the detector.
The speckles are modeled as a linear combination of modes that are equivalent to fitting a spline.
The spectral features of the star are first imprinted in these modes by multiplying them by the continuum normalized spectrum of the star $\mathcal{S}$. There are $K$ modes for each row, making the model ${\bm M}_{\mathrm{spl}}$ an approximately $(R\,N) \times (R\,K) \sim 35,000 \times 900$ matrix where $(R\,K)$ is the number of free parameters modeling the starlight. The dimensions are approximate because there is always a number of rows and columns that are removed at each location based on identified bad pixels and FOV edges.
We get
\begin{align}
    {\bm M}_{\mathrm{spl}} & =
    \begin{bmatrix}
    {\bm S}_{1}{\bm M}_{\mathrm{row,1}}  & 0 &    0&\dots& 0   \\
     0     & {\bm S}_{2}{\bm M}_{\mathrm{row,2}} &0&\dots& 0 \\
     0     &0  &&\ddots& \vdots\\
     \vdots     & \vdots & \ddots &\ddots& 0\\
     0     & 0 &\dots&0& {\bm S}_{R}{\bm M}_{\mathrm{row,R}}
    \end{bmatrix} \nonumber \\
    {\bm M}_{\mathrm{row,i}} & =
    \begin{bmatrix}
    {\bm c}_{i1} & \dots & {\bm c}_{iK} 
    \end{bmatrix}, \nonumber  \\
    {\bm S}_{i} & = \mathrm{diag} \left[ \mathcal{S}({\bm w}_{i}) \right],
\end{align}
with ${\bm w}_{i}$ the vector of wavelength for each pixel in row $i$, $ \mathrm{diag} \left[ \mathcal{S}({\bm w}_{i}) \right]$ representing the diagonal matrix with diagonal element $\mathcal{S}({\bm w}_{i})$.

\subsection{Additional components to model residuals}
\label{sec:pcmodel}

\begin{figure*}
  \centering
\gridline{\fig{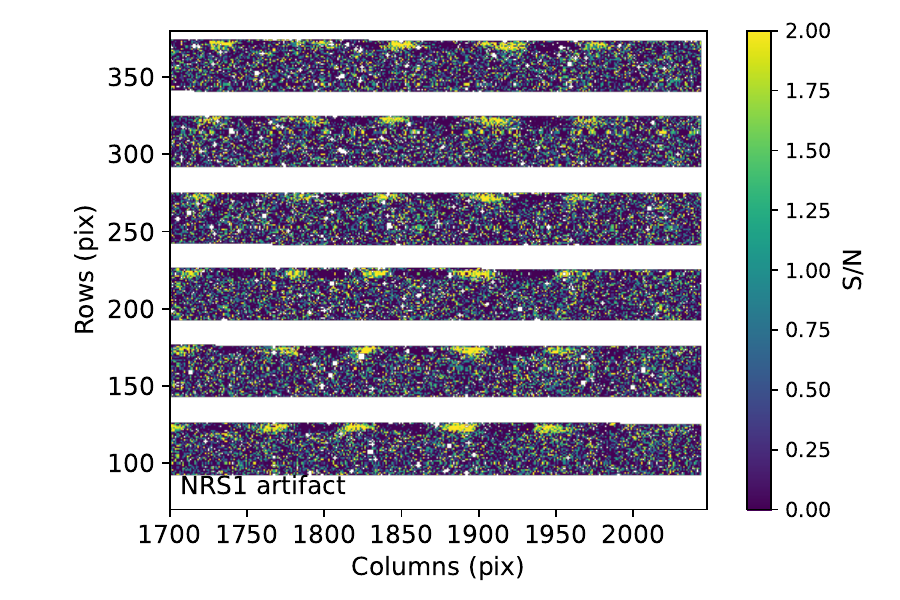}{0.49\textwidth}{(a) Residual S/N image}
		\fig{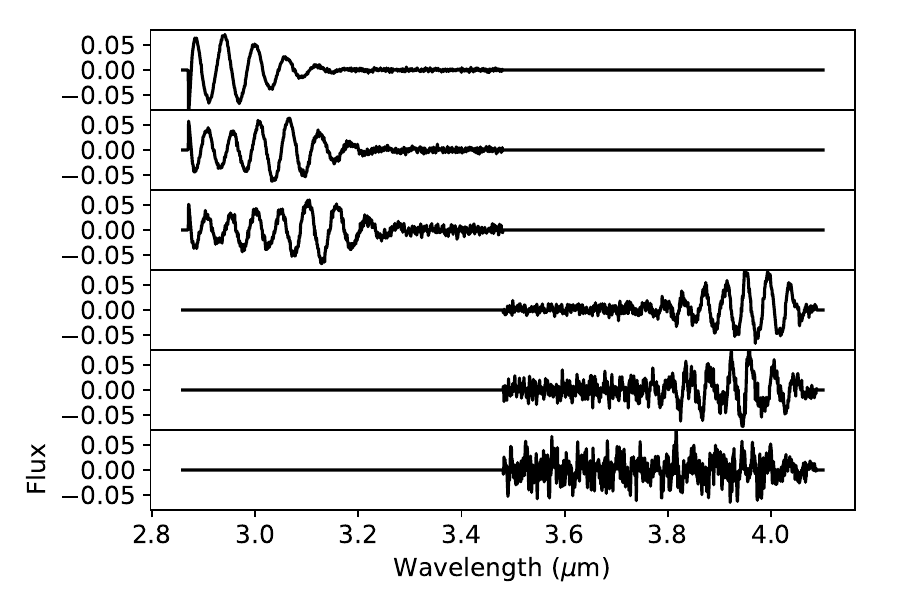}{0.49\textwidth}{(b) Principal components}
          }
  \caption{Identification and modeling of a possible stray light in NIRSpec. (Left) Artifact in the bottom right of the NRS1 NIRSpec detector images (file \texttt{jw01414004001\_02101\_00001\_nrs1\_cal.fits}). The artifact was highlighted by subtracting the best-fit starlight model and dividing the residuals by the pixel flux errors resulting in a pixel signal-to-noise ratio (S/N) map. (Right) Individual left and right principal components (PC) used in the forward model for the same file. The left PCs highlight some edge instabilities of the spline model of the continuum. The right PCs feature the detector artifact shown in the left panel.}
  \label{fig:ghost}
\end{figure*}
 
The joint planet and starlight model is a fair representation of the data, but it can be improved further by modeling residual artifacts. Notably, we identify a quasi-periodic pattern at the $10\,\mu\mathrm{Jy}$ level towards the bottom right quadrant of the NRS1 NIRSpec detector that appears at the top of each slice (Figure \ref{fig:ghost}). These oscillations are not an artifact of the data reduction because they are visible in the stage 1 rate maps from the JWST science calibration pipeline. The origin of the this artifact is unknown, although its amplitude would be consistant with stray light.

Another source of residual artifacts causes oscillations at the edges of the detectors. They appear on the left side of NRS1 and the right side of NRS2 as shown in the third and fourth panel in Figure \ref{fig:fm}. We believe that these oscillations are caused by the steeper curvature of the traces toward the edges of the filter. The stronger curvature leads to more rapid variations of the continuum that the spline model is not able to follow due to the fixed separation between nodes. Future work is necessary to explore new continuum modeling schemes that would better account for the curvature of the trace.

We model those artifacts using principal component analysis of the residuals using similar methods to \citet{Hoeijmakers2018A&A...617A.144H} or \citet{Ruffio2021AJ....162..290R}. First, the residuals from fitting the starlight-only model to the image from Section \ref{sec:contnormspec} are normalized by the noise. Then, each row is interpolated on a $R\sim10,800$ wavelength grid which is separated into the left and right half of the detector. The principal components of the rows in each half are then computed. Finally, we interpolate the principal components on the wavelength sampling of the relevant rows. The reason for separating the two halves of the detector is to decouple the edge issue from the possible stray light when fitting the principal components in the forward model. We found that 3 principal components for each side was enough and improved the fit substantially. There are therefore $L=6$ additional sub-components vectors (${\bm r}_{i1},\dots, {\bm r}_{iL}$) to model each row $i$ of the data. These are plotted for one NRS1 image in Figure \ref{fig:ghost}. The corresponding matrix for the forward model is defined as

\begin{align*}
     {\bm M}_{\mathrm{pca}}  =
     \begin{bmatrix}
    {\bm r}_{11} & \dots & {\bm r}_{1L}   & 0     & \dots & 0      &        &0     & \dots & 0       \\
    0     & \dots & 0      & {\bm r}_{21} & \dots & {\bm r}_{2L}   &        &\vdots&       & \vdots  \\
    \vdots&       &\vdots  &       &       &        & \ddots &0     &  \dots & 0       \\
    0     & \dots & 0      &  0    & \dots & 0      &        &{\bm r}_{R1} & \dots & {\bm r}_{RL}
    \end{bmatrix}.
\end{align*}

\subsection{Companion detection and sensitivity}

\begin{figure*}
  \centering
  \includegraphics[trim={0cm 0cm 0cm 0cm},clip,width=1\linewidth]{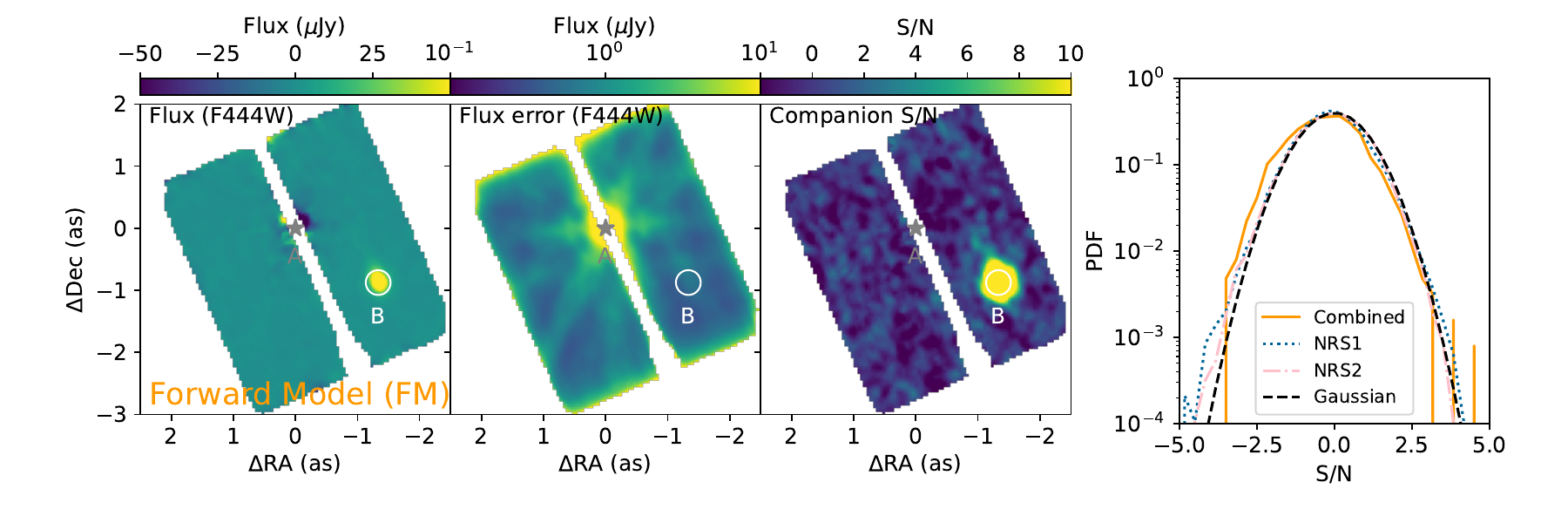}
  \includegraphics[trim={0cm 0cm 0cm 0cm},clip,width=1\linewidth]{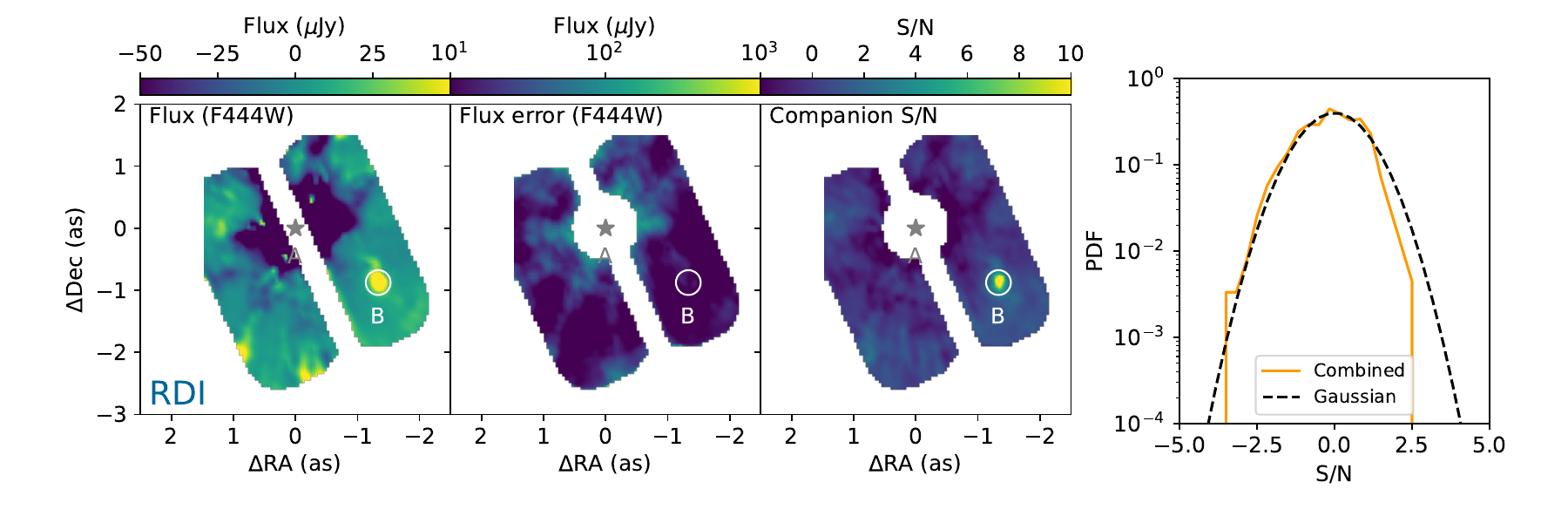}
  \caption{Companion detection maps. From left to right: companion flux map, flux error map, S/N map, and S/N value histogram. The top panels are products of the forward model (FM) and the bottom panels are obtained from reference-star differential imaging (RDI). The absolute fluxes are expressed for the F444W filter but the fits include the entire $2.9-5.3\,\mu\mathrm{m}$ NIRSpec spectral range. The scales and colorbars for the flux and S/N maps are identical between the FM and RDI to allow for cross-comparison. The flux error maps illustrate the spatial variations of the sensitivity due to the 2D profile of the JWST PSF such as the diffraction spikes. The similarity of S/N histograms to a Gaussian distribution is a way to evaluate the meaningfulness of the S/N values, and the validity of the $5\sigma$ detection limits in Figure \ref{fig:summarydetec}.}
  \label{fig:detecmaps}
\end{figure*}

This framework can be used to detect a companion by fitting the model on a grid of $\Delta \mathrm{RA}$, $\Delta \mathrm{Dec}$. We choose to sample the sky coordinate every $0.05^{\prime\prime}$ in both directions. At each position, the model returns a best-fit companion amplitude $\phi_{\mathrm{comp}}$ and its uncertainty. 
This provides best-fit flux, flux error, and S/N maps for each detector and each dither position. These maps are then combined using a weighted mean, resulting in the final maps shown in Figure \ref{fig:detecmaps}. For HD~19467, the companion is detected with a S/N of $232$.
We confirm the accuracy of the noise model by plotting the histogram of the S/N map, which is very similar to a normal distribution with zero mean and unit standard deviation. 
The flux extraction is validated with an injection and recovery test described in Appendix \ref{app:injectionrecovery}. We also confirm that the flux of HD~19467~B that is estimated from the forward model is in good agreement with Table~\ref{tab:photometry}. We find $54.5\pm0.5\,\mu\mathrm{Jy}$ in F360M from NRS1, and $90.6\pm0.4\,\mu\mathrm{Jy}$ in F460M from NRS2. We therefore confirm that the NRS2 detector, which is also the most sensitive, yields estimated fluxes in excellent agreement with the original RDI spectrum. The NRS1 detector is slightly inconsistent with a $\sim13\%$ relative difference, which could be explained by the worse sampling at shorter wavelengths and the simplication made in the scaling of the WebbPSF model. This level of inconsistency is however negligible for the purpose of defining detection limits.
% Photometry in F360M: 54.5+-0.5 uJy from FM  compared to 63+-8 uJy from RDI spectrum (Table 2)
% Photometry in F460M: 90.6+-0.4 uJy from FM  compared to 88+-8 uJy from RDI spectrum (Table 2)

Direct imaging observations often require an ad-hoc normalization of the S/N maps by estimating the standard deviation in concentric annuli to correct for unaccounted covariances in the noise \citep[e.g.][]{Cantalloube2015A&A...582A..89C,Ruffio2017ApJ...842...14R}. 
This step leads to several issues including the assumption that the noise is uniform at a constant radius. It also requires a small sample statistics correction due to the limited number of independent samples of the noise at small inner working angles \citep{Mawet2014ApJ...792...97M}.

An interesting property of this forward model thanks to the quality of the NIRSpec data is that the flux error can be interpreted without such corrections. This means that the sensitivity can be independently measured at each location in the FOV.
The lower sensitivity on top of the diffraction spikes can be seen in the flux error panel in Figure \ref{fig:detecmaps}. 
More importantly, there is no need for a small sample statistics correction at a small inner working angle. Indeed, a small sample statistics correction is only warranted when the standard deviation of the companion fluxes has to be estimated from the flux map itself, which is typically done in concentric annuli in the final image. With this forward model, the companion flux errors are fully derived from the pixel-level errors that are produced by the JWST science calibration pipeline.
Small sample statistics can otherwise decrease the sensitivity by more than an order of magnitude for classical high-contrast imaging instruments.
The companion sensitivity in terms of flux ratio (a.k.a., contrast curve) is defined at each location as $5\times$ the flux error divided by the brightness of the star in the F444W filter. We use the stellar photometry from \citet{Greenbaum2023} to estimate the F444W brightness of HD~19467 at $1.3\,\mathrm{Jy}$. The $5\sigma$ sensitivity as a function of the projected separation to the star is shown in Figure \ref{fig:summarydetec}. The sensitivity is however not valid within $0.3^{\prime\prime}$ because the current implementation of the forward model assumes that there is no companion signal contamination in the starlight spectrum. 
This assumption breaks down if the putative companion is too close to the star. 
One way to address this limitation is to compute a new starlight spectrum after masking the hypothetical companion area at each location in the FOV, similar to the method implemented in \citet{Agrawal2023AJ....166...15A}. However, the only way to push the forward model validity down to zero separation would be to use a theoretical starlight model instead of an empirical one. This could be used for studying planets that are neither far enough from the star to be spatially resolved, nor close enough to leverage their radial velocity variations \citep{Snellen2010Natur.465.1049S,Finnerty2023AJ....166...31F}.

\subsection{Dependence on the companion's effective temperature }
\label{sec:contvsteff}
\begin{figure*}
\gridline{\fig{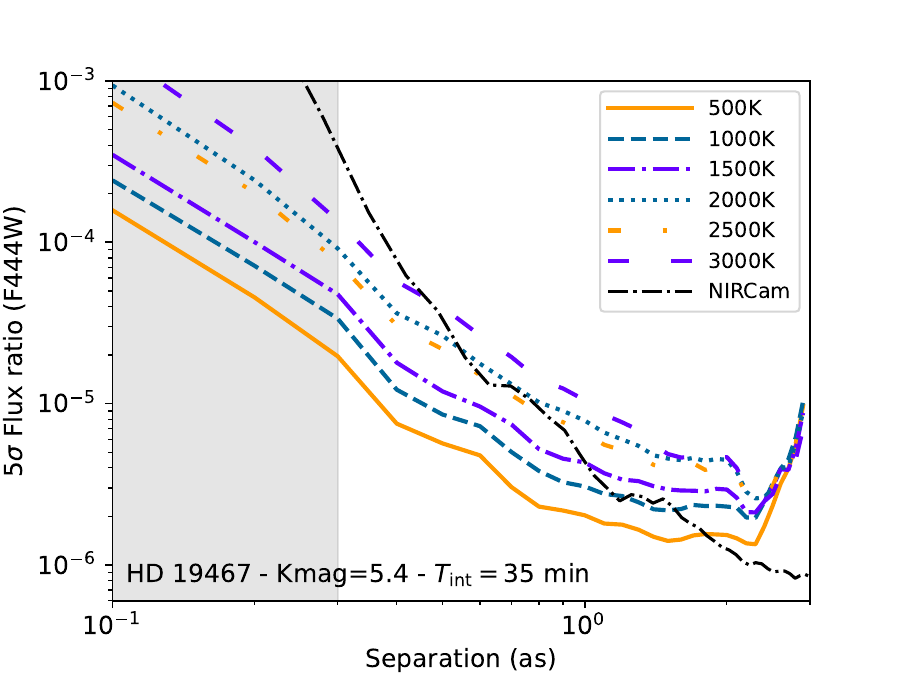}{0.49\textwidth}{(a) Detection sensitivity}
		\fig{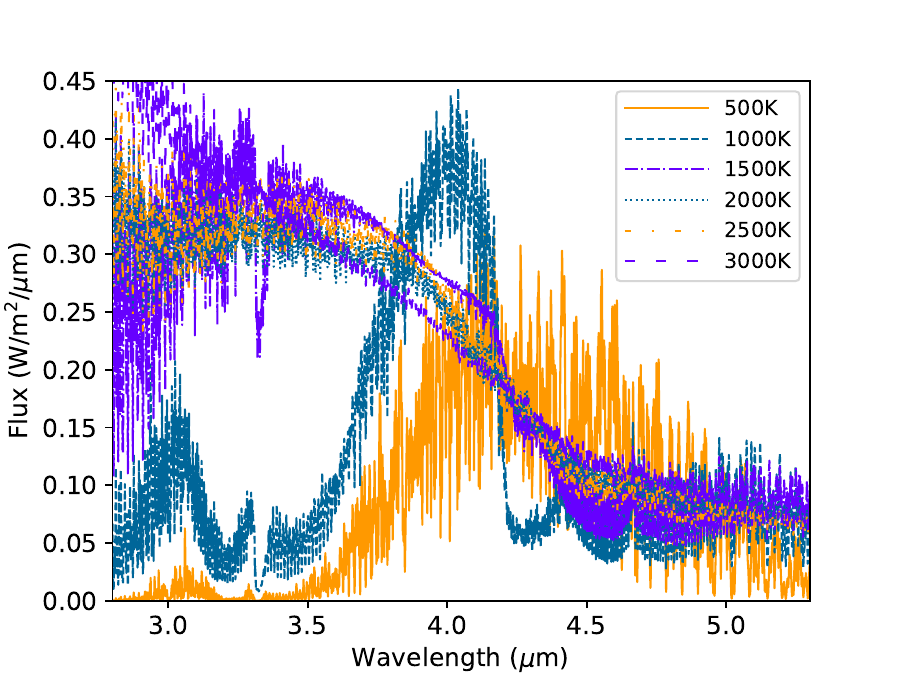}{0.49\textwidth}{(b) BT-Settl models}
          }
\caption{Effect of the effective temperature of a companion on the high-contrast sensitivity of the NIRSpec IFU for HD~19467. (Left) $5\sigma$ detection limits as a function of the assumed effective temperature for the model. While lower effective temperatures achieve better detection limits in terms of flux ratio, this does not mean that cooler companions are easier to detect. This is because they will also be significantly fainter than their warmer counterpart, everything else being equal. The limits are expressed in terms of the companion-to-star flux ratio and they are taken as the median value at each separation.
This uses the HD~19467 dataset, which means a stellar K-band apparent magnitude of $K=5.4$, and a total integration time of $\sim35\,\mathrm{min}$. 
The $5\sigma$ companion-to-star flux ratio detection limits are here defined as the median value at each separation. The region within $0.3^{\prime\prime}$ is greyed out because the current implementation of the method is not valid there, but it may theoretically be achieved. 
(Right) BT-Settl atmospheric models at each temperature used to fit for the companion \citep{Allard2003IAUS..211..325A}. Each model is normalized to $1\,\mathrm{Jy}$ in F444W. The NIRCam contrast curve is from \citet{Carter2023} scaled by the square root of the exposure time and stellar brightness. 
A machine-readable table of the detection limits as a function of companion effective temperature is available in the following filters: F356W, F444W, Lp, and Mp.
}
\label{fig:contrastteff}
\end{figure*}

\begin{figure*}
\gridline{\fig{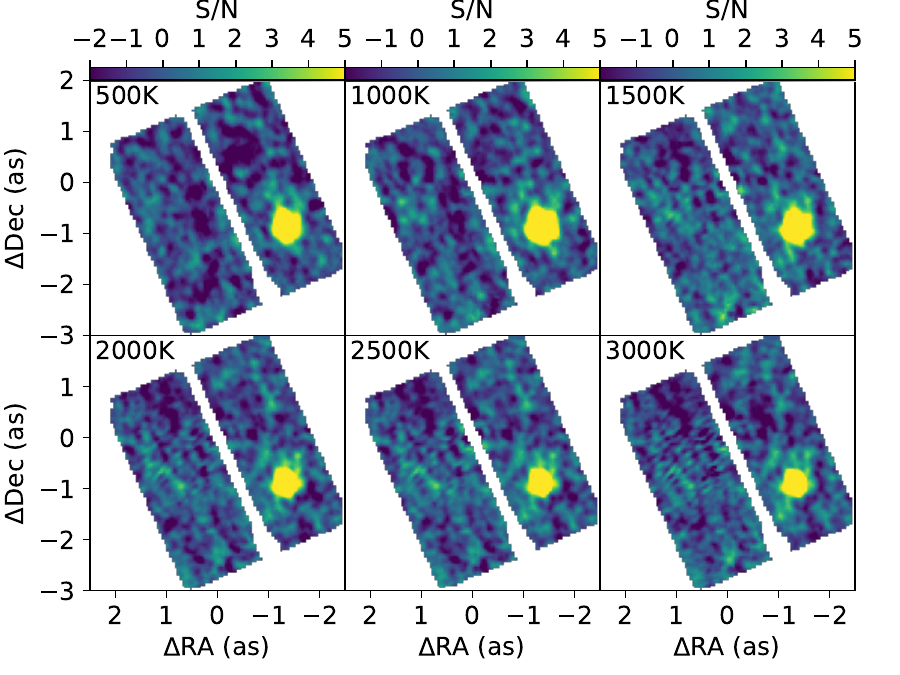}{0.49\textwidth}{(a) Detection maps}
		\fig{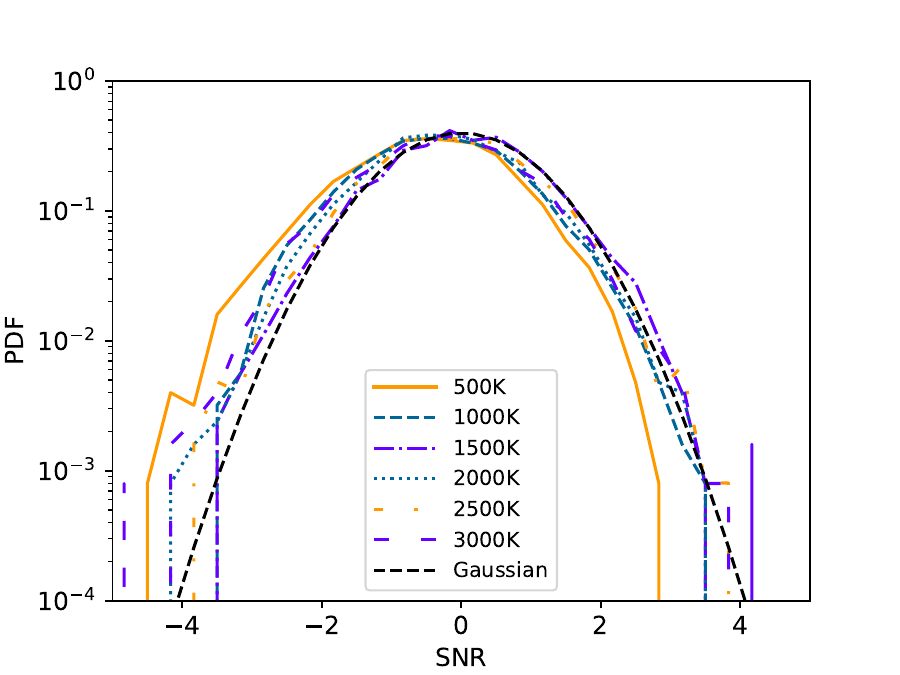}{0.49\textwidth}{(b) S/N histograms}
          }
\caption{Detection maps as a function of model effective temperature corresponding to Figure \ref{fig:contrastteff}. (Left) Companion S/N maps. (Right) Histograms of the S/N values after masking HD~19467~B.
}
\label{fig:snrVSteff}
\end{figure*}

As discussed previously, the flux ratio detection sensitivity of the forward model is dependent on the atmospheric model assumed for the companion. 
To illustrate this dependence, we reduce the HD~19467 dataset using different BT-Settl atmospheric models \citep{Allard2003IAUS..211..325A} with effective temperature $T_{\mathrm{eff}} \in \{500,1000,1500,2000,2500,3000\}\,\mathrm{K}$ and a fixed surface gravity of $\mathrm{log}(g)=5.0$, which are characteristic of directly imaged substellar companions.
The spectral models are broadened with a Gaussian kernel of full width at half maximum $\Delta \lambda$ matching the spectral resolution of the instrument $R=\lambda/\Delta \lambda =2700$. 
Using $K=40$ nodes to fit the continuum led to large systematics in the S/N maps for effective temperatures larger than $1500\,\mathrm{K}$. To mitigate this issue, we increased the number of nodes to $K=60$ in this section. This gave the continuum model enough additional flexibility to avoid using the companion signal as a way to fit the speckles.
The companion to star flux ratios $5\sigma$ detection thresholds and the spectra are shown in Figure \ref{fig:contrastteff}. The corresponding combined S/N maps and histograms for each temperature are shown in Figure \ref{fig:snrVSteff}. The contrast limits are deepest for the lowest temperature atmosphere model. This is as expected because cooler atmospheres have deeper spectral features (ie, peak to valley) relative to the continuum intensity and also make a companion's atmosphere more spectrally distinct from the host star spectrum.

We also obtain a detection S/N of HD~19467~B for each BT-Settl model.
Changing the effective temperature of the model from $500-3000\,\mathrm{K}$ only changes the S/N from 110 to 90, with a peak $S/N=206$ at $1000\,\mathrm{K}$. This demonstrates that getting the companion perfectly right is not necessary for detection purposes as the S/N will not be strongly impacted. This is because the detection mostly leverages the moderate-resolution spectral features of the common molecules such as CO and H$_2$O. 
However, as discussed in Section~\ref{sec:compmodel}, the fluxes and contrast curves estimated from the forward model are only as accurate as the atmospheric model used for the companion template. This caveat is important when interpreting the detection limits from Figure~\ref{fig:contrastteff}.

\section{Reference star PSF subtraction (RDI)}
\label{sec:RDI}

\subsection{Context}

In the previous Section \ref{sec:FM}, we demonstrated how forward modeling the starlight and the companion signal in the detector images could be used to detect high-contrast objects. The power of the method relies on the moderate spectral resolution that can resolve the distinct spectral features of a cool atmosphere compared to the host star. We chose a very flexible spline model of the starlight that can reproduce any chromatic and spatial fluctuations of the speckle field, because it can be difficult to physically model or predict the stellar PSF at the required level of precision.
The main advantage of the forward model is that it can be applied to an individual exposure alone without relying on classical observing strategies such as angular, spectral, or reference differential imaging (ADI, SDI, or RDI).
However, the price being paid for not requiring a prior model of the stellar PSF is that the continuum of the companion spectrum is mostly lost in the process. 
Even though the moderate-resolution spectral features are informative, the ability to independently measure the absolute continuum level of a planet spectrum is also very important to characterize an atmosphere. Another downside of the forward model is the difficulty of identifying unexpected spectral features that are not included in the atmospheric models.
However, recovering a companion spectrum with its continuum requires PSF subtraction. Due to the failed observations in the program (cf Section \ref{sec:context}), the only strategy that can be tested with this dataset is using RDI with a reference star or a simulated PSF. 
On the one hand, the advantage of a simulated PSF is that it has no noise and no interpolation errors, but it might not be the best match to the data due to imperfections in the models. On the other hand, an empirical PSF should be the most accurate, but it is subject to interpolation errors and it has limited S/N.
Without a more substantial investment in more accurately simulating the NIRSpec PSF with \texttt{WebbPSF}, using the reference star observation is the only viable pathway in the context of this work. 

We implement this PSF subtraction on the sampled point cloud, once again in order to avoid as much as possible the systematics inherent in spatial interpolation into datacubes. However, it is difficult to entirely avoid spatial interpolation in this case because of the need to spatially align the reference and science observations. 

\subsection{Fitting a reference star}
\label{sec:refstarfit}

We use the reference star HD~18511 (Observation 5 in Table \ref{tab:observations}), which is taken with the same instrument configuration and dithering strategy as science observation 4. The only difference is a shorter exposure time to account for the brighter reference star.
Fitting the reference star to the science data is done in a similar fashion as the spectral extraction described in Section \ref{sec:webbpsffit}. For flux extraction, the centroid and flux scaling of a model PSF (\texttt{WebbPSF}) was fitted to a 2D point cloud at each wavelength.
For PSF subtraction, we replace the simulated PSF by a reference star, and we fit it separately in concentric annuli instead of the entire PSF at once.
The reference star fitting and subtraction is illustrated in Figure \ref{fig:fitrefpsf}.

\begin{figure*}
  \centering
  \includegraphics[trim={0cm 0cm 0cm 0cm},clip,width=0.5\linewidth]{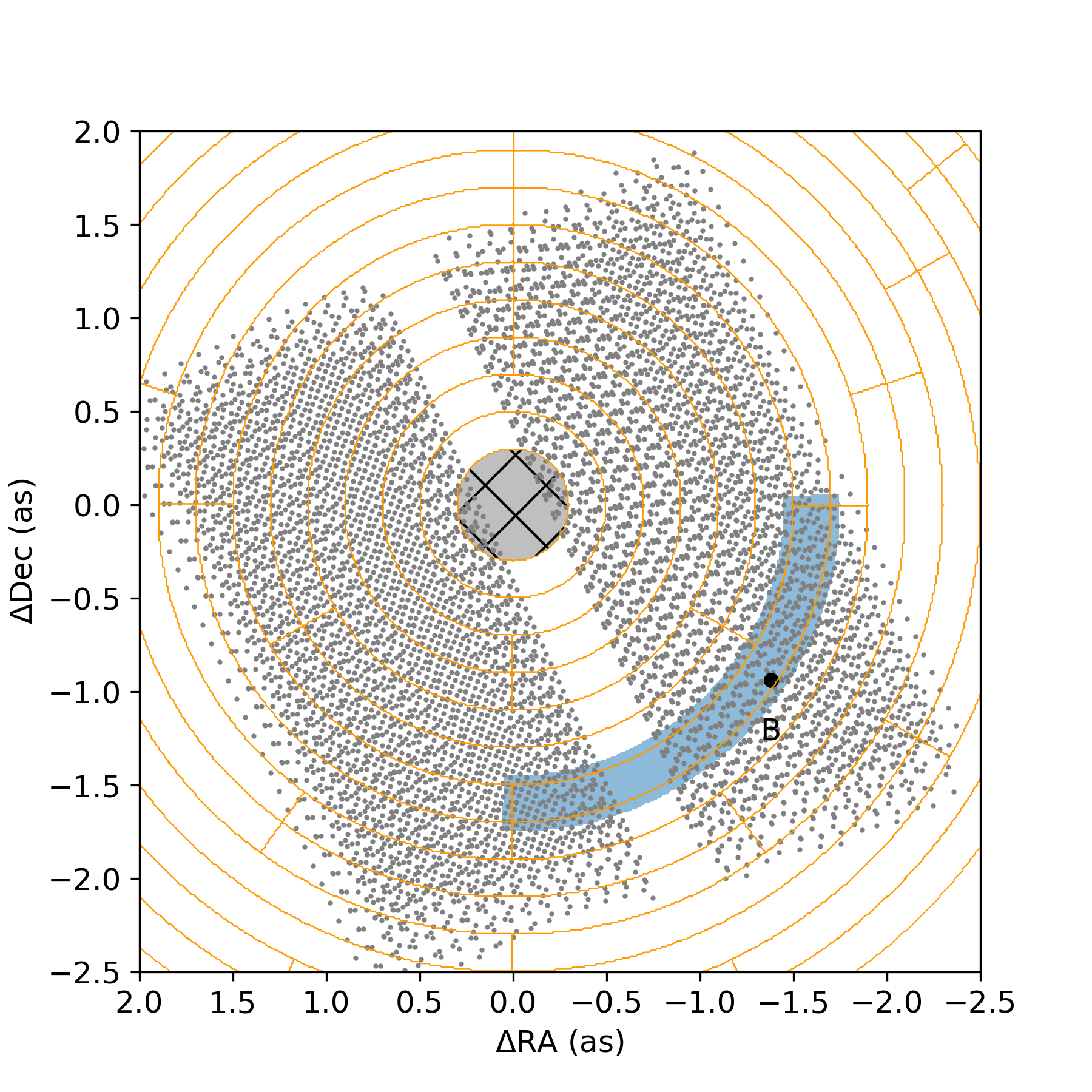}
  \caption{Definition of the sectors used for reference-star differential imaging (RDI). The dots represent the spatial sampling of the 9-dither HD~19467 dataset. The blue-shaded region is the sector that is shown in Figure \ref{fig:fitrefpsf}, which includes HD~19467~B at the marked location. The data within a $0.3^{\prime\prime}$ radius of the star is not included in RDI and shown as the hatched region.}
  \label{fig:sectors}
\end{figure*}

\begin{figure*}
  \centering
  \includegraphics[trim={0.75cm 0.75cm 1cm 0cm},clip,width=1\linewidth]{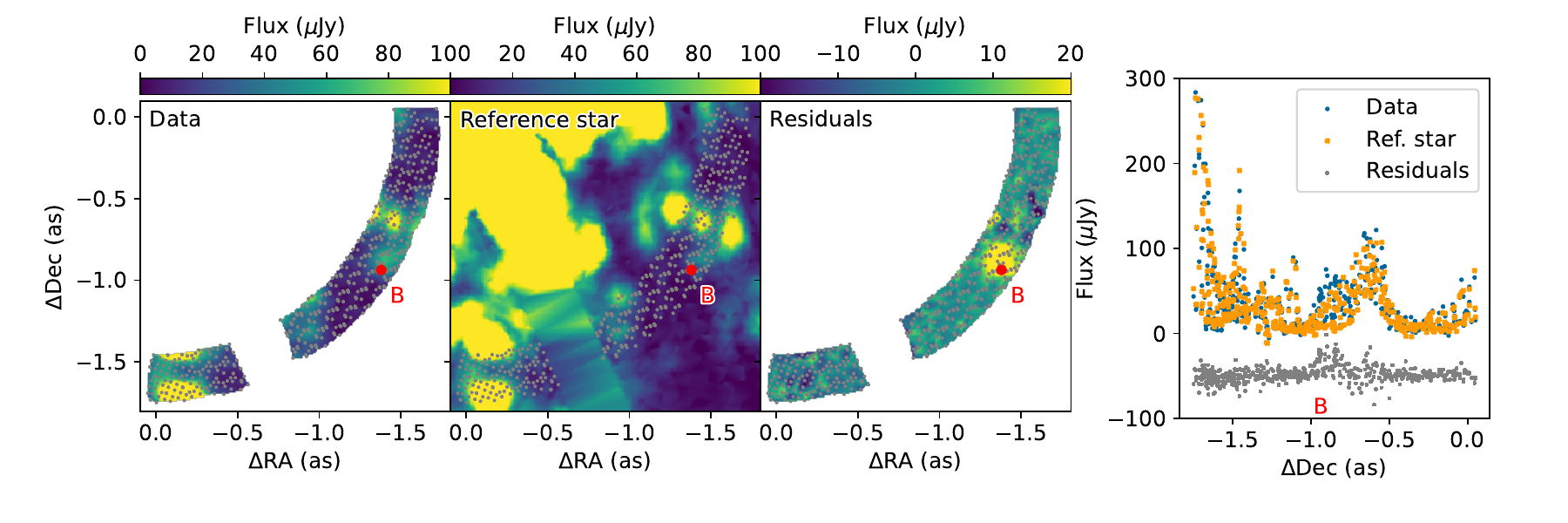}
  \caption{Reference Differential Imaging PSF subtraction in the point cloud. Similar to Figure \ref{fig:fitwebbpsf}, but for fitting of a reference star observation instead of a simulated PSF at $4.705\,\mu\mathrm{m}$. (Left) 
  The fit is done one sector and one wavelength at a time. The sector illustrated here includes the companion HD~19467~B. As in Figure \ref{fig:sectors}, the dots show the spatial sampling locations. (Middle Left) Linear interpolation of the 2D point cloud of the reference star dataset, which is sampled onto the 2D point cloud of the science data. (Middle right) Linear interpolation of the 2D point-cloud residuals, showing the data minus the best-fit model. The companion HD~19467~B is visible in the residual map as a bright yellow point source. (Right) Scatter plot of the data, model, and residuals that better represent the nature of the fit. The residuals were offset for clarity. }
  \label{fig:fitrefpsf}
\end{figure*}

The model reference PSF is built by interpolating each detector image onto a regular wavelength grid and then combining the sets of point cloud values from the nine dithers together, similar to the science data. The spatial sampling of the reference PSF is therefore irregular and a function of the dithering strategy, while the sampling of the simulated PSF was uniform and user-defined. However, the nature of the 2D point-cloud sampling makes no difference in the implementation of the linear interpolation of the model PSF.
In order to improve the flexibility of the PSF subtraction, we fit the PSF in sectors that are defined in $0.2^{\prime\prime}$-wide concentric annuli around the star as shown in Figure \ref{fig:sectors}. Annuli are divided to ensure that the area of each sector is as close to $0.5\,\mathrm{arcsec}^2$ as possible. The one sector highlighted in Figure \ref{fig:sectors} includes HD~19467~B, which is the same sector illustrated in Figure \ref{fig:fitrefpsf}. The companion can be seen in the PSF subtracted image, or residuals, of a single wavelength slice.
We fit the reference star PSF for each sector and at each wavelength to obtain a dataset of speckle subtracted rectified detector images. In this context, the rectification refers to the fact that the images have been interpolated on a regular wavelength grid.

\subsection{Spectral extraction}
\label{sec:RDIextract}

The format of the dataset is unchanged after the speckle subtraction so we use the PSF-fitting photometry method described in Section \ref{sec:webbpsffit} and Section \ref{sec:specextract} to extract the spectrum of a putative point source everywhere in the field of view.
We define a grid in sky coordinates $\Delta \mathrm{RA}$, $\Delta \mathrm{Dec}$ with a sampling of $0.05^{\prime\prime}$ in both directions. A flux-normalized PSF model from \texttt{WebbPSF} is then fitted at each position and each wavelength resulting in a cube of best-fit fluxes and associated errors. 
We extract the spectra at the position of HD~19467~B as well as in an $0.3-0.4^{\prime\prime}$ annulus around the companion to sample the speckle field. 
Due to the imperfect RDI PSF subtraction, the speckle spectra featured a small non-zero mean, which was subtracted from the companion spectrum. The flux uncertainties were computed by taking the standard deviation of the speckle spectra in the annulus at each wavelength. 
The extracted NIRSpec spectrum of HD~19467~B is shown above in Figure \ref{fig:summaryRDIspec}, including a comparison to the NIRCam photometry from Section \ref{sec:nircam}.  Synthetic photometry values derived from the NIRSpec spectrum are reported in Table \ref{tab:photometry}. Some photometric filters overlap with the wavelength gap in the NIRSpec spectrum, so the photometry was computed after interpolating a BTSettl model \citep{Allard2003IAUS..211..325A} with $T_{\mathrm{eff}}=800\,\mathrm{K}$ and $\mathrm{log}(g)=5.0$ just in the region of the gap in order to fill in the missing flux.

This implementation of RDI for NIRSpec is limited by interpolation systematics from spatially sampling the reference star point cloud onto the science data, even with the 9 dither positions. 
Indeed, the wavefront measurements taken before and after these science data indicate the telescope was very stable during this time period (negligible wavefront drift, consistent with zero within the sensing precision of $\sim 7$ nm), which rules out any variation with time of the PSF properties between the science and reference targets. Therefore the imperfect subtraction must be due to factors other than change in wavefront error.
This likely explains why the statistical photometric errors from NIRCam are an order of magnitude smaller than the ones derived from NIRSpec RDI. The S/N per spectral bin is $\sim10$ in the CO$_2$ and CO band heads ($4.2-5.2\,\mu\mathrm{m}$) and peaks at $\sim20$ around $4\,\mu\mathrm{m}$.

The systematic error due to the variable centroid of the companion across the spectral range of each detector should be $<3\%$ as explained in Section \ref{sec:specextract}. We prefer to fix the centroid given the lower S/N and the risk of the centroid to be biased by overlapping speckles. The 9 dither positions available in this dataset compared to the 4 dithers for TYC 4433-1800-1 in Section \ref{sec:specextract} should reduce the effect of the $\sim4\%$ peak-to-valley oscillations seen at the shortest wavelength. In any case, these oscillations remain within the  RDI residual speckle noise, so we only assign the overall $5\%$ flux calibration systematic uncertainty of the NIRSpec IFU in Table \ref{tab:photometry}. 

An important issue with the uncertainties on the RDI spectrum is that they are correlated. Therefore, we decompose the total variance into two different noise terms: the noise correlated on small spectral scales ($<2$ pixels) from interpolations and the noise correlated on larger scales from residuals speckles. This decomposition is motivated by the auto-correlation of the speckle spectra illustrated in Appendix~\ref{app:speccov}. We estimate the standard deviation of the small-scale noise from the speckle spectra simply by high-pass filtering them first with a 25-pixel ($0.017\,\mu\mathrm{m}$) sliding window median filter. Using this decomposition, we build a semi-empirical model of the covariance in Appendix~\ref{app:speccov}. The small-scale and large-scale variance add up to the diagonal matrix of the covariance of the RDI spectrum.

\subsection{Companion detection and sensitivity}

We compare the detection sensitivity of this RDI implementation with the forward model from Section \ref{sec:FM}.
To do so, we compute similar flux maps normalized in the NIRCam photometric band F444W, associated error, and S/N in Figure \ref{fig:detecmaps}. The companion flux map and error map were derived from a matched filter by fitting the 1D RDI spectrum at each spatial location in the RDI cube of extracted fluxes without subtracting the continuum. The noise of the RDI cube is derived directly from the WebbPSF model fit. We are however not modeling the covariance of the noise in the matched filter, which means that the speckle noise will dominate the residuals in the S/N map. The noise and S/N map were rescaled by the standard deviation of the S/N map after masking the central $0.3^{\prime\prime}$ disk, which feature very negative fluxes.
We show the $5\sigma$ flux ratio sensitivity (ie, contrast curve) in Figure \ref{fig:summarydetec} which is one to two orders of magnitude worse than the forward model. 

The lesser detection sensitivity of the RDI method can be explained by the fact that the residual speckle errors are dominating the noise budget. The moderate-resolution spectral features are also effectively buried in the noise. Using cross-correlation techniques on a high-pass filtered RDI cube, or alternatively including an accurate noise covariance, would undoubtedly yield better detection sensitivity than RDI alone. However, this approach would be worse than the forward model presented previously, which is why it is not attempted in this work.
This highlights how challenging classical PSF subtraction is with the NIRSpec IFU due to its spatial undersampling. The best way to overcome the interpolation systematics would be to fit an accurate simulated PSF instead of real observations in the future.

\section{Using RDI as a speckle prior in the forward model}
\label{sec:FMRDI}

\subsection{Implementation}

A potential route to further improve the forward model would be to incorporate information on the stellar PSF from RDI as a prior.
A prior on the speckle intensity can be implemented through the regularization of the starlight parameters introduced in Section \ref{sec:FM}. 
This addresses the main limitation of the forward model, which is the effective loss of information of the companion continuum. This approach can in theory benefit both the detection sensitivity and the accuracy of atmospheric inferences.

\begin{figure*}
  \centering
  \includegraphics[trim={0cm 0cm 0cm 0cm},clip,width=1\linewidth]{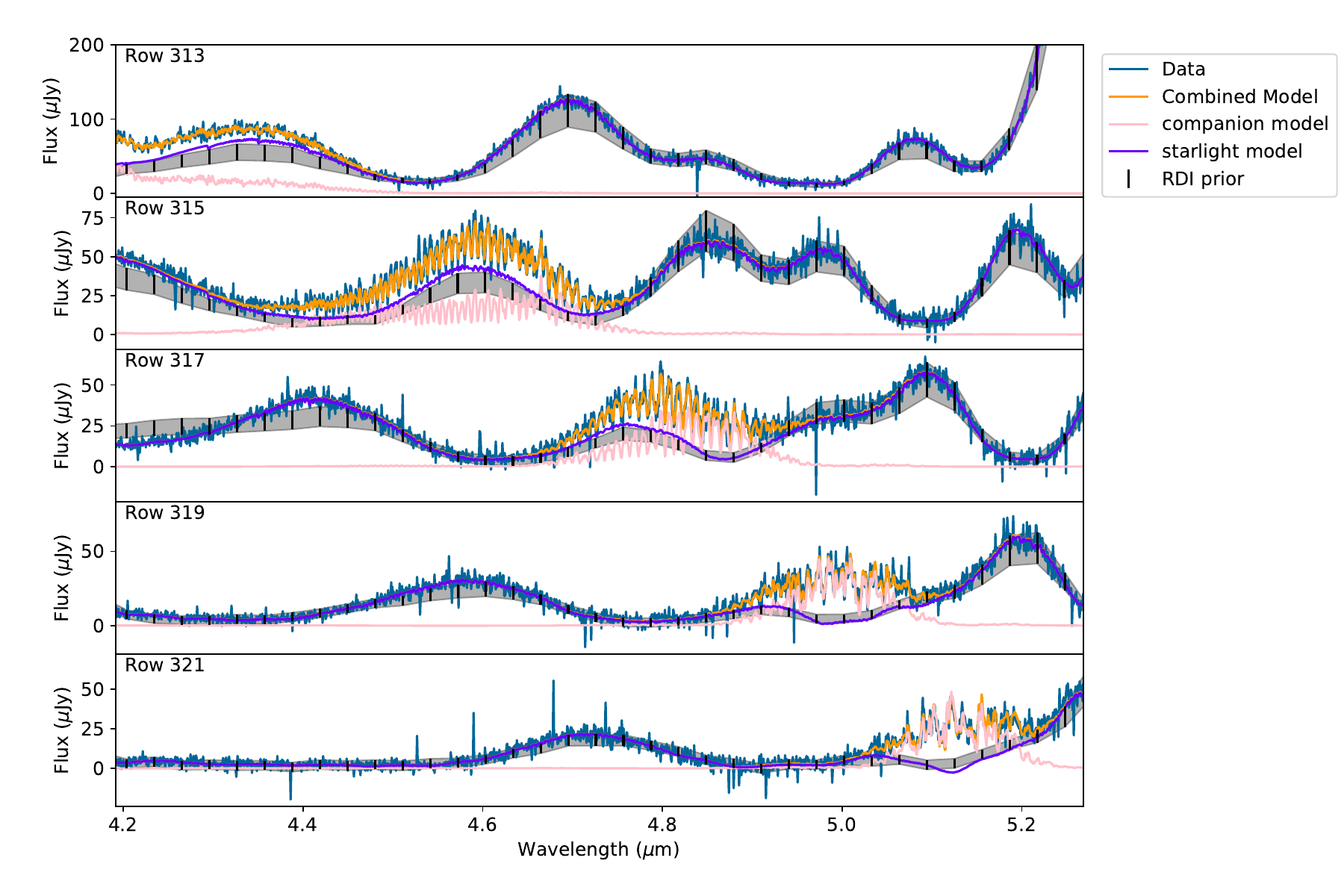}
  \caption{ RDI prior for the speckle intensity in different rows of the NIRSpec detector. Each panel is similar to the top panel of Figure \ref{fig:fm}, but the $1\sigma$ speckle intensity prior is represented as the shaded grey region. The row numbers match those in Figure \ref{fig:FM_im}. Only rows intersecting the companion trace on the detector were chosen to highlight the deviations of the continuum caused by the companion signal.}
  \label{fig:fmrdi}
\end{figure*}

This prior can be simply implemented by setting the value of the ${\bm \mu}_{{\bm \phi},i}$ (see Section \ref{sec:FMdef}) to the speckle continuum level of the best-fit reference model of the PSF that was derived in Section \ref{sec:RDI} for RDI. Based on the RDI residuals, we also set the standard deviations of the continuum prior ${\bm s}_{{\bm \phi},i}$ to 20\% of the continuum, but they are clipped to a minimum value of $3\,\mu\mathrm{Jy}$.
The speckle intensity prior is illustrated in Figure \ref{fig:fmrdi} with different rows of the NIRSpec detector including the signal from HD~19467~B. 
Unlike the forward model used in Section \ref{sec:FM}, this new model of the data can optimally leverage the continuum level of the companion. It could theoretically even detect a planet with muted moderate-resolution spectral features by detecting deviations of the data from the reference PSF. The continuum contribution of HD~19467~B to the data is clearly visible in Figure \ref{fig:fmrdi}.

\subsection{Application to atmospheric characterization}

The goals of this section are 1) to demonstrate, in the context of NIRSpec, that the characterization of a companion's atmosphere can be performed in detector space and 2) to show that it is possible to leverage the continuum information in the high-resolution-specific forward modeling framework.

Moderate to high resolution spectroscopy has proven to be a powerful tool to study the atmosphere of directly imaged exoplanets and measure their composition (see Section \ref{sec:intro}). 
It is a common practice to remove the continuum of the planetary spectrum using a high-pass filter because the continuum contribution of the planet cannot typically be accurately separated from the continuum of the speckles. The high-pass filter minimally changes the high resolution spectral features. 
When the continuum of a moderate-resolution spectrum is retrieved \citep[e.g.][]{Wilcomb2020AJ....160..207W}, the error on the continuum level from residual speckles are typically much larger and highly correlated compared to the noise of the high spectral resolution features. This can also be seen in Figure \ref{fig:summaryRDIspec}. 

Fitting atmospheric models to such spectra therefore requires a careful model of the covariance of the noise to appropriately weigh the correlated uncertainty of the continuum compared to the well-resolved spectral features \citep{Greco2016ApJ...833..134G}. For example, we compute the covariance for the RDI spectrum of HD~19467~B in Appendix~\ref{app:speccov}.
Beyond the correlated noise coming from the residual speckles, each data processing step used to extract the signal of a high-contrast companion can significantly impact the accuracy of atmospheric retrievals \citep{Nasedkin2023arXiv230801343N}. One example is interpolations used to sample the data on regular spatial or wavelength grids, which is visible in the autocorrelation of the speckle residuals illustrated in Appendix~\ref{app:speccov}. Modeling the covariance of the noise can help \citet{Greco2016ApJ...833..134G,Nasedkin2023arXiv230801343N}, but it can be a challenging task to do accurately. Indeed, fully-empirical covariance matrices can be noisy, and analytically-defined covariances might not accurately account for the various sources of correlation, each of which has its own correlation length and amplitude. 
The best way to address this issue is to avoid including correlated noise in the spectra in the first place, which is the main motivation behind implementing a detector-level forward model of the data. 
The detector noise is indeed dominated by uncorrelated pixel-to-pixel noise. Additionally, the linear model of the starlight used in this work allows for a direct marginalization of the speckle subtraction without the need for a non-diagonal covariance.
For example, the RDI spectrum derived in Section \ref{sec:RDI} will come with its own additional noise floor due to the use of interpolated detector images in the wavelength direction.
Combining the forward model with speckle priors from RDI could alleviate these issues while keeping the best aspects of both techniques.

As a proof of concept for atmospheric characterization, we fit a BT-Settl model grid \citep{Allard2003IAUS..211..325A} using the forward model likelihood including the RDI prior. Such atmospheric characterization has for example already be done in \citet{Ruffio2021AJ....162..290R} or \citet{Agrawal2023AJ....166...15A} with a very similar forward model, albeit not in detector space and without RDI priors.
The grid is defined with an effective temperature ranging from $500\,\mathrm{K}$ to $1600\,\mathrm{K}$ in steps of $100\,\mathrm{K}$ and a surface gravity ranging from 3.5 to 5.0 in steps of 0.5. The models are linearly interpolated from this grid for any given value of the effective temperature and the surface gravity.
We fit each detector and each dither independently varying the effective temperature and the surface gravity as the non-linear parameters with the spatial coordinates being fixed. The 2D log-posterior of a each detector is shown in Figure \ref{fig:logprob_FMRDI}. The definition of the posterior probability is given in Appendix~\ref{app:newmaths}. The fit favors larger surface gravity ($\log(g)>5$) than this BT-Settl model grid allows in all cases. We show the best-fit temperature for each detector and each dither position in Figure \ref{fig:logprob_FMRDI}. 
We find a combined best-fit effective temperature of $T_{\mathrm{eff}}=952\pm3\,\mathrm{K}$ ($T_{\mathrm{eff}}=931\pm2\,\mathrm{K}$) for NRS1 (NRS2). These error bars do not include model grid interpolation systematics \citep{Czekala2015}, which would be limited to $\sim100\,\mathrm{K}$.
We compared the best-fit model to the RDI spectrum in Figure \ref{fig:FMRDI_bestfit}. However, it is important to note that the RDI spectrum was not the object of the fit. The best-fit absolute brightness of the model was derived directly from the detector images.

We demonstrated that the best-fit model derived from the forward modeling framework with RDI prior is not only a reasonable fit to the spectral features at moderate spectral resolution, but also that the inferred absolute brightness is consistent with the RDI-extracted spectrum from Section \ref{sec:RDI}. We emphasize that the usual loss of the continuum in moderate to high-resolution spectral analysis is not a fundamental limitation of the technique, and that the continuum can be recovered with dedicated observing strategies.
However, the coarseness of the model grid used in this analysis preclude a more accurate atmospheric inference, which is outside the scope of this work. The atmospheric characterization of HD~19467~B is the topic of a companion paper led by Hoch et al.

\begin{figure*}
  \centering
  \includegraphics[trim={0cm 0cm 0cm 0cm},clip,width=1\linewidth]{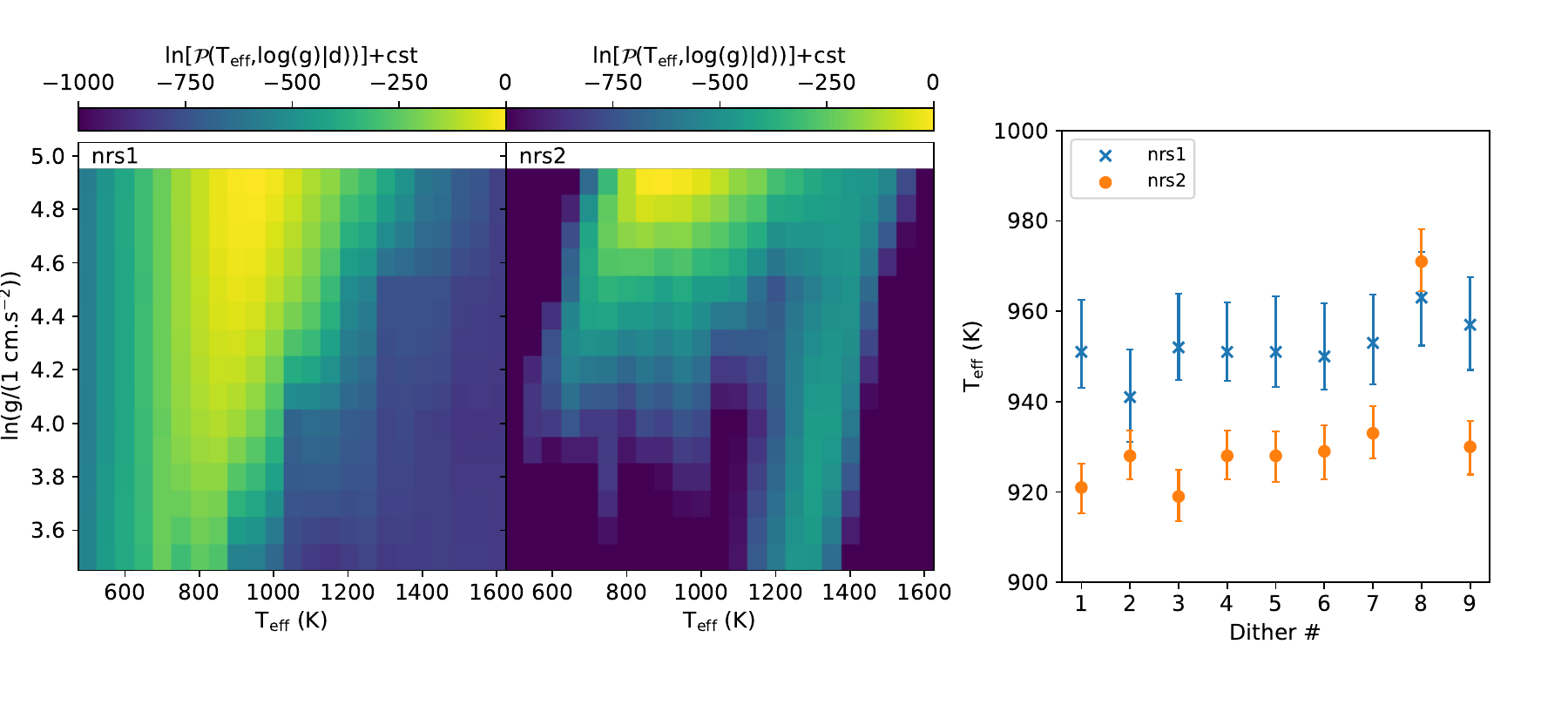}
  \caption{  Proof of concept for the atmospheric characterization of HD~19467~B using the detector forward model and a RDI-based prior on the speckle fluxes. We use the BT-Settl model grid \citep{Allard2003IAUS..211..325A}. (Left) Natural logarithm of the marginalized posterior of the effective temperature and surface gravity for a single dither position. (Right) Best-fit effective temperature for each detector image and each dither position in the observing sequence.}
  \label{fig:logprob_FMRDI}
\end{figure*}

\begin{figure*}
  \centering
  \includegraphics[trim={0cm 0cm 0cm 0cm},clip,width=1\linewidth]{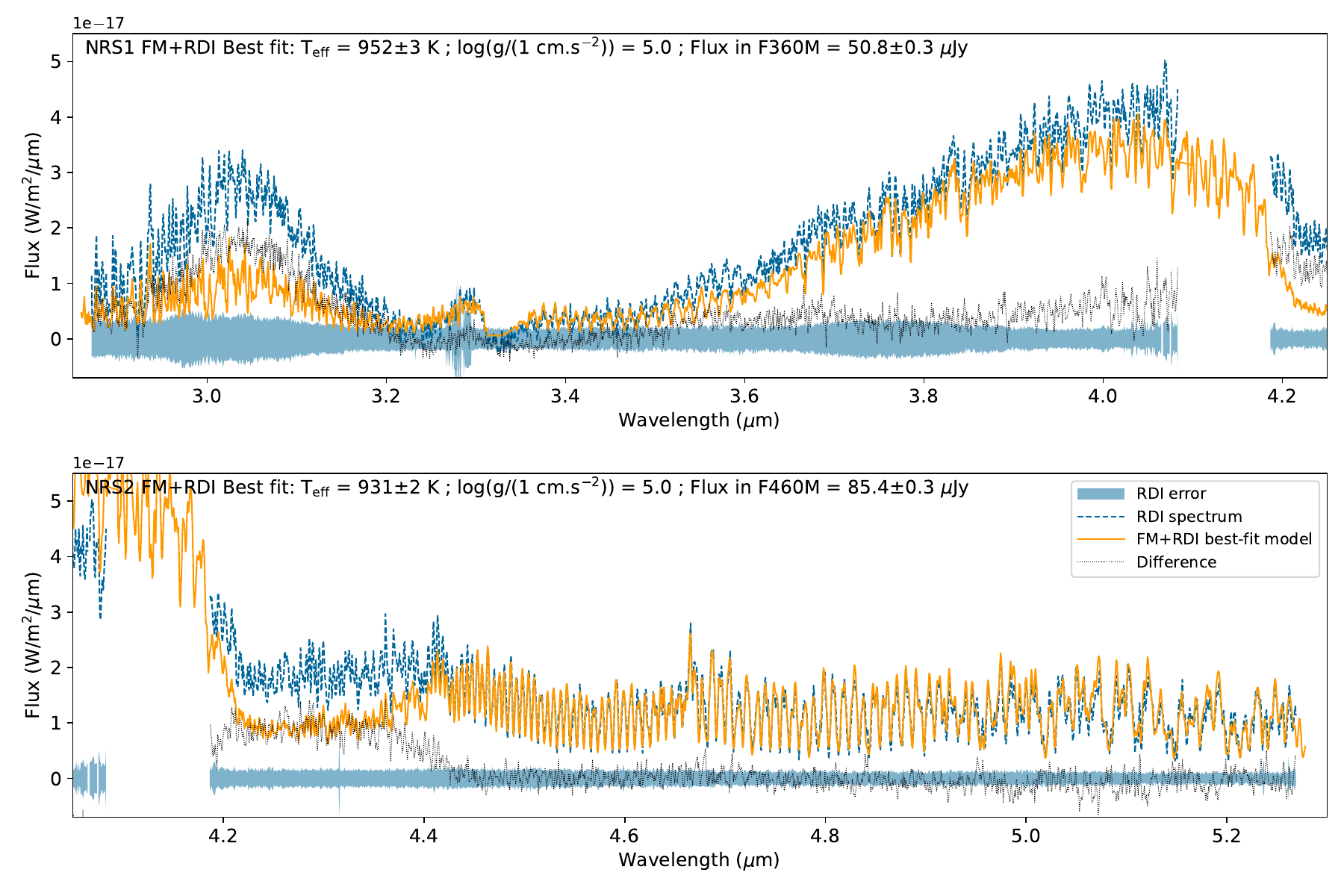}
  \caption{  Best-fit BT-Settl model \citep{Allard2003IAUS..211..325A} derived from the forward model with RDI priors (FM+RDI) compared to the RDI-extracted spectrum. The model was fitted directly to the detector images (NRS1 and NRS2 separately) so the difference between the model and the RDI spectrum is not the residuals of a fit.}
  \label{fig:FMRDI_bestfit}
\end{figure*}

\subsection{Application to exoplanet detection}

For high-contrast imagers and low spectral resolution IFUs, planet detection algorithms rely on the continuum of the planet using observing strategies such as ADI, SDI or RDI. On the other hand, for moderate to high spectral resolution spectrographs (IFU or not) and cross-correlation-inspired techniques \citep[e.g.][]{Hoeijmakers2018A&A...617A.144H},  the spectral continuum including the starlight is often high-pass filtered, as mentioned before. This means that either method is missing part of the available information. The approach proposed in this section provides a way to optimally combine all the information available, continuum and spectral features alike.

Assuming a companion template ${\bm t} = \mathcal{P}(\{ \lambda_i \})$, the ideal detection S/N of a companion is the root sum squared of the signal such that $\mathrm{S/N} = \sqrt{{\bm t}^\top{\bm t}} / \sigma$ using the matched filter formula \citep{Landman2023A&A...675A.157L}. This formula assumes a perfect host star subtraction and uniform noise $\sigma$ in the wavelength bins. By comparing this toy model S/N for a high-pass filtered template ${\bm t}_\mathrm{HPF}$ to the original model ${\bm t}$, we can estimate how much signal has been lost in the high pass filtering process. For HD~19467~B, the S/N would be 5 times higher with the continuum up to S/N$\sim1300$ for our observations. This ratio increases to $\sim36$ for a $3000\,\mathrm{K}$ BT-Settl model as the spectral features get shallower compared to the continuum. Using a RDI prior in the forward model can help tap into this otherwise lost signal. The detection sensitivity curves derived in this work (Figure \ref{fig:summarydetec} and Figure \ref{fig:contrastteff}) are therefore not fundamental limits and could still be improved significantly. This is however not a straightforward task.

We tried computing S/N maps using the forward model with RDI priors, but they were filled with artifacts negatively affecting the S/N histogram to the point of effectively degrading the sensitivity instead of improving it. The implementation of RDI using a reference star observation presented in this work is not accurate enough to leverage the potential of this method yet. If simulated NIRSpec PSFs are made more accurate in the future, they could be used for RDI instead and would not be limited by spatial interpolation errors. This could be the most promising avenue for improving NIRSpec high-contrast sensitivity.

\section{Discussion}
\label{sec:discussion}

\subsection{Strategies for high contrast with the NIRSpec IFU}

The goal of this work was to assess the performance and optimal observing strategies for high-contrast imaging of exoplanets with the NIRSpec IFU. We demonstrated that the NIRSpec IFU is a powerful instrument for both detecting and characterizing directly imaged exoplanets at moderate spectral resolution, achieving sensitivities impossible from the ground at these wavelengths. 
We presented three approaches to the data analysis that are discussed below.

First, we implemented a forward model of the detector images (Section \ref{sec:FM}), which is less prone to introducing systematics compared to other classical high-contrast strategies. The speckle model is flexible enough to fully subtract the starlight down to the photon noise limit. It does so while retaining the distinct spectral signature of the cool planet atmosphere thanks to the moderate spectral resolution. The resulting likelihood can be used for both detection and atmospheric characterization of a high-contrast companion. The most significant downside of this approach is the effective loss of the continuum information of the companion spectrum. While the companion continuum is included in the forward model, it is in practice degenerate with the starlight continuum. 
Using this implementation of a forward model, Figure \ref{fig:summarydetec} demonstrates that the NIRSpec IFU is very sensitive to detect high-contrast companions at small projected separations ($<2^{\prime\prime}$) in the $2.9-5.2\,\mu\mathrm{m}$ wavelength range. For example, we derived a sensitivity to planets that are $2\times10^{-6}$ fainter than their stars at $1^{\prime\prime}$ in the HD~19467 dataset (Figure \ref{fig:summarydetec}).
The detection sensitivity of the forward model is however dependent on the spectrum of the companion. The deeper the spectral features relative to the continuum, the deeper the sensitivity in terms of companion-to-star flux ratio. The sensitivity varies by a factor $\sim5$ between $500\,\mathrm{K}$ and $3000\,\mathrm{K}$ as shown in Figure \ref{fig:contrastteff}. A planet embedded in a circumplanetary disk might however not be detectable at all with this method if its spectral features are fully muted.

The achieved sensitivity will enable spectroscopy of planets in very scientifically interesting parts of mass/separation parameter space. With JWST/NIRCam, \citet{Lawson2023arXiv230802486L} demonstrated a sensitivity to $\sim0.1\,\mathrm{M}_\mathrm{Jup}$ beyond $\sim2^{\prime\prime}$ ($\sim20\,\mathrm{au}$) around AU Microscopii, which is estimated to be $20\pm10\,\mathrm{Myr}$. NIRSpec can achieve a similar sensitivity in the same spectral band (F444W) at closer separation $\sim1^{\prime\prime}$, where planets are more common.
Assuming the $2\times10^{-6}$ flux ratio sensitivity achieved in 35 min around HD~19467 at $\sim1^{\prime\prime}$, NIRSpec could detect sub-Jupiters at ages up to $1\,\mathrm{Gyr}$ at $10\,\mathrm{pc}$ using the COND evolutionary model \citep{Baraffe2003A&A...402..701B} at the same separation and around a similar star. Adolescent sub-Jupiters are therefore within reach of the NIRSpec IFU.
Based on pre-launch simulations, \citet{LlopSayson2021AJ....162..181L} also showed that NIRSpec IFU could detect the planet $\epsilon$ Eridani b which is $0.57-0.78\,\mathrm{M}_\mathrm{Jup}$ at $400-800\,\mathrm{Myr}$ and $3.2\,\mathrm{pc}$.
By demonstrating the feasibility of photon noise limited planet detection with real observations, we also confirmed the validity of these simulations.
NIRSpec will therefore enable the direct detection of planets that are more similar to the solar system.

Secondly, in order to  measure the full spectrum of a companion including its continuum, we implemented a PSF subtraction routine using observations of a reference star (Section \ref{sec:RDI}). This RDI implementation successfully enabled the extraction of the spectrum of HD~19467~B (Figure \ref{fig:summaryRDIspec}). The S/N$\sim10$ of the RDI spectrum is however limited by speckle noise residuals due to interpolation errors when resampling the reference star point cloud to match the science data. 
In its current implementation, RDI is not competitive for detection purposes compared to the forward model, but remains the only way to measure the continuum of a companion. 
RDI is generally not the only way to subtract the host star PSF and recover the companion continuum. ADI and SDI have otherwise been powerful observing strategies for high-contrast imaging. However, their performance for the NIRSpec IFU is expected to be limited by interpolation errors similar to RDI. 
Future efforts to more accurately model the NIRSpec IFU PSF should significantly improve RDI reductions.

Finally, we presented a proof of concept for a combined data reduction strategy (Section \ref{sec:FMRDI}) that augments the forward model framework with an independent prior of the speckle intensity from RDI. This approach allows the forward model to leverage the spectral continuum of the companion, but in practice the expected benefits are negated by the higher level of systematics in the current RDI implementation. This however remains a highly promising strategy, but will require further effort to accurately model the stellar PSF free from systematics and interpolation errors.

\subsection{Recommendations, limitations, and path forward}
We highlight some specific recommendations for designing high-contrast observations with the NIRSpec IFU.
We also identify the following limitations in achieving more accurate or more precise high-contrast science, which can be used to prioritize future NIRSpec calibration efforts.

\begin{description}
    \item[Modeling detector images] 
    We recommend modeling the data directly in the detector images whenever possible.
    The main limitations of the NIRSpec IFU are currently its spatial undersampling and the management of bad pixels. As discussed throughout this work, these systematics are better addressed by directly forward modeling the detector images.
    All interpolations come with a price by raising the systematic floor of the data so they should be used sparingly.
    \item[Dithering strategy] 
    We recommend using as many dither positions as possible to improve the spatial sampling of the dataset unless otherwise required by the science case. The RDI subtraction in Section \ref{sec:RDI} is limited by spatial interpolation errors of the point cloud despite using a 9-position dither with the cycling pattern. The \texttt{WebbPSF} spectral extraction described in Section \ref{sec:specextract} featured systematic oscillations around $\sim3\,\mu\mathrm{m}$ with 4 dither positions due to the smaller PSF and worse spatial sampling at the shortest wavelengths. Future improvements in PSF modeling are expected to at least partially mitigate these issues.
    This issue is compounded by the current NIRSpec IFU dithering strategy that does not optimally sample the spatial coordinates as shown in Section \ref{sec:sampling}. To avoid distortion effects between IFU slices, we suggest the implementation of a sub-pixel dithering strategy similar to the small grid dithers used for NIRCam coronagraphy. 
    \item[Shift in source position] We identified a large centroid trend of $0.02^{\prime\prime}$ when fitting a point source over the $2.9-5.2\,\mu\mathrm{m}$ wavelength range in Figure \ref{fig:centroid}. The field dependence of this distortion remains to be characterized so the extent to which it might affect the sampling of the point cloud from a combined dither sequence is also unknown. This could for example contribute to the systematics of the RDI subtraction.
    \item[Accurate PSF models] Obtaining more accurate model PSFs is the most important path forward to improving the spectral extraction and the stellar PSF subtraction for high-contrast imaging.
    Unless the reference and science stars sample the same pixel phases, fully empirical PSFs are subject to interpolation errors. Simulated PSFs can mitigate interpolation errors, but they also require accurate modeling of the NIRSpec instrument and wavefront errors. Simulated PSFs from \texttt{WebbPSF} that are calibrated from empirical observations are a promising avenue for NIRSpec IFU data analysis. A more accurate model of the NIRSpec IFU PSF could enable the implementation of the RDI prior in the forward model for detection purposes and further improve NIRSpec high-contrast sensitivity.
    \item[Saturation, charge diffusion, and stray light] Many host stars of directly imaged exoplanets saturate the NIRSpec detector in two groups even when dispersing the light at moderate spectral resolution. The saturation and charge transfer on the NIRSpec detectors does not appear to be a limitation unless the exoplanet signal directly overlaps with the affected regions. IFU observations should be designed to ensure the planet is not placed in the same IFU slices as the stellar PSF core. Placing the star outside the field of view can be a simple way to mitigate the risks associated with the charge transfer, although it may limit our ability to precisely determine the position of the star.
    In general, modeling the charge transfer would improve the field of view coverage of the IFU, which would be beneficial for exoplanet searches. Another issue related to observations of bright stars is stray light in the optical path. We identified a possible ghost visible in the NRS1 detector, but were able to mitigate its effects by including principal components of the residuals in the forward model. A better characterization of stray light in the NIRSpec IFU could be used to better optimize future observations and limit the probability of aligning a faint companion with such artifacts.
    \item[Photometric calibration] The absolute flux measurements of HD~19467~B remain limited by residuals speckles from RDI and the $5\%$ absolute flux calibration of the NIRSpec instrument in the JWST science calibration pipeline. The latter is likely to improve from the ongoing absolute flux calibration program \citep{2022AJ....163..267G}. 
    Additionally, the simulated PSF flux calibration used in this work (Section \ref{sec:specextract}) did not accurately predict the flux of the photometric reference star. This is not unexpected given the mismatch between the model \texttt{WebbPSF} and the real PSF. It led to a $10\%$ systematic error that needed to be corrected a posteriori. A more detailed characterization of the PSF will be necessary to retire this issue in the future.
\end{description}

\subsection{Relevance to Future High-Contrast Missions}

We demonstrated that moderate-resolution spectroscopy can enable high-contrast sensitivity close to the star.
It is important to consider the effect of the spectral resolution in the design of the future Habitable World Observatory (HWO) building on existing studies such as \citet{Wang2018JATIS...4c5001W} and \citet{Landman2023A&A...675A.157L}. 
Defining a metric to evaluate the spectral resolution trade space will be an important aspect of these studies. As an example, \citet{Batalha2017AJ....153..151B} evaluated the best observing modes, or combination of modes, to characterize transiting exoplanets with JWST by optimizing their information content. Planet or molecule detections need to be based on a template matching S/N, not the S/N per resolution element. Higher spectral resolution will by definition result in a lower S/N per resolution element, but can still lead to an overall better detection. 
We emphasize that it is possible to retain the continuum information of the companion spectrum with dedicated observing and data analysis strategies even at moderate to high spectral resolution.
There are many advantages to the higher spectral resolution to be considered.
For example, it is a more robust planet detection strategy because it is insensitive to speckle chromaticity and stability. It is also more robust to confusion from other astrophysical sources such as disk features, possible features in the exozodiacal dust, and background objects because it uses the distinct spectral signature of cool planet atmospheres.
As demonstrated in this work, the companion detection statistics (\textit{i.e.}, S/N histogram) is well-behaved and its implementation should not require a small sample statistics correction at small inner working angles.
Higher spectral resolution is also an efficient observing strategy because it allows for the joint detection and the atmospheric characterization of a planet.
All these factors should be taken into account when exploring the design trade space for HWO. 
The main downside of a higher spectral resolution spectrograph compared to an imager is the larger negative impact of detector level noise like dark current, so the benefits of high resolution spectroscopy might also be a function of the planet's brightness.
It is also possible to combine higher resolution spectroscopy with coronagraph designs like for the future SCALES instrument at the W. M. Keck Observatory \citep{Skemer2022SPIE12184E..0IS}.

\section{Conclusion}
\label{sec:conclusion}

We analyzed the data from the first high-contrast observations with the JWST NIRSpec IFU as part of the cycle 1 GTO program 1414 using the moderate spectral resolution G395H ($\lambda/\diff\lambda\sim2,700$) and F290LP filter ($2.9-5.3\,\mu\mathrm{m}$). 

\begin{itemize}
    \item Forward modeling the starlight and the companion signal in the detector images enabled a photon noise-limited planet detection sensitivity despite NIRSpec spatial undersampling. In 35 min of integration, the T-dwarf substellar companion HD~19467~B was detected with a $S/N\sim232$ at $1.6^{\prime\prime}$ with a flux ratio varying around $10^{-5}-10^{-4}$ between $3-5\,\mu\mathrm{m}$. Given the achieved $3\times10^{-6}$ flux ratio sensitivity at $\sim1^{\prime\prime}$, NIRSpec could detect sub-Jupiters at ages up to $1\,\mathrm{Gyr}$ at $10\,\mathrm{pc}$.
    \item Using a reference-star PSF subtraction, we extracted the first moderate-resolution spectrum of HD~19467~B between $2.9-5.3\,\mu\mathrm{m}$. The continuum S/N$\sim10$ is limited by interpolation errors in our implementation of RDI.  
    The atmospheric characterization of this brown dwarf based on this spectrum is performed in a separate work (Hoch et al.).
    \item As a proof of concept, we tested a strategy to combine the power of the forward model with that of RDI. Our proposed method would optimally combine the higher resolution spectral features with the continuum information of the companion. This approach should further improve the detection and characterization sensitivity of the NIRSpec IFU for high-contrast imaging of exoplanets.
    \item We suggest to continue the efforts to model NIRSpec IFU data directly in the detector images. Alternatively, interpolations of the data should be minimized as they are the largest source of error in this highly undersampled instrument.
    \item Improvements to the NIRSpec dithering strategy, distortion calibrations, flux calibrations, modeling of the charge diffusion, and most importantly simulated PSFs would all greatly benefit high-contrast science cases with NIRSpec.
\end{itemize}

In conclusion, the NIRSpec instrument on-board JWST is a powerful instrument to study faint planets around bright stars. It will enable the direct detection and spectral characterization of older, cooler, and more closely separated exoplanets that are similar to the solar system gas giants. 

The Python scripts developed for this work, both for data reduction and figure creation, are publicly available in a Github repository\footnote{\url{https://github.com/jruffio/HD_19467_B} commit hash b853a66} \citep{HD19467Bscripts}.

%% file: appendix.tex
\clearpage
\appendix

\section{Linear model statistic with a prior on the linear parameters}
\label{app:newmaths}

In this appendix, we explain the changes to the linear model formalism from \citet{Ruffio2019AJ....158..200R} (Appendix D) to include the regularization, meaning prior, on the linear parameters.

In \citet{Ruffio2019AJ....158..200R} Appendix D, we defined a linear model as 
\begin{equation}
    {\bm d} = {\bm M}{\bm \phi} + {\bm n},
\end{equation}
where ${\bm d}$ is the data vector of length $N$, ${\bm n}$ is a Gaussian random vector with zero mean and covariance matrix ${\bm \Sigma}$, ${\bm M}$ is a matrix representing the linear model, and ${\bm \phi}$ represents the $N_{\phi}$ linear parameters. 

The corresponding Gaussian likelihood is given by
\begin{equation}
   \mathcal{P}({\bm d} \vert {\bm \phi}) =  \frac{1}{\sqrt{(2\pi)^{N}\vert{\bm \Sigma}\vert}} \exp\left\lbrace-\frac{1}{2}({\bm d}-{\bm M}{\bm \phi})^{\top}{\bm \Sigma}^{-1}({\bm d}-{\bm M}{\bm \phi})\right\rbrace,
    \label{eq:applike}
\end{equation}

It is possible to regularize the problem by adding a Gaussian prior on the parameters ${\bm \phi}$ with expected values $\mu_\phi$ and covariance ${\bm \Sigma}_{\mathrm{reg}}$. 
The posterior including the prior is given by:
\begin{align}
     \mathcal{P}({\bm \phi} \vert {\bm d}) &\propto \mathcal{P}({\bm d} \vert {\bm \phi}) \mathcal{P}({\bm \phi}) ,\nonumber \\
     &\propto \frac{1}{\sqrt{(2\pi)^{N}\vert{\bm \Sigma}\vert}} \exp\left\lbrace-\frac{1}{2}({\bm d}-{\bm M}{\bm \phi})^{\top}{\bm \Sigma}^{-1}({\bm d}-{\bm M}{\bm \phi})\right\rbrace \times \nonumber \\
     & \frac{1}{\sqrt{(2\pi)^{N_{\phi}}\vert{\bm \Sigma}_{\mathrm{reg}}\vert}} \exp\left\lbrace-\frac{1}{2}(\mu_\phi-{\bm \phi})^{\top}{\bm \Sigma}_{\mathrm{reg}}^{-1}(\mu_\phi-{\bm \phi})\right\rbrace ,\nonumber \\
     &\propto  \frac{1}{\sqrt{(2\pi)^{N+N_{\phi}}\vert{\bm \Sigma}^{\prime}\vert}} \exp\left\lbrace-\frac{1}{2}({\bm d}^{\prime}-{\bm M}^{\prime}{\bm \phi})^{\top}{\bm \Sigma}^{\prime-1}({\bm d}^{\prime}-{\bm M}^{\prime}{\bm \phi})\right\rbrace
     \label{eq:appprobwithprior}
\end{align}
where we defined
\begin{equation}
     {\bm d}^{\prime} = 
    \begin{bmatrix}
    {\bm d} \\
    \mu_\phi 
    \end{bmatrix},
    {\bm M}^{\prime} = 
    \begin{bmatrix}
    {\bm M}   \\
     {\bm M}_{\textbf{reg}}
    \end{bmatrix},
    {\bm \Sigma}^{\prime} = 
    \begin{bmatrix}
    {\bm \Sigma} &0   \\
     0& {\bm \Sigma}_{\textbf{reg}}
    \end{bmatrix}.
\end{equation}
If the prior is defined for all the parameters ${\bm \phi}$, then ${\bm M}_{\textbf{reg}}$ is simply the identity matrix of size $N_{\phi} \times N_{\phi}$. When the prior only affect a subset of the parameters, ${\bm M}_{\textbf{reg}}$ is effectively what is called a selection matrix that needs to be adapted, which is the case in Equation \ref{eq:fm}. A selection matrix is a matrix of ones and zeros whose purpose is to select a subset of parameters from a vector.
We therefore demonstrated that the problem can be rewritten in the same form as Equation \ref{eq:applike}.
This means that the maximum a posteriori (MAP) will also be given by the pseudo inverse (see Equation 16 in \citealt{Ruffio2019AJ....158..200R}), but with updated vectors and matrices:
\begin{equation}
\label{eq:apppseudoinv}
\tilde{{\bm \phi}} = ({\bm M}^{\prime\top}{\bm \Sigma}^{\prime-1}{\bm M}^{\prime})^{-1}{\bm M}^{\prime\top}{\bm \Sigma}^{\prime-1}{\bm d}^{\prime}
\end{equation}

An important element of the analysis is the derivation of the errors on the linear parameters, which is given by the covariance of $\tilde{{\bm \phi}}$.
However, the derivation of $\textrm{cov}(\tilde{{\bm \phi}})$ needs to be updated compared to Equation 23 in \citet{Ruffio2019AJ....158..200R}. Following the method in \citet{Ruffio2019AJ....158..200R} Appendix D.3, we can replace ${\bm d}^{\prime}$ in Equation \ref{eq:apppseudoinv} as

\begin{equation}
\tilde{{\bm \phi}} = ({\bm M}^{\prime\top}{\bm \Sigma}^{\prime-1}{\bm M}^{\prime})^{-1}{\bm M}^{\prime\top}{\bm \Sigma}^{\prime-1}\begin{bmatrix}
    {\bm M}{\bm \phi} + {\bm n} \\
    \mu_\phi 
    \end{bmatrix},
\end{equation}
The only random variable is ${\bm n}$ and everything else is constant, therefore,
\begin{equation}
\tilde{{\bm \phi}} = \mathrm{cst} + ({\bm M}^{\prime\top}{\bm \Sigma}^{\prime-1}{\bm M}^{\prime})^{-1}{\bm M}^{\top}{\bm \Sigma}^{-1}{\bm n}.
\end{equation}

In order to calculate the covariance of $\tilde{{\bm \phi}}$, we rely on the fact that the linear transformation of a Gaussian random vector is also a Gaussian random vector. Indeed, 
if ${\bm X}\sim\mathcal{N}({\bm \mu_x},{\bm \Sigma_x})$ is a Gaussian random vector, ${\bm A}$ a matrix, then ${\bm Y}={\bm A}{\bm X}$ also follows a normal distribution with vector mean,
\begin{equation}
    {\bm \mu_y} = {\bm A}{\bm \mu_x}
\end{equation}
and covariance matrix,
\begin{equation}
    {\bm \Sigma_y} = {\bm A}{\bm \Sigma_x}{\bm A}^{\top}.
\end{equation}

This means that the covariance matrix of $\tilde{{\bm \phi}}$ with a prior is given by,
\begin{align}
    \textrm{cov}(\tilde{{\bm \phi}}) &= \left[ ({\bm M}^{\prime\top}{\bm \Sigma}^{\prime-1}{\bm M}^{\prime})^{-1}{\bm M}^{\top}{\bm \Sigma}^{-1} \right]{\bm \Sigma}\left[ ({\bm M}^{\prime\top}{\bm \Sigma}^{\prime-1}{\bm M}^{\prime})^{-1}{\bm M}^{\top}{\bm \Sigma}^{-1} \right]^{\top}, \nonumber \\
    &=({\bm M}^{\prime\top}{\bm \Sigma}^{\prime-1}{\bm M}^{\prime})^{-1}{\bm M}^{\top}{\bm \Sigma}^{-1}
    {\bm \Sigma} {\bm \Sigma}^{-1} {\bm M}({\bm M}^{\prime\top}{\bm \Sigma}^{\prime-1}{\bm M}^{\prime})^{-1}, \nonumber \\
    \textrm{cov}(\tilde{{\bm \phi}}) &=({\bm M}^{\prime\top}{\bm \Sigma}^{\prime-1}{\bm M}^{\prime})^{-1}{\bm M}^{\top}{\bm \Sigma}^{-1}{\bm M}({\bm M}^{\prime\top}{\bm \Sigma}^{\prime-1}{\bm M}^{\prime})^{-1}.
\label{eq:appcovphi}
\end{align}
We used the fact that the inverse of symmetric matrix is also symmetric. This formula cannot generally be simplified further, unlike the simpler Equation 23 in \citep{Ruffio2019AJ....158..200R} that was derived without a prior:
\begin{equation}
\textrm{cov}(\tilde{{\bm \phi}}) = ({\bm M}^{\top}{\bm \Sigma}^{-1}{\bm M})^{-1}.
\end{equation}

% We can also decompose ${\bm M}^{\prime\top}{\bm \Sigma}^{\prime-1}{\bm M}^{\prime}$ as
% \begin{equation}
%     {\bm M}^{\prime\top}{\bm \Sigma}^{\prime-1}{\bm M}^{\prime} = {\bm M}^{\top}{\bm \Sigma}^{-1}{\bm M}+{\bm M}_{\mathrm{reg}}^{\top}{\bm \Sigma}_{\mathrm{reg}}^{-1}{\bm M}_{\mathrm{reg}}
% \end{equation}
% Using that fact that ${\bm M}_{\mathrm{reg}}$ is only a selection matrix of zeros and ones, we note that ${\bm M}_{\mathrm{reg}}^{\top}{\bm \Sigma}_{\mathrm{reg}}^{-1}{\bm M}_{\mathrm{reg}}$ is simply resizing ${\bm \Sigma}_{\mathrm{reg}}$ to match the dimensions of ${\bm \Sigma}^{-1}$ while placing the prior covariance elements in their matching position. This means that if ${\bm \Sigma}_{\mathrm{reg}}$ is diagonal, which is usually the case, then ${\bm M}_{\mathrm{reg}}^{\top}{\bm \Sigma}_{\mathrm{reg}}^{-1}{\bm M}_{\mathrm{reg}}$ is also a diagonal matrix.

Finally, there is another deviation from \citet{Ruffio2019AJ....158..200R} required in this work for the calculation of the posterior of the non-linear parameters ${\bm \psi}$, which can be the planet position, radial velocity, atmospheric parameters, etc.
\citet{Ruffio2019AJ....158..200R} was able to marginalize the posterior over the linear parameters ${\bm \phi}$ as well as a noise scaling factor (called $s$ therein) to account for inaccurate data errors. The regularization of the linear model does not allow a similar marginalization over the noise scaling factor, but we do keep the most important marginalization over the linear parameters. However, not including the marginalization does not mean that we cannot empirically normalize the noise based on the residuals. We note that the covariance of the linear parameter was never marginalized over the noise scaling parameter. It is always possible to scale the data covariance ${\bm \Sigma}$ to ensure that the data errors are empirically scaled to the amplitude of the residuals of the fit.

Similarly to \citet{Ruffio2019AJ....158..200R} and writing ${\bm M}^{\prime} = {\bm M}^{\prime}_{{\bm \psi}}$, we then write the marginalized posterior of ${\bm \psi}$ as,
\begin{align}
    \mathcal{P}({\bm \psi} \vert {\bm d}) &=\int_{{\bm \phi}} \mathcal{P}({\bm \psi}, {\bm \phi}\vert {\bm d}),\nonumber \\
    &=  \frac{1}{\mathcal{P}({\bm d})} \int_{{\bm \phi}} \mathcal{P}({\bm d} \vert {\bm \psi}, {\bm \phi})\mathcal{P}( {\bm \psi}, {\bm \phi}),\nonumber \\
    &=  \frac{\mathcal{P}( {\bm \psi})}{\mathcal{P}({\bm d})} \int_{{\bm \phi}} \mathcal{P}({\bm d} \vert {\bm \psi}, {\bm \phi})\mathcal{P}( {\bm \phi}),\nonumber \\
    \mathcal{P}({\bm \psi} \vert {\bm d}) &\propto  \mathcal{P}( {\bm \psi})
     \sqrt{\frac{\left\vert\textrm{cov}(\tilde{{\bm \phi}})\right\vert}{\vert{\bm \Sigma}^{\prime}\vert }}
    \exp\left\lbrace-\frac{1}{2} \chi^{\prime 2}_{{\bm \phi}=\tilde{{\bm \phi}},{\bm \psi}} \right\rbrace,
    \label{eq:apppost}
\end{align}
where $\textrm{cov}(\tilde{{\bm \phi}})$ is given by Equation \ref{eq:appcovphi}, and 
\begin{equation}
    \chi^{\prime 2}_{{\bm \phi}=\tilde{{\bm \phi}},{\bm \psi}} = ({\bm d}^{\prime}-{\bm M}^{\prime}\tilde{{\bm \phi}})^{\top}{\bm \Sigma}^{\prime-1}({\bm d}^{\prime}-{\bm M}^{\prime}\tilde{{\bm \phi}})
\end{equation}
We used Equation \ref{eq:appprobwithprior} in this Appendix combined with Equation 36 in \citet{Ruffio2019AJ....158..200R}.
The determinant of the covariance $\left\vert\textrm{cov}(\tilde{{\bm \phi}})\right\vert $ can suffer from numerical instabilities. In those cases, we found that approximating $\textrm{cov}(\tilde{{\bm \phi}})$ by a diagonal matrix such that $\mathrm{Ln}\left\vert\textrm{cov}(\tilde{{\bm \phi}}) \right\vert  \sim  \sum_i \mathrm{Ln} \left[ \textrm{cov}(\tilde{{\bm \phi}})[i,i] \right]$ can provide a reasonable alternative. It effectively neglects the correlation between the linear parameters. 

\begin{figure*}
  \centering
  \includegraphics[trim={0cm 0cm 0cm 0cm},clip,width=0.5\linewidth]{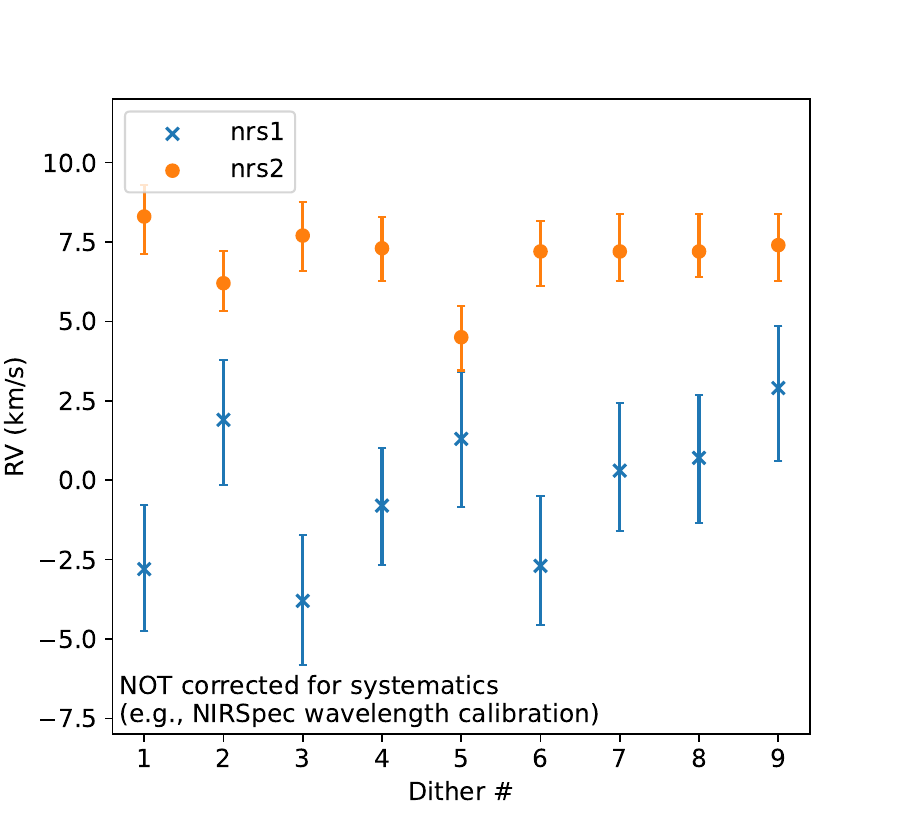}
  \caption{Radial velocity measurements of HD~19467~B for individual exposures. At the time of writing, systematics in the NIRSpec wavelength solution has not been characterized or calibrated at this level of precision, which could explain the discrepancy between the NRS1 and NRS2 detectors.}
  \label{fig:HD19467B_RVs}
\end{figure*}

We test Equation \ref{eq:apppost} with this approximation in Figure \ref{fig:HD19467B_RVs} where we fit for the radial velocity of HD~19467~B using the forward model from Section \ref{sec:FM} with a BT-Settl atmospheric model ( $T_{\mathrm{eff}}=940\,\mathrm{K}$ and $\mathrm{log}(g)=5.0$) for the companion \citep{Allard2003IAUS..211..325A}. For the NRS1 detector, the scatter appear in relative good agreement with the estimated error bars from the RV posterior. The error bars appear higher than the exposure-to-exposure scatter for NRS2. We think that this is due to the residuals being dominated by model systematics.
% https://jwst-docs.stsci.edu/jwst-data-calibration-considerations/jwst-calibration-uncertainties#JWSTCalibrationUncertainties-Wavelengthcalibration.2

\section{NIRCam sensitivity}
\label{app:NIRCamContrast}

\begin{figure*}
  \centering
  \includegraphics[trim={0cm 0cm 0cm 0cm},clip,width=0.5\linewidth]{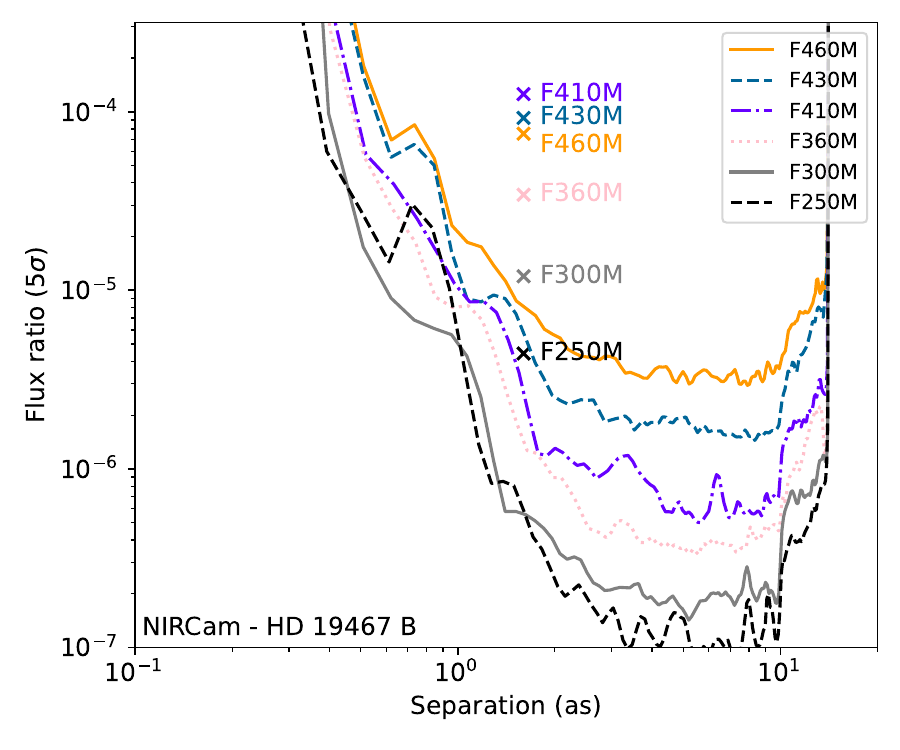}
  \caption{Revised companion detection sensitivity of the NIRCam coronographic data originally presented in \citet{Greenbaum2023}. The $5\sigma$ detection limits are defined in terms of the companion-to-star flux ratio in different NIRCam photometric filters. The crosses indicate the flux ratio of HD 19467 B in the various filters.}
  \label{fig:NIRCam_contrast}
\end{figure*}

We re-reduced the NIRCam observations of HD 19467 B that were presented in \citet{Greenbaum2023}. The companion sensitivity curves for each NIRCam filter are shown at $5\sigma$ in Figure \ref{fig:NIRCam_contrast} including small sample statistics correction \citep{Mawet2014ApJ...792...97M}. Note that the S/N derived from the statistical uncertainty of the photometry in Table~\ref{tab:photometry} is slightly better than the S/N read directly from Figure \ref{fig:NIRCam_contrast} because the photometry was estimated with a forward model of the PSF \citep{Pueyo2016} while the detection limits used simple aperture photometry to speed up the computation.

\section{Background subtraction for NIRSpec IFU rate files}
\label{app:cleanrate}

\begin{figure*}[ht]
  \centering
  \includegraphics[trim={0cm 0cm 0cm 0cm},clip,width=1\linewidth]{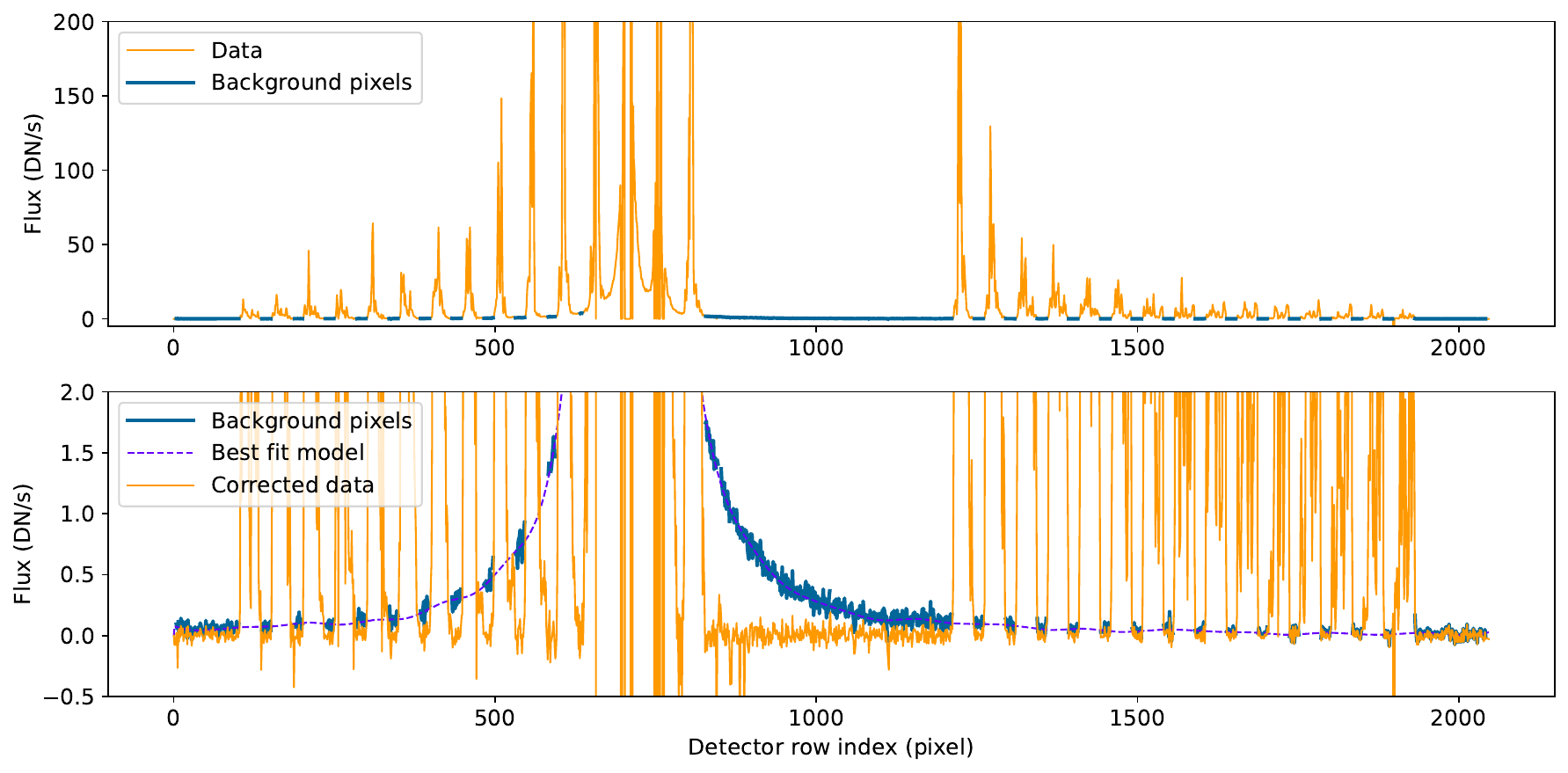}
  \caption{Removal of the background flux in NIRSpec rate maps, in particular the diffuse background created by charge transfer from the very highly saturated stellar PSF core. This illustrates a single column of the NRS1 detector (index 300 of \texttt{jw01414004001\_02101\_00001\_nrs1\_rate.fits}). The region of the detector corresponding to illuminated IFU slice data is masked. The spline continuum model (``Best fit model'') is fit to the ``background pixels'' only and then subtracted from the data (``Corrected data''). Note the two panels above show the same data but with the Y axis ranges different by a factor of 100$\times$. The resulting image is shown in Figure \ref{fig:ratecleanimages}.}
  \label{fig:rateclean}
\end{figure*}

\begin{figure*}[ht]
  \centering
  \includegraphics[trim={0cm 0cm 0cm 0cm},clip,width=1\linewidth]{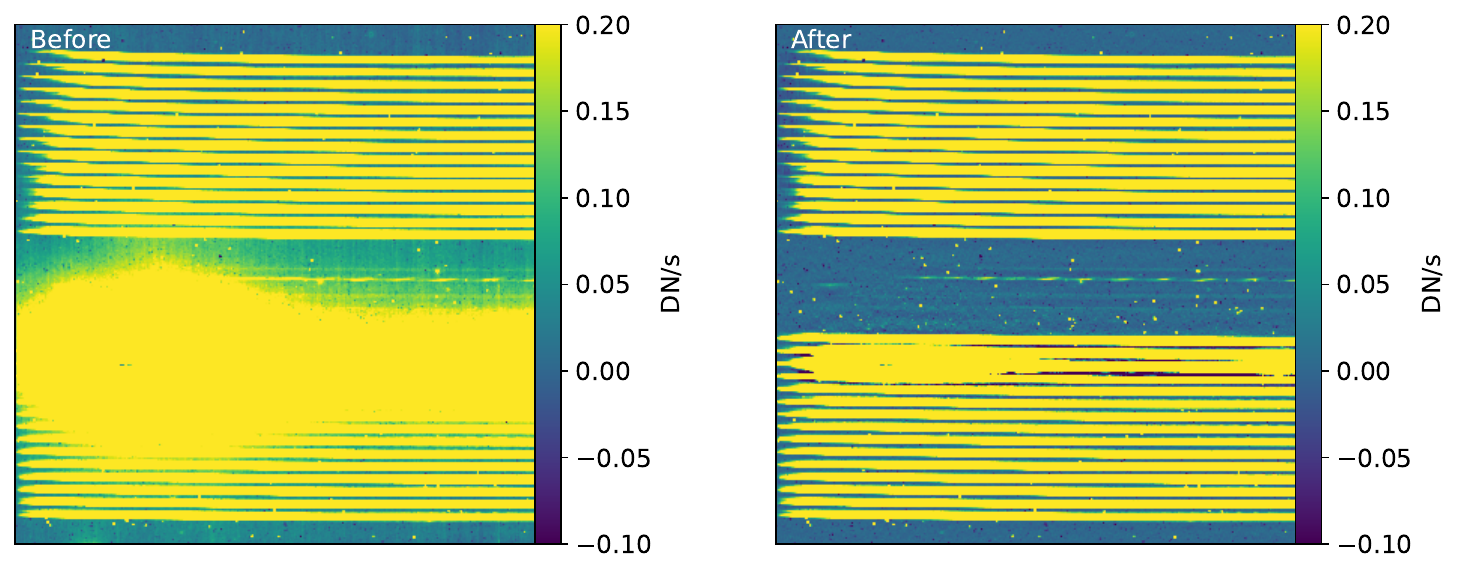}
  \caption{NIRSpec IFU rate map before and after background subtraction for \texttt{jw01414004001\_02101\_00001\_nrs1\_rate.fits}. Both vertical striping from 1/f noise and the diffuse signal excess extending away from the very saturated stellar PSF core are effectively removed. }
  \label{fig:ratecleanimages}
\end{figure*}

We added an intermediate step between the stage 1 (uncal to rate) and stage 2 (rate to cal) pipeline reduction to remove the vertical striping in NIRSpec images (1/f noise) as well as the additional diffuse charges caused by the saturated central star.
The method is conceptually similar to the NSClean algorithm \citep{Rauscher2024PASP..136a5001R}, although we implemented our own model of the background flux using the spline model used throughout this work (see Section \ref{sec:contnormspec}). We fit the rate maps column by column. In each column, we mask the pixels that belong to an IFU slice. We then fit a spline to the background pixels using 40 nodes equidistant across the 2048 pixels. The background model is then subtracted as illustrated in Figure \ref{fig:rateclean}. We manage bad pixels by fitting the model twice and removing the outliers larger than $5\times$ the median absolute deviation column-wise in the residuals after the first fit. The before and after rate maps are shown in Figure \ref{fig:ratecleanimages}. The new rate maps are then processed normally with the stage 2 JWST calibration pipeline to produce cleaned flux calibrated images. 

We note in passing that this level of charge transfer in extremely saturated NIRSpec data has not previously been discussed in detail, to the best of our knowledge. This is the detector-level view of the effect that creates the bright signal throughout the IFU slices containing the stellar PSF core (e.g. as in Figure 7 of \citealp{Boker2022}), and indeed we see that the charge transfer away from the saturated region extends far outside of just a single IFU slice. In this case an increased signal level can be seen extending for \textit{hundreds of pixels} on either side of the rows containing the highly saturated stellar PSF core. This is well beyond the typical subtle changes in PSF properties typically referred to as the ``brighter-fatter effect''.
We remind readers that the host star HD 19467 is a $K=5.4$ magnitude star, extremely bright compared to typical NIRSpec targets. Different physical effects within the detector may be responsible for this large-scale charge transfer effect distinct from the usual brighter-fatter effect. In particular we speculate that this level of extreme oversaturation may be sufficient to drive the semiconductor diodes within affected pixels into the forward bias regime. Similar effects have previously been produced intentionally in detector ground testing of a NIRSpec flight candidate detector (E. Bergeron, private communication). That mechanism may also help explain charge transfer effects seen in exceptionally saturated NIRCam images, for instance deep images of Mars and Jupiter far beyond saturation in programs 1373 and 2787.  That said, the particular details of the detector physics are not necessary for the spline model to empirically fit the resulting diffuse background and allow its subtraction.

\section{Injection and recovery test for the forward model}
\label{app:injectionrecovery}

\begin{figure*}
  \centering
  \includegraphics[trim={0cm 0cm 0cm 0cm},clip,width=0.5\linewidth]{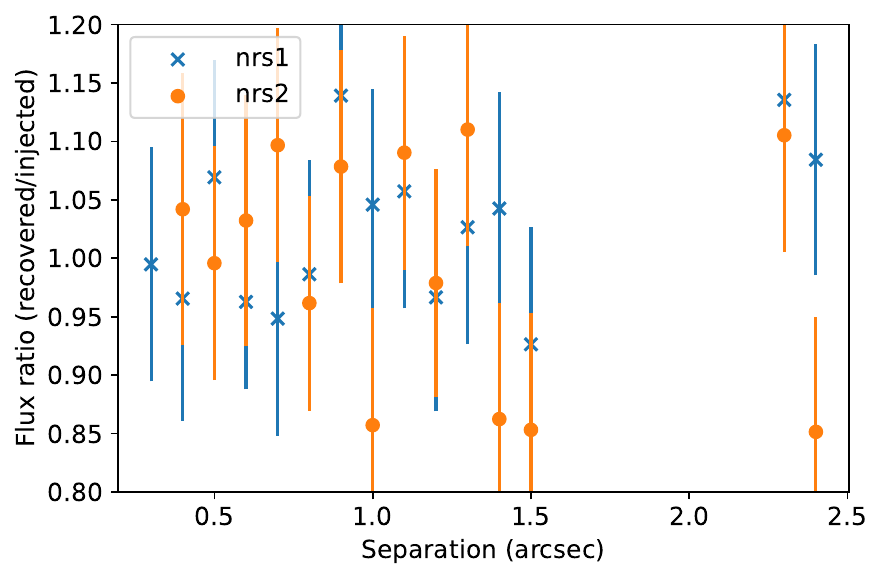}
  \caption{Simulated companion injection and recovery test for the forward model in Section \ref{sec:FM}. The recovered fluxes are consistent with the original simulated companion showing no significant signs of self or over subtraction in the algorithm.  }
  \label{fig:injectionrecovery}
\end{figure*}

Speckle subtraction algorithms for high-contrast imaging often suffer from over- and self-subtraction, which can bias the detection limits when they are not corrected for. This is usually done with an injection and recovery test of simulated companions with known fluxes. The ratio between the recovered flux after speckle subtraction and the injected flux is the algorithmic throughput.
We test the flux extraction of the forward model by injecting a simulated companion signal in a single exposure of both NIRSpec detectors using the simulated PSF from WebbPSF and the RDI spectrum of HD~19467~B in a similar fasion to Section \ref{sec:compmodel}. The fluxes of the simulated companions are chosen to be S/N$\sim10$. The ratios of the recovered fluxes to the injected one is shown in Figure \ref{fig:injectionrecovery}.

\section{Covariance matrix estimation for the RDI spectrum of HD19467B}
\label{app:speccov}

In Section~\ref{sec:FMRDI}, we discussed a proof of concept for atmospheric inference using the detector forward model and a RDI-based prior on the speckle intensity to constrain the continuum. While such forward modeling is statistically the most accurate way to model the data, it is also a lot more complicated than fitting a 1D spectrum. However, correlated noise in the 1D spectrum from the imperfect speckle subtraction can significantly bias atmospheric inference if it is not accounted for with a covariance in the definition of the $\chi^2$ \citep{Greco2016ApJ...833..134G}. In this section, we therefore compute an semi-empirical noise covariance for the RDI spectrum of HD~19467~B extracted in Section~\ref{sec:RDIextract} and illustrated in Figure~\ref{fig:summaryRDIspec}.
 
\begin{figure}[h]
    \centering
    \includegraphics[width=\textwidth]{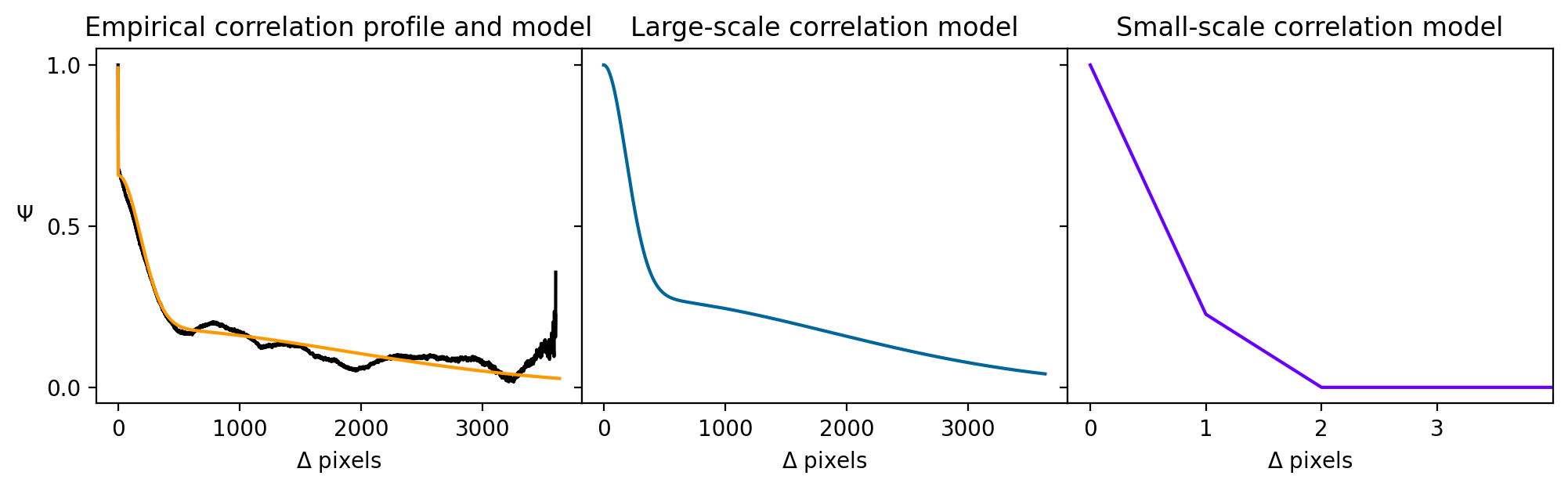}
    \caption{Modeling the correlation of the noise for the RDI spectrum of HD~19467~B shown in Figure~\ref{fig:summaryRDIspec}. Left: The empirical correlation profile in black and the model fit to this profile in orange. Middle: the double Gaussian model for the correlated component of the empirical profile. Right: The model for the uncorrelated noise component which is only nonzero in the first two pixels, note the scaling on the x-axis.}
    \label{fig:empirical_correlation}
\end{figure}

The covariance is estimated from the correlation of the residual speckle spectra in an annulus from $0.4$--$0.5^{\prime\prime}$ in radius from the companion. We construct the covariance assuming that the correlation between two spectral bin is only a function of the wavelength difference. 
Due to difficulties with inverting a noisy covariance matrix constructed directly with the empirical correlation profile, we chose to model the correlation profile with a functional form before constructing the covariance matrix and inverting it.
The empirical auto-correlation is computed as follow. 
First, the speckle spectra are normalized by the flux uncertainty, so that the standard deviation over all positions at each wavelength is equal to one. Then, the auto-correlation profile is calculated as the average over wavelengths and spatial positions of the sliding window product over wavelengths of the normalized speckle spectra. So, if $r_{x,l}$ is the normalized residual speckle spectra at location $x$ and wavelength $l$, then the empirical speckle correlation $\Psi$ for wavelength bins $\Delta l$ pixels apart is calculated as
\begin{equation}
    \Psi_{\Delta l} = \langle r_{x,l} r_{x,l+\Delta l} \rangle_{x,l},
\end{equation}
where the average $\langle ... \rangle$ is over spatial positions $x$ and wavelengths $l$ and NaNs are implicitly ignored in the sliding window average by using \texttt{numpy.nanmean}. 
We identify four different regimes in the auto-correlation that feature correlation at different scales, which are illustrated in Figure~\ref{fig:empirical_correlation}. The white noise fraction of $\Psi_{\Delta l = 0}$ is the result of the irreducible noise sources: photon noise, read noise, etc. We then identify correlation between neighboring pixels ($\Psi_{\Delta l <=1}$) which we attribute to the interpolation of the detector images on a regular wavelength grid. Improperly subtracted speckles lead to correlations on scales corresponding to $\Psi_{\Delta l <=500}$, and also identify a trend-like correlation across the entire spectrum. We refer to the white noise and interpolation effects as the ``small-scale'' correlation ($\Psi_{\mathrm{small}}$), and the residual speckle effects as ``large-scale'' correlation ($\Psi_{\mathrm{large}}$). The latter is modeled with a double Gaussian in $\Delta l$ to account for correlated noise at the different scales. Both $\Psi_{\mathrm{small},\Delta l=0}$ and $\Psi_{\mathrm{small},\Delta l=0}$ are normalized to unity. The empirical profile for the correlation over $\Delta l $ and the model which are fit to this profile are shown in Figure (\ref{fig:empirical_correlation}).

The final covariance matrix is the combination of the small-scale and large-scale covariances. In Section~\ref{sec:RDIextract}, we computed the respective standard deviation contribution of each scale at each wavelength which we label as $\sigma_{\mathrm{small},l}$ and $\sigma_{\mathrm{large},l}$, which are illustrated in Figure~\ref{fig:summaryRDIspec}. We therefore define the covariance as
\begin{equation}
    \mathrm{Cov}(l_1,l_2) = (\Psi_{\mathrm{small},|l_2-l_1|}\sigma_{\mathrm{small},l_1}\sigma_{\mathrm{small},l_2}) + (\Psi_{\mathrm{large},l_2-l_1}\sigma_{\mathrm{large},l_1}\sigma_{\mathrm{large},l_2}).
\end{equation}

\begin{figure}[h]
    \centering
    \includegraphics[width=\textwidth]{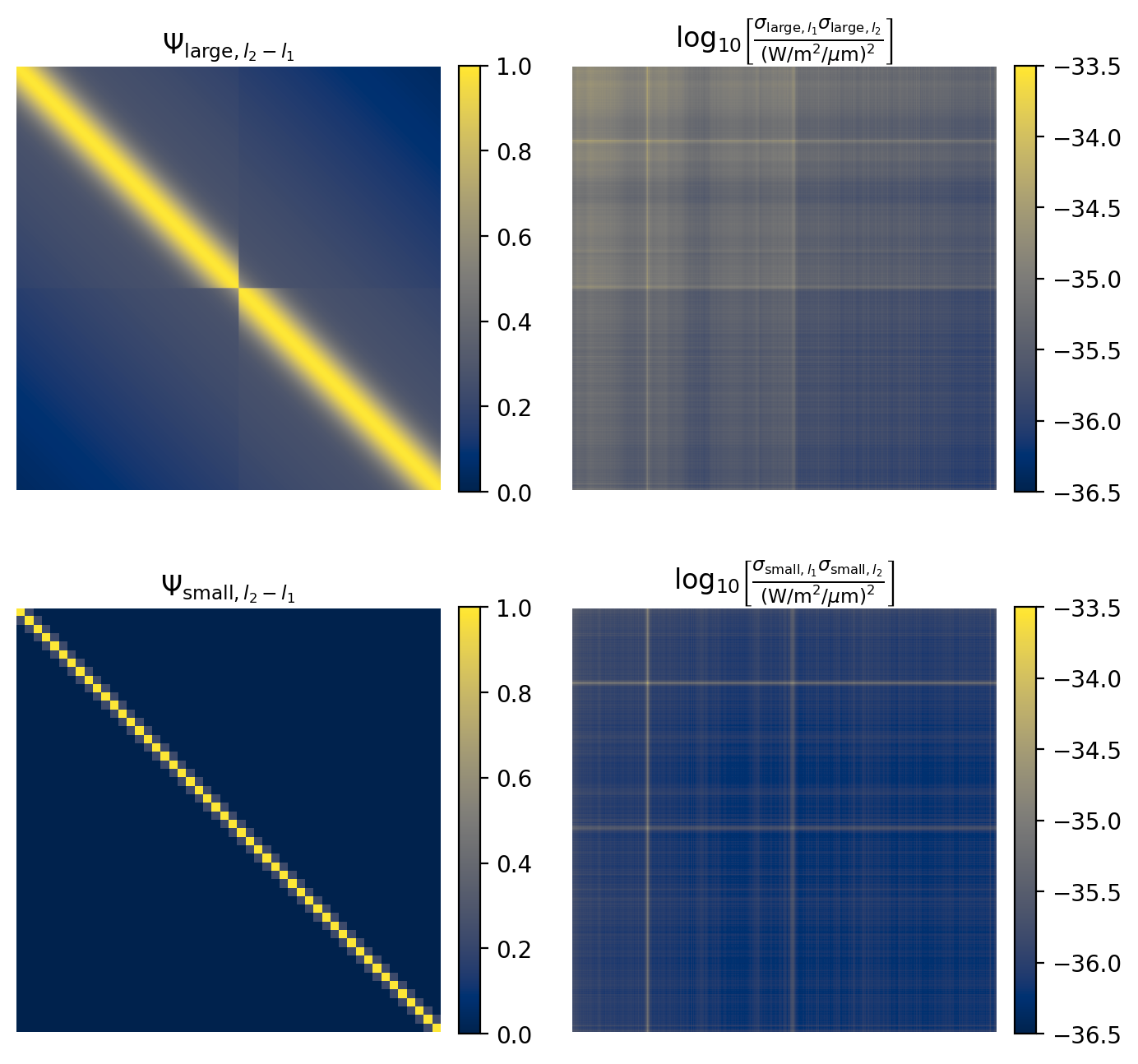}
    \caption{Definition of the model covariance. Top: the large-scale 2D correlation model ($\Psi_{\mathrm{large},l_2-l_1}$) and the outer product of the large-scale errors ($\sigma_{\mathrm{large},l_1}\sigma_{\mathrm{large},l_2}$). Bottom: the small-scale 2D correlation model ($\Psi_{\mathrm{small},l_2-l_1}$) and the outer product of the small-scale errors ($\sigma_{\mathrm{small},l_1}\sigma_{\mathrm{small},l_2}$). The small-scale correlation matrix only shows a 50x50 pixel area so that the fine structure along the diagonal is visible.
    The final covariance is shown in Figure~\ref{fig:covariance_inverse}.}
    \label{fig:covariance_model}
\end{figure}

The two dimensional covariance models and the outer product of the two noise profiles ($\sigma_{l_1}\sigma_{l_2}$) are shown in Figure (\ref{fig:covariance_model}).

\begin{figure}[h]
    \centering
    \includegraphics[width=\textwidth]{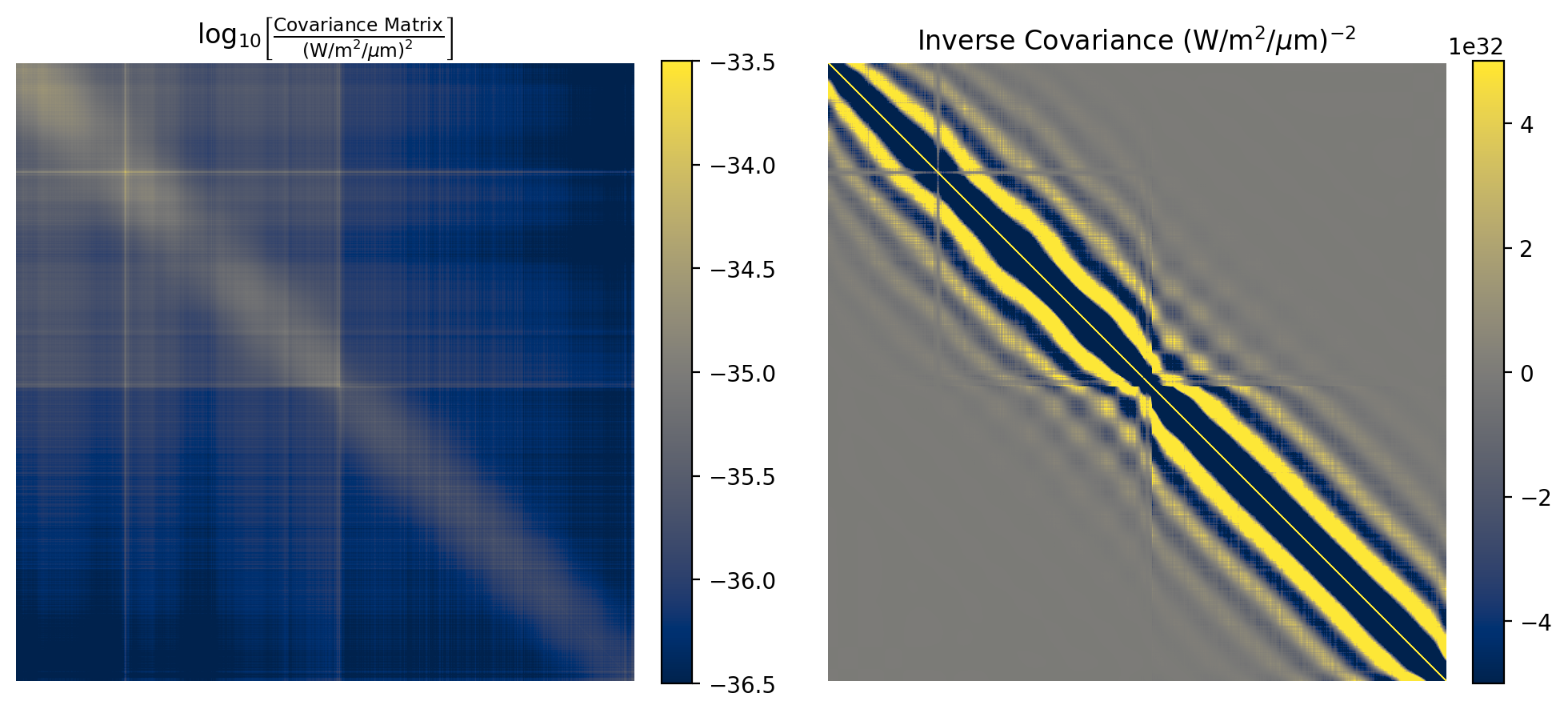}
    \caption{The final covariance matrix and its inverse. The logarithmic scale is used on the left to show fine structure but cannot used on the right due to negative values in the inverse. The code to generate these matrices is available in the Github repository dedicated to this work. }
    \label{fig:covariance_inverse}
\end{figure}

The inverse of this matrix is computed using \texttt{numpy.linalg.inv}. Both of these matrices are shown in Figure (\ref{fig:covariance_inverse}). The covariance can be obtained by running a single python notebook in the Github repository dedicated to this work\footnote{\url{https://github.com/jruffio/HD_19467_B}}. There is also an example for calculating the $\chi^{2}$ of an atmospheric model.

\begin{figure}[h]
    \centering
    \includegraphics[width=\textwidth]{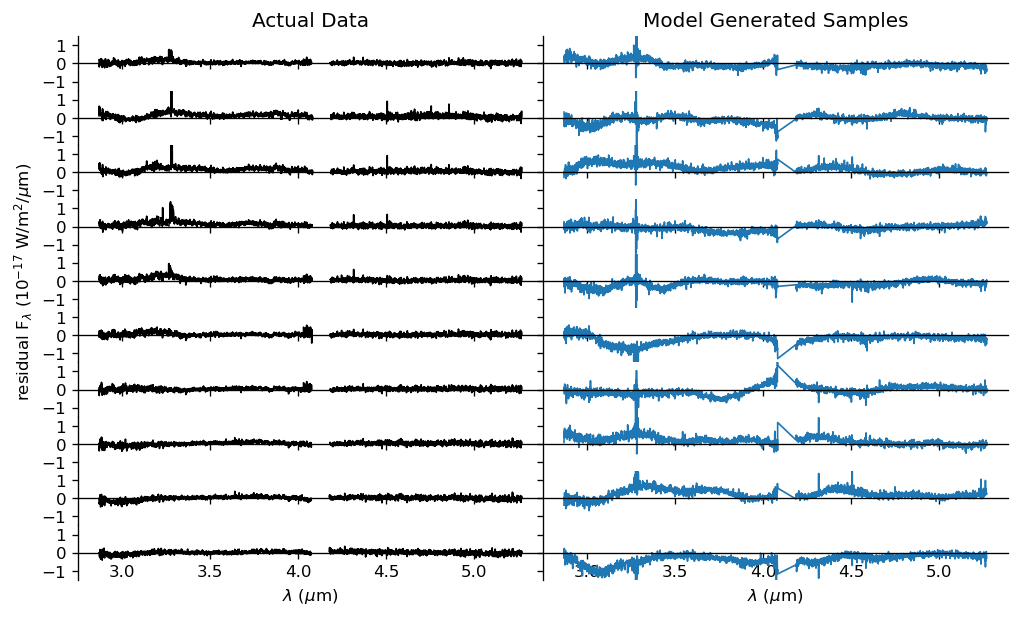}
    \caption{Visual verification of the model covariance. Left: 10 examples of residual speckle spectra used to calculated the empirical profile of the correlation. Right: samples of residual speckles generated by querying the model covariance matrix for comparison.}
    \label{fig:generated_speckle_spectra}
\end{figure}

In order to validate this model covariance matrix, we generate samples from it using \texttt{numpy}. In Figure (\ref{fig:generated_speckle_spectra}) we show an example of 10 random extracted speckle spectra which are used to generate the model, and 10 random samples generated from the model covariance. It can be seen that the general statistical behavior of the speckles is reproduced, although there is some minor discrepancies between the speckle statistics in each detector.

